\documentclass[a4paper,fleqn]{cas-sc}

\usepackage{subcaption}
\usepackage{graphicx}
\usepackage{doi}
\usepackage{multirow}
\usepackage{amssymb}
\usepackage{amsmath}
\usepackage{amsthm}
\usepackage{amssymb}
\usepackage{placeins}
\usepackage{xcolor}
\usepackage{hyperref}
\usepackage[numbers]{natbib} 

\def\tsc#1{\csdef{#1}{\textsc{\lowercase{#1}}\xspace}}
\tsc{WGM}
\tsc{QE}
\tsc{EP}
\tsc{PMS}
\tsc{BEC}
\tsc{DE}

\begin{document}
\let\WriteBookmarks\relax
\def\floatpagepagefraction{1}
\def\textpagefraction{.001}

\shorttitle{State-Space Dynamic Functional Regression for Multicurve}

\shortauthors{Peilun He et~al.}

\title [mode = title]{State-Space Dynamic Functional Regression for Multicurve Fixed Income Spread Analysis and Stress Testing}                      
\tnotemark[1]

\tnotetext[1]{The original bond yield data was obtained from \url{https://www.tradingview.com/}. A sample data and the codes are available on GitHub at: \url{https://github.com/peilun-he/State-Space-Dynamic-Functional-Regression-for-Multicurve}.}


%
\author[1]{Peilun He}[orcid=0000-0002-2740-3390]



\ead{peilun.he@mq.edu.au}


\credit{Software, Validation, Investigation, Formal analysis, Writing - Original Draft, Writing - Review \& Editing, Visualization}

\affiliation[1]{organization={Department of Actuarial Studies and Business Analytics, Macquarie University},
    city={Macquarie Park},
    citysep={}, 
    postcode={2109}, 
    state={NSW},
    country={Australia}}

\author[2]{Gareth W. Peters}[orcid=0000-0003-2768-8979]

\ead{garethpeters@ucsb.edu}

\ead[url]{https://www.qrslab.com/}

\credit{Conceptualization, Investigation, Methodology, Formal Analysis, Software, Writing - Original Draft, Writing - Review \& Editing, Supervision}

\affiliation[2]{organization={Department of Statistics and Applied Probability, University of California Santa Barbara},
    city={Santa Barbara},
    citysep={}, 
    postcode={93106}, 
    state={CA},
    country={United States}}

\author[3]{Nino Kordzakhia}[orcid=0000-0002-7853-4550]

\ead{nino.kordzakhia@mq.edu.au}


\credit{Software, Investigation, Methodology, Writing - Review \& Editing, Supervision}

\affiliation[3]{organization={School of Mathematical and Physical Sciences, Macquarie University},
    city={Macquarie Park},
    citysep={}, 
    postcode={2109}, 
    state={NSW},
    country={Australia}}

\author[1]{Pavel V. Shevchenko}[orcid=0000-0001-8104-8716]

\cormark[1]


\ead{pavel.shevchenko@mq.edu.au}


\credit{Investigation, Methodology, Writing - Review \& Editing, Supervision}

\cortext[cor1]{Corresponding author}



\begin{abstract}
    The Nelson-Siegel model is widely used in fixed income markets to produce yield curve dynamics. The multiple time-dependent parameter model conveniently addresses the level, slope, and curvature dynamics of the yield curves. In this study, we present a novel state-space functional regression model that incorporates a dynamic Nelson-Siegel model and functional regression formulations applied to multi-economy setting. This framework offers distinct advantages in explaining the relative spreads in yields between a reference economy and a response economy. To address the inherent challenges of model calibration, a kernel principal component analysis is employed to transform the representation of functional regression into a finite-dimensional, tractable estimation problem. A comprehensive empirical analysis is conducted to assess the efficacy of the functional regression approach, including an in-sample performance comparison with the dynamic Nelson-Siegel model. We conducted the stress testing analysis of yield curves term-structure within a dual economy framework. The bond ladder portfolio was examined through a case study focused on spread modelling using historical data for US Treasury and UK bonds.
\end{abstract}




\maketitle

\section{Introduction}

In the multi-economy environment, bond yields serve as barometers of global economic conditions, providing invaluable insights into the fixed income markets of specific economies, risk assessment, investment valuations, and income generation. In the insurance sector, interest rates play a pivotal role in the valuation of life insurance policies \cite{gaillardetz2008valuation, bernard2005market}, annuities \cite{fontana2023valuation, gunther2024analyzing}, and the optimisation of investment strategies \cite{han2017optimal, wang2018robust, peng2023optimal}. A significant challenge for any model used in this field is the inverted bond yield curve, where short-term bonds offer higher yields than long-term bonds. The tendency of the yield curve to invert indicates an economic recession. Therefore, understanding the factors affecting adverse movements in bond yields is crucial for making informed decisions.

The study of the term structure of interest rates has evolved significantly over several decades, with early efforts focusing on two primary classes of approaches. The first class comprises no-arbitrage models, exemplified in \cite{hull1990pricing} and \cite{heath1992bond}, while the second class includes equilibrium term structure models, as in \cite{vasicek1977equilibrium}, \cite{cox1985theory} and \cite{duffie1996yield}. A good overview of interest rate models can be found in the monograph by Cairns \cite{cairns2004interest}. 

In \cite{dobbie1978ft}, Dobbie and Wilkie introduced an exponential model in which the yields were modelled as a sum of two exponentials. This class of exponential models became widely used in fixed income markets. In \cite{nelson1987parsimonious}, Nelson and Siegel introduced a new model, influenced by a polynomial regression approach on a grid, addressing the level, slope, and curvature of the yield curve. Svensson \cite{svensson1994estimating} expanded upon Nelson and Siegel’s model by introducing the fourth term and a new loading parameter for enhanced flexibility. Then Cairns \cite{cairns1998descriptive} further refined these models by introducing additional exponential terms, reducing the risk of catastrophic jumps seen in the models studied in \cite{dobbie1978ft} and \cite{svensson1994estimating} due to multiple maxima. This extended model has been applied to UK and Germany bond in \citep{cairns1998descriptive} and \citep{cairns2001stability}, respectively. Furthermore, in \cite{diebold2006forecasting}, Diebold and Li took the Nelson-Siegel model a step further by incorporating AR(1) dynamics, resulting in accurate out-of-sample forecasts of US Treasury yields over long horizons. Next, Diebold et al. \cite{diebold2006macroeconomy} identified the three key macroeconomic variables, manufacturing capacity utilisation, the federal funds rate, and annual price inflation, establishing a crucial link between macroeconomy and yield curves which reversely affect these three key variables. In \cite{diebold2008global}, Diebold et al. further refined the interconnection of the key macroeconomic variables with the bond yields in a multi-economy framework accommodating the global and country-specific factors. In \cite{koopman2010analyzing}, Koopman et al. introduced a time-varying loading parameter as the fourth hidden factor and incorporated a generalised autoregressive conditional heteroscedasticity (GARCH) dynamics for measurement error terms. In \cite{yu2011forecasting}, Yu and Zivot applied the dynamic Nelson Siegel model to US Treasury and corporate yields, demonstrating the superior out-of-sample forecasting performance of the model with AR(1) factors. In \cite{andersson2013optimal}, Andersson and Lager{\aa}s considered a portfolio of rolling horizon bonds with fixed time to maturity, where the driving stochastic processes are generalised Ornstein–Uhlenbeck processes. For an investment portfolio, the authors discussed the performance of Nelson Siegel model relative to the affine model, even though the Nelson Siegel model is not arbitrage-free \cite{filipovic1999note}. The Nelson-Siegel model and its extensions are thoroughly discussed in \cite{diebold2013yield}. 

The limitations of the Nelson-Siegel family of models have been addressed in the literature. Firstly, the lack of accuracy of yields for the long maturity bonds, e.g. with maturities over 10 years. In \cite{christensen2009arbitrage} and \cite{dubecq2011analysis}, it was shown that the additional factors are required for improving the fit for long maturity bonds' yields. Secondly, the Nelson-Siegel models do not account for the interdependencies between different economies. However, for the dynamic Nelson-Siegel state-space model, a stochastic probabilistic principal component analysis (PCA) set of regression factors, that accounted for global economic, financial and inter-country relationships, was considered in \cite{toczydlowska2018financial}.

In this paper, we propose a novel parametric model that combines elements of the dynamic Nelson-Siegel model with functional data analysis. Specifically, we represent the relative spread between reference and response countries’ yield curves, respectively, through a functional regression framework. To overcome the challenges associated with parameter estimation in this model, we employ a kernel principal component analysis (kPCA) method, which transforms the functional regression model calibration into a finite-dimensional estimation problem. Relative to traditional principal component analysis (PCA) and functional principal component analysis (FPCA), kPCA offers distinct advantages. Firstly, while traditional PCA is limited to capturing linear relationships between variables, which may not adequately represent complex datasets, kPCA excels at capturing nonlinear relationships by mapping the original data into a high-dimensional space using a kernel function. This capability allows kPCA to capture intricate structures within the dataset, enhancing its ability to model complex phenomena. Secondly, kPCA provides flexibility in selecting different kernel functions based on the characteristics of the data under consideration. By choosing an appropriate kernel function, kPCA can effectively represent the underlying features of the data, leading to more accurate and robust results. The utility of kPCA in nonlinear feature extraction has been extensively discussed in the literature, e.g., \cite{scholkopf1998nonlinear}, \cite{mika1998kernel}, \cite{rosipal2001kernel}, and \cite{hoffmann2007kernel}. 

Bond yields' term structure plays an important role in risk management. For instance, the prediction power of yield curves in forecasting recessions has been studied in \cite{wright2006yield} and \cite{chinn2015predictive}. Gerhart and L{\"u}tkebohmert \cite{gerhart2020empirical} highlighted the importance of the slope factor variability in pre- and post-crisis periods. In this paper, we modelled the risks which arise in the bond markets from two perspectives. First, we conducted a stress testing analysis, which plays a crucial role as a risk management tool. For example, Christensen et al. \cite{christensen2015probability} employed probability-based stress tests to analyse interest rate risk, while Karimalis et al. \cite{karimalis2017multi} investigated the effects of liquidity and credit shocks on the yield curves in Italy and Spain. In this paper, we evaluate the impacts of the shocks applied to the bond yields in reference country propagating through the bond market in the response country. Our analysis defines two stress testing scenarios that represent either short-term disruptions or long-term structural changes. These scenarios are applied to reference country yield curves, and we evaluate their impact on response country yield curves. In this study, we analyse the interdependencies between two economies and explore the risks arising in response country bond market. Secondly, in a case study of a bond ladder portfolio, we evaluate the changes in portfolio values and the 5\% value-at-risk (VaR) under various shocks. This comprehensive analysis allows us to assess the risks from an investment strategy perspective.

The implementations of the Nelson-Siegel model and its extensions are available in R (``YieldCurve'' and ``NMOF'' packages, see \cite{guirreri2010yieldcurve} and \cite{schumann2016nmof}, respectively) and Python (``PyCurve''). 

\subsection{Contributions}

This work makes significant contributions to both the methodological and applied aspects of yield curve modelling and risk assessment.

Firstly, we introduce a novel state-space functional regression model that integrates the dynamic Nelson-Siegel model's state-space formulation with a functional regression framework. This innovative approach not only merges time series and functional regression variables but also enhances the accuracy and applicability of the Nelson-Siegel model in modelling yield curves. To address the inherent challenges of estimation, we utilise kernel principal component analysis (kPCA), which transforms the dynamic functional regression representation into a finite-dimensional, tractable estimation problem.

Secondly, we apply this functional regression model to the yield curves of eight countries and regions, comparing its performance with the traditional dynamic Nelson-Siegel model. By incorporating the relative spread of US yields, our functional regression framework demonstrates robust estimation capabilities across all response countries. This is evident both in the overall in-sample estimation and in accurately capturing different yield curve structures, particularly at the long end of the yield curves.

Thirdly, we conduct stress testing to assess the risks under extreme market conditions. Specifically, we apply two distinct shocks to the US Treasury yields: a temporary shock, representing short-term disruptions, and a permanent shock, indicating long-term structural changes. By evaluating the impacts of these shocks on the estimation of the UK yield curve, we gain a deeper understanding of the interdependencies between these two bond markets. This analysis not only reveals the immediate effects of the shocks but also uncovers the persistent influences of sustained changes in US economic conditions. Such insights are crucial for identifying potential risks, enhancing the robustness of risk mitigation strategies, and ensuring the stability of financial institutions in the face of both temporary and permanent market changes.

Finally, we contribute by constructing a bond ladder portfolio using out-of-sample forecasting over a 12-month horizon. We quantify the maximum potential loss using the 5\% VaR and further evaluate how VaR changes under different stress testing scenarios. This analysis provides a comprehensive understanding of the potential risks associated with shocks to this investment strategy, thereby informing better risk management and investment decisions.

\subsection{Notations and Structures}

Throughout this paper, we adopt the following multicurve notations: $Y_t(\tau_i)$ represents the complete bond yields of the response country (dependent curve observation process to be modelled), and $Y_t^{(r)}(\tau_i)$ denotes the complete bond yields of the reference country, i.e. the independent curve dynamic that characterises part of the explanatory factors for the response country, where $t \in \{ 1, 2, \dots, T \}$ represents the current time, and $\tau_1 < \tau_2 < \dots < \tau_N$ are times to maturity. Both $Y_t(\tau_i)$ and $Y_t^{(r)}(\tau_i)$ are annualised rates. In the real data over time, we may not observe a complete term structure, so we denote by $\tilde{Y}_t(\tau_i)$ and $\tilde{Y}_t^{(r)}(\tau_i)$ the observed data which may contain missing values in the term structure over time. Additionally, we use $F_{i, t}$ for $i \in \{1, 2, 3\}$ to denote Nelson-Siegel factors and $U_{j,t}$ for $j \in \{1, \dots, Q\}$ to represent US factors extracted using kPCA. Bold font is used to represent matrices and vectors. $\boldsymbol{Y} = \{Y_t(\tau_i)\}$ is the matrix of all observations of the response country, while $\boldsymbol{Y}_t^\top$ and $\boldsymbol{Y}(\tau_i)$ are the $t$-th row and $i$-th column of $\boldsymbol{Y}$, respectively. Similarly, we define $\boldsymbol{Y}^{(r)} = \{Y_t^{(r)}(\tau_i)\}$ as the matrix of all observations of the reference country, and $\boldsymbol{Y}_t^{(r)\top}$ and $\boldsymbol{Y}^{(r)}(\tau_i)$ are the $t$-th row and $i$-th column of $\boldsymbol{Y}^{(r)}$.

This paper is structured to provide a comprehensive exploration of the dynamic Nelson-Siegel functional regression (DNS-FR) model and its applications. In Section \ref{sec:model}, we first define the dynamic Nelson-Siegel (DNS) model, and then extend it by adding a functional regression component to incorporate the relative spread of a reference country. Section \ref{sec:functional_representation} focuses on the functional transformation that transfers the functional regression to a finite-dimensional estimation problem through kPCA. Subsequently, in Sections \ref{sec:estimation} and \ref{sec:forecast}, we delve into the estimation and forecasting methods, respectively. Section \ref{sec:data} offers an overview of the data utilised in this study. Empirical results, including the comparison of the DNS model and the DNS-FR model in terms of in-sample estimation and out-of-sample forecasting, stress testing, and a case study of a bond ladder portfolio, are presented in Section \ref{sec:empirical_analysis}. Finally, Section \ref{sec:conclusion} concludes.

\section{Dynamic Nelson-Siegel Functional Regression Model}
\label{sec:model}

We model the complete bootstrapped bond yields $Y_t(\tau_i)$ at time $t$ with time to maturity $\tau_i$ (in months) for a set $\{ \tau_1, \dots, \tau_N \}$ as follows: 
\begin{align}
    &Y_t(\tau_i) | Y_t^{(r)}(\tau_{1:N}) = \underbrace{F_{1,t} + F_{2,t} \left( \frac{1 - e^{-\lambda \tau_i}}{\lambda \tau_i} \right) + F_{3,t} \left( \frac{1 - e^{-\lambda \tau_i}}{\lambda \tau_i} - e^{-\lambda \tau_i} \right)}_{\text{Nelson-Siegel model}} + \underbrace{\int_0^{\tau_N} \gamma_i(s) Y_{t}^{(r)}(s) ds }_{\text{Functional regression model}} + \epsilon_t(\tau_i) \label{eq:DNS_FR_m} \\
    &F_{j,t} = \psi_{j,0} + \psi_{j,1} F_{j, t-1} + \eta_{j, t}, \;\; j \in \{ 1,2,3 \}.
    \label{eq:DNS_FR_s}
\end{align}
Here, $\tau_N$ is the maximum maturity of the reference country's bonds. $Y_t^{(r)}(\tau_{1:N})$ is the yield curve of reference country with maturities from $\tau_1$ to $\tau_N$, where $\tau_i$. The parameter $\lambda$ governs the decay rate. $\gamma_i(s)$ represents the functional coefficient. $\psi_{j,0} \in \mathbb{R}$ and $\psi_{j,1} \in (-1, 1)$ are parameters to be estimated. $\epsilon_t(\tau_i)$ and $\eta_{j, t}$ are independent and identically distributed normal noises. We will discuss the structures of the covariance matrices considered for both $\epsilon_t(\tau_i)$ and $\eta_{j, t}$ later in Section \ref{sec:trans_fr}. 

As demonstrately, this model consists of two parts: a Nelson-Siegel model and a functional regression component. In Section \ref{sec:DNS}, we first introduce the Nelson-Siegel model, and then in Section \ref{sec:fr_representation} we extend it by incorporating a functional regression component that captures the relative spread of a reference country's bonds.

\subsection{Dynamic Nelson-Siegel Model}
\label{sec:DNS}

We begin by introducing the latent three-factor model for the yield curve, initially proposed in \cite{nelson1987parsimonious} and extended in \cite{diebold2006forecasting} to allow time-varying factors. The efficiency of this model has been demonstrated in numerous previous works, such as \cite{diebold2008global}, \cite{yu2011forecasting} and \cite{karimalis2017multi}. The Diebold and Li \cite{diebold2006forecasting} formulation for the yield curve is given as:
\begin{equation}
    Y_t(\tau_i) = F_{1,t} + F_{2,t} \left( \frac{1 - e^{-\lambda \tau_i}}{\lambda \tau_i} \right) + F_{3,t} \left( \frac{1 - e^{-\lambda \tau_i}}{\lambda \tau_i} - e^{-\lambda \tau_i} \right) + \tilde{\epsilon}_t(\tau_i), 
    \label{eq:DNS_measurement}
\end{equation}
and
\begin{equation}
    F_{j,t} = \psi_{j,0} + \psi_{j,1} F_{j, t-1} + \tilde{\eta}_{j, t}, \;\; j \in \{ 1,2,3 \}
\end{equation}
for $i \in \{1, 2, \dots, N\}$ and $t \in \{1, 2, \dots, T\}$. Here, $F_{1,t}, F_{2,t}, F_{3,t}$ are three latent factors typically interpreted as level, slope, and curvature. The parameter $\lambda$ determines the rate of exponential decay. $\tilde{\epsilon}_t(\tau_i)$ and $\tilde{\eta}_{j, t}$ are independent and identically distributed normal noises for the measurement and state equations, respectively. Additionally, we assume that all three factors $F_{j,t}$ are uncorrelated. Given the dynamic nature of the hidden factors and for consistency with other works, we refer to this model as the dynamic Nelson-Siegel (DNS) model.

\begin{figure}[h]
    \centering
    \includegraphics[width=0.6\textwidth]{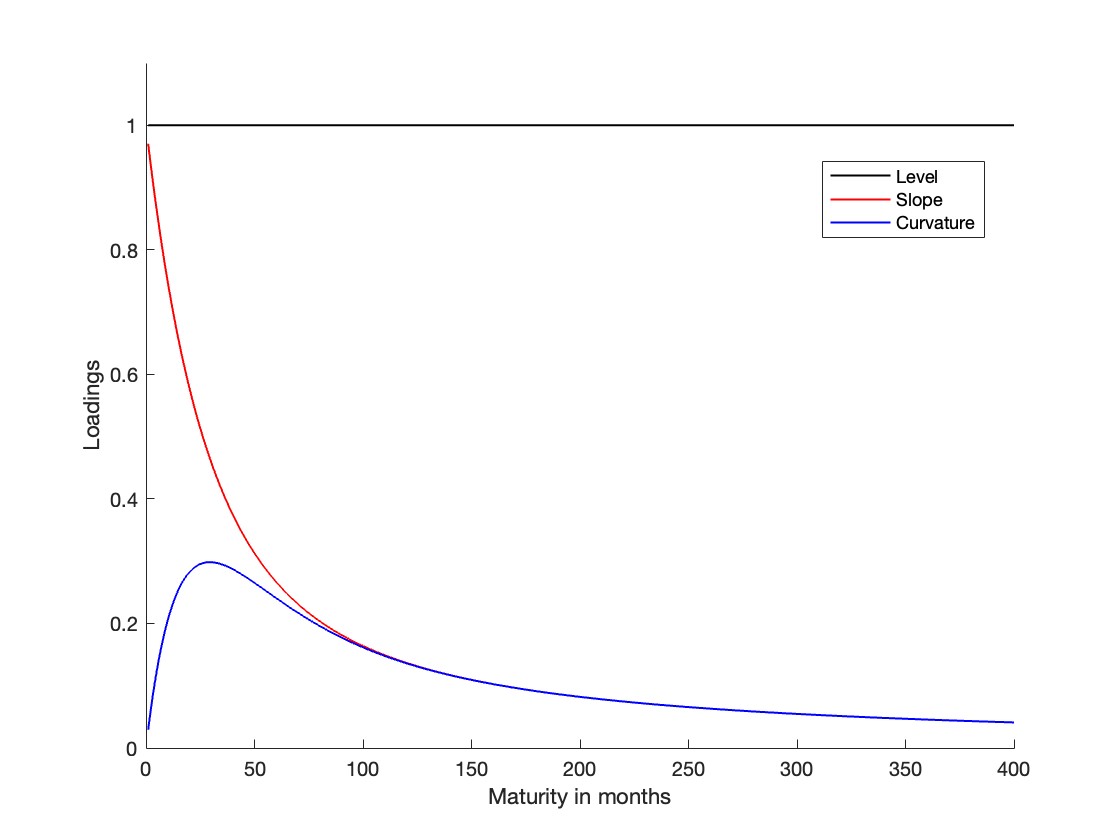}
    \caption{Factor loadings of the DNS model with $\lambda = 0.0609$. }
    \label{fig:DNS_loading}
\end{figure}

The loading on the first factor $F_{1,t}$ is always 1 and is usually called the level. It affects all yields equally, and thus it is viewed as a long-term factor. The loading on the second factor $F_{2,t}$ is $(1 - e^{-\lambda \tau_i})/(\lambda \tau_i)$, which starts from 1 and then decays monotonically to 0 as maturity goes to infinity, hence its called the slope or short-term factor. The loading on the third factor $F_{3,t}$ is $(1 - e^{-\lambda \tau_i})/(\lambda \tau_i) - e^{-\lambda \tau_i}$. It starts from 0, first increases, and then decays to 0 again. Therefore, it contributes mostly to medium maturities and is called the curvature or medium-term factor. Figure \ref{fig:DNS_loading} shows the factor loadings with a fixed $\lambda = 0.0609$ as in \cite{diebold2006forecasting}. 

\subsection{Functional Regression Representation with Missing Data}
\label{sec:fr_representation}

To address the limitation that the Dynamic Nelson-Siegel (DNS) model does not account for conditional relationships between different economies, we extend the DNS model by incorporating a functional regression component into the measurement equation. In recent years, functional data analysis (FDA) has gained popularity in time series modelling. In \cite{hyndman2009forecasting} such an approach was pioneered by introducing a weighted functional principal component regression method designed for forecasting functional time series. They successfully applied this method to Australian fertility rate data, representing time series curves as a finite sum of weighted hidden factors. Building on this methodology, Hays et al. \cite{hays2012functional} introduced the functional dynamic factor model, which estimates both the hidden factor time series and the functional factor loadings simultaneously. This approach was applied to US Treasury yields, demonstrating superior performance compared to the dynamic Nelson-Siegel model. In a related study, Mart{\'\i}nez-Hern{\'a}ndez et al. \cite{martinez2022nonparametric} proposed a nonparametric three-factor model, showing that the estimated factor loadings align with the level, slope, and curvature of the Nelson-Siegel model, highlighting the versatility and adaptability of FDA in capturing the complex dynamics of yield curves. The applications of FDA in the interest rate market was explored in \cite{bowsher2008dynamics} and \cite{jiang2014multivariate}. 

It is noteworthy that, in reality, data is often incomplete. In the presence of missing data, bootstrapping methods are commonly used to handle the gaps. Different bootstrapping methods were discussed in \cite{ametrano2013everything} and \cite{peters}. In this paper, missing data can occur in both the reference country yields $Y_t^{(r)}(\tau_i)$ and the response country yields $Y_t(\tau_i)$. For the former, given the small amount of missing data in our datasets, we use linear interpolation to fill in the missing values for simplicity. For the latter, we define a binary variable $S_{ti}$ to track the missing data: 
\begin{equation*}
S_{ti} = \begin{cases} 
1 \quad \text{if} \; Y_t(\tau_i) \; \text{is recorded} \\ 
0 \quad \text{otherwise.}  
\end{cases} 
\end{equation*}
The extended model is given as:
\begin{equation}
    \tilde{Y}_t(\tau_i) | Y_{t}^{(r)}(\cdot) = \left[ F_{1,t} + F_{2,t} \left( \frac{1 - e^{-\lambda \tau_i}}{\lambda \tau_i} \right) + F_{3,t} \left( \frac{1 - e^{-\lambda \tau_i}}{\lambda \tau_i} - e^{-\lambda \tau_i} \right) + \int_0^{\tau_N} \gamma_i(s) Y_{t}^{(r)}(s) ds + \epsilon_t(\tau_i) \right] \cdot S_{ti}, 
    \label{eq:DNS_FR_m_missing}
\end{equation}
and the state equation remains unchanged: 
\begin{equation}
    F_{j,t} = \psi_{j,0} + \psi_{j,1} F_{j, t-1} + \eta_{j, t}, \;\; j \in \{ 1,2,3 \}. 
    \label{eq:DNS_FR_s_missing}
\end{equation}
If data is complete, this model reduces to Equations \eqref{eq:DNS_FR_m} and \eqref{eq:DNS_FR_s}. Here, $Y_{t}^{(r)}(s)$ represents the yield curve of a reference country at time $t$, and $\gamma_i(s)$ are the functional coefficients. In this study, we assume that the yield curve of the response country at time $t$ depends only on the yield curve of the reference country at the current time $t$, and not on previous times. Additionally, we assume that the functional coefficient $\gamma_i(s)$ is time-invariant and varies across different maturities of bonds from the response country. We refer to this model as the dynamic Nelson-Siegel functional regression (DNS-FR) model.

However, estimating the functional coefficient $\gamma_i(s)$ is challenging. Therefore, in the next section, we will transform the functional regression component into a weighted sum of a finite number of factors.

\section{Functional Representation and Transformation}
\label{sec:functional_representation}

In order to transform the integral representation of functional regression to a vector operation which is then applicable for a linear estimation procedure, in Section \ref{sec:kpca} we introduce the kernel principal component analysis (kPCA) method, which is used to extract factors from the yields of the reference country. The transformation is completed in Section \ref{sec:trans_fr}.

\subsection{Kernel Principal Component Analysis}
\label{sec:kpca}

The kPCA method is employed to reduce the dimensionality of the data, with advantages over traditional PCA in capturing the non-linear structure of the data. Consider the bootstrapped input matrix of bond yields for the reference country: 
\begin{equation*}
    \boldsymbol{Y}_{N \times T}^{(r)} = \left[ \begin{matrix}
    Y_1^{(r)}(\tau_1) & Y_2^{(r)}(\tau_1) & \cdots & Y_T^{(r)}(\tau_1) \\
    Y_1^{(r)}(\tau_2) & Y_2^{(r)}(\tau_2) & \cdots & Y_T^{(r)}(\tau_2) \\
    \vdots & \vdots & \ddots & \vdots \\
    Y_1^{(r)}(\tau_N) & Y_2^{(r)}(\tau_N) & \cdots & Y_T^{(r)}(\tau_N) \\ 
\end{matrix} \right]. 
\end{equation*}
We denote $\boldsymbol{Y}^{(r)}(\tau_i) = \left[ Y_1^{(r)}(\tau_i), Y_2^{(r)}(\tau_i), \dots, Y_T^{(r)}(\tau_i) \right]^\top$ as the time series of the yields of the bonds with maturity $\tau_i$. Assume $\boldsymbol{\phi}: \mathcal{Y} \to \mathcal{F}$ is a non-linear mapping from the observed input space $\mathcal{Y} \subset \mathbb{R}^T$ to the feature space $\mathcal{F} \subset \mathbb{R}^T$ such that $\boldsymbol{\phi}(\boldsymbol{Y}^{(r)}(\tau_i)) = \left[ \phi_1(\boldsymbol{Y}^{(r)}(\tau_i)), \dots, \phi_T(\boldsymbol{Y}^{(r)}(\tau_i)) \right]^\top$. Let $\boldsymbol{\Phi}$ be an $N \times T$ matrix:
\begin{equation*}
    \boldsymbol{\Phi}_{N \times T} = \left[ \begin{matrix}
    \phi_1(\boldsymbol{Y}^{(r)}(\tau_1)) & \cdots & \phi_T(\boldsymbol{Y}^{(r)}(\tau_1)) \\
    \vdots & \ddots & \vdots \\ 
    \phi_1(\boldsymbol{Y}^{(r)}(\tau_N)) & \cdots & \phi_T(\boldsymbol{Y}^{(r)}(\tau_N)) 
\end{matrix} \right]. 
\end{equation*}
Define $\boldsymbol{C}_{T \times T} = \boldsymbol{\Phi}^\top \boldsymbol{\Phi}$ which is a positive definite matrix. The kernel function $k: \mathcal{Y} \times \mathcal{Y} \to \mathcal{F}$ defines the inner product in the feature space $\mathcal{F}$ and is given by:
\begin{equation}
    k(\boldsymbol{Y}^{(r)}(\tau_i), \boldsymbol{Y}^{(r)}(\tau_j)) = \boldsymbol{\phi}(\boldsymbol{Y}^{(r)}(\tau_i))^\top \boldsymbol{\phi}(\boldsymbol{Y}^{(r)}(\tau_j)) = \sum_{k=1}^T \phi_k(\boldsymbol{Y}^{(r)}(\tau_i)) \phi_k(\boldsymbol{Y}^{(r)}(\tau_j))
    \label{eq:kernel_function}
\end{equation}
for $i, j \in \{ 1, \dots, N \}$. Define $\boldsymbol{K}_{N \times N} = \boldsymbol{\Phi} \boldsymbol{\Phi}^\top$. The objective of kPCA is to find a linear projection that projects $\boldsymbol{\Phi}$ onto uncorrelated components denoted $\boldsymbol{A}$, with lower dimensionality. Each point $\boldsymbol{\phi}(\boldsymbol{Y}^{(r)}(\tau_i))$ can be expressed as a linear combination of $Q \le T$ vectors of dimension $T$: 
\begin{equation*}
    \boldsymbol{\phi}(\boldsymbol{Y}^{(r)}(\tau_i)) = \sum_{q=1}^Q \alpha_{i,q} \boldsymbol{v}_q, 
\end{equation*}
where $\boldsymbol{v}_q$ are orthonormal vectors of dimensions $T$ such that: 
\begin{equation*}
\boldsymbol{v}_q^\top \boldsymbol{v}_k = \sum_{t = 1}^{T} v_{q,t} v_{k,t} = \begin{cases} 
1 \quad \text{if} \; q = k \\ 
0 \quad \text{otherwise} 
\end{cases}
\end{equation*}
Vectors ${\boldsymbol{\alpha}}_q = [ \alpha_{1,q}, \dots, \alpha_{N,q}]^\top$ and ${\boldsymbol{\alpha}}_k = [ \alpha_{1,k}, \dots, \alpha_{N,k}]^\top$ are orthogonal: 
\begin{equation*}
{\boldsymbol{\alpha}}_q^\top {\boldsymbol{\alpha}}_k = \sum_{i=1}^{N} \alpha_{i,q} \alpha_{i,k} = \begin{cases} 
\lambda_q \quad \text{if} \; q = k \\ 
0 \quad \text{otherwise}  
\end{cases} 
\end{equation*}
where $\lambda_q$ is the $q$th eigenvalue of $\boldsymbol{C}$. Define $N \times Q$ matrix $\boldsymbol{A}$ and $Q \times T$ matrix $\boldsymbol{V}$ as:
\begin{equation*}
\boldsymbol{A}_{N \times Q} = [{\boldsymbol{\alpha}}_1, \dots, {\boldsymbol{\alpha}}_Q ] = 
\left[ \begin{matrix} \alpha_{1,1} & \dots & \alpha_{1,Q} \\ 
\vdots & \ddots &  \vdots \\ 
\alpha_{N,1} & \dots & \alpha_{N,Q}  
\end{matrix} \right]_{N \times Q}
\end{equation*}
and
\begin{equation*}
\boldsymbol{V}_{Q \times T} = 
\left[ \begin{matrix} 
\boldsymbol{v}_1^\top \\ 
\vdots \\ 
\boldsymbol{v}_Q^\top
\end{matrix} \right] = 
\left[ \begin{matrix} 
v_{1,1} & \dots & v_{1,T} \\ 
\vdots & \ddots & \vdots \\ 
v_{Q,1} & \dots & v_{Q,T} 
\end{matrix} \right]_{Q \times T}. 
\end{equation*}
Given the assumptions of uncorrelation and orthonormality, we have $\boldsymbol{A}^\top \boldsymbol{A} = \boldsymbol{\Lambda}_{Q \times Q}$ and $\boldsymbol{V} \boldsymbol{V}^\top = \boldsymbol{I}_{Q}$. Therefore, our solution can be written as
\begin{equation}
    \boldsymbol{\Phi}_{N \times T} = \boldsymbol{A}_{N \times Q} \boldsymbol{V}_{Q \times T}. 
\end{equation}

If the non-linear mapping $\boldsymbol{\phi}(\cdot)$ is known, the matrix $\boldsymbol{C}$ is also known and can be rewritten as
\begin{equation*}
    \boldsymbol{C} = \boldsymbol{\Phi}^\top \boldsymbol{\Phi} = \boldsymbol{V}^\top \boldsymbol{A}^\top \boldsymbol{A} \boldsymbol{V} = \boldsymbol{V}^\top \boldsymbol{\Lambda} \boldsymbol{V}. 
\end{equation*}
Therefore, matrix $\boldsymbol{V}$ can be obtained by applying the eigen-decomposition on $\boldsymbol{C}$, and the new representation $\boldsymbol{A}$ of the sample matrix $\boldsymbol{\Phi}$ is obtained by:
\begin{equation}
    \boldsymbol{A} = \boldsymbol{A} \boldsymbol{V} \boldsymbol{V}^\top = \boldsymbol{\Phi} \boldsymbol{V}^\top 
\end{equation}
as the result of the orthonormality of rows in $\boldsymbol{V}$. $\boldsymbol{A}$ is of lower dimension and represents the matrix of the principal components. 

However, the mapping $\boldsymbol{\phi}(\cdot)$ is usually unknown. In this case, the matrix $\boldsymbol{C}$ and the matrix of eigenvectors $\boldsymbol{V}$ are also unknown. One possible solution is given by the employment of the kernel function $k(\cdot, \cdot)$ given in Equation \eqref{eq:kernel_function}. Since $\boldsymbol{C} = \boldsymbol{\Phi}^\top \boldsymbol{\Phi} = \boldsymbol{V}^\top \boldsymbol{\Lambda} \boldsymbol{V}$, we have
\begin{align}
    \boldsymbol{C} &= \boldsymbol{V}^\top \boldsymbol{\Lambda} \boldsymbol{V} \nonumber \\
    \boldsymbol{V} \boldsymbol{C} &= \boldsymbol{V} \boldsymbol{V}^\top \boldsymbol{\Lambda} \boldsymbol{V} \nonumber \\
    \boldsymbol{V} \boldsymbol{\Phi}^\top \boldsymbol{\Phi} &= \boldsymbol{V} \boldsymbol{V}^\top \boldsymbol{\Lambda} \boldsymbol{V} \nonumber \\ 
    \underbrace{ \boldsymbol{V} \boldsymbol{\Phi}^\top }_{=\boldsymbol{A}^\top} \underbrace{ \boldsymbol{\Phi} \boldsymbol{\Phi}^\top }_{=\boldsymbol{K}} &= \underbrace{ \boldsymbol{V} \boldsymbol{V}^\top }_{ = \boldsymbol{I}_{Q}} \boldsymbol{\Lambda} \underbrace{ \boldsymbol{V} \boldsymbol{\Phi}^\top }_{=\boldsymbol{A}^\top} \nonumber
\end{align}
As $\boldsymbol{A} = \boldsymbol{\Phi} \boldsymbol{V}^\top$ and $\boldsymbol{K} = \boldsymbol{\Phi} \boldsymbol{\Phi}^\top$, therefore, by simplifying the last equation, we have
\begin{equation}
    \boldsymbol{A}^\top \boldsymbol{K} = \boldsymbol{\Lambda} \boldsymbol{A}^\top. 
    \label{eq:pc}
\end{equation}
So far, the matrix $\boldsymbol{A}$ is only orthogonal but not orthonormal as $\boldsymbol{A}^\top \boldsymbol{A} = \boldsymbol{\Lambda}$. We define the matrix $\boldsymbol{Z}$ as $\boldsymbol{Z}_{N \times Q} = \boldsymbol{A} \boldsymbol{\Lambda}^{-\frac{1}{2}}$. Since
\begin{equation*}
    \boldsymbol{Z}^\top \boldsymbol{Z} = \boldsymbol{\Lambda}^{-\frac{1}{2}} \boldsymbol{A}^\top \boldsymbol{A} \boldsymbol{\Lambda}^{-\frac{1}{2}} = \boldsymbol{\Lambda}^{-\frac{1}{2}} \boldsymbol{\Lambda} \boldsymbol{\Lambda}^{-\frac{1}{2}} = \boldsymbol{I}_Q, 
\end{equation*}
we obtained orthonormal eigenvectors $\boldsymbol{Z}$. By multiplying both sides of Equation \eqref{eq:pc} by $\boldsymbol{\Lambda}^{-\frac{1}{2}}$, we have 
\begin{align}
    \boldsymbol{\Lambda}^{-\frac{1}{2}} \boldsymbol{A}^\top \boldsymbol{K} &= \boldsymbol{\Lambda}^{-\frac{1}{2}} \boldsymbol{\Lambda} \boldsymbol{A}^\top \nonumber \\
    \boldsymbol{Z}^\top \boldsymbol{K} &= \boldsymbol{\Lambda} \boldsymbol{Z}^\top \nonumber
\end{align}
Therefore, by applying the eigen-decomposition on the matrix $\boldsymbol{K}$, we obtain the matrix $\boldsymbol{Z}$ and $\boldsymbol{\Lambda}$, and $\boldsymbol{A}$ is calculated as $\boldsymbol{A} = \boldsymbol{Z} \boldsymbol{\Lambda}^{\frac{1}{2}}$. 

Next, we discuss the out-of-sample problem either when the feature mapping $\boldsymbol{\phi}(\cdot)$ is known or unknown. In the first case, since $\boldsymbol{\phi}(\cdot)$ is known, the principal components of the new sample $\boldsymbol{\phi}(\boldsymbol{Y}^{(r)}(\tau^*))$, where $\tau^* \notin \{ \tau_1, \dots, \tau_N \}$ and $\tau^* \in [0, \tau_{max}]$, can be obtained as 
\begin{equation*}
    \boldsymbol{\alpha}^* = \boldsymbol{\phi}(\boldsymbol{Y}^{(r)}(\tau^*)) \boldsymbol{V}^\top. 
\end{equation*}
However, in the second case, we need to define the new observation $\boldsymbol{\phi}(\boldsymbol{Y}^{(r)}(\tau^*))$ in terms of the decomposition of $\boldsymbol{K}$. We have
\begin{align}
    \boldsymbol{\Phi} &= \boldsymbol{A} \boldsymbol{V} \nonumber \\
    \boldsymbol{A}^\top \boldsymbol{\Phi} &= \boldsymbol{A}^\top \boldsymbol{A} \boldsymbol{V} \nonumber \\
    \boldsymbol{A}^\top \boldsymbol{\Phi} &= \boldsymbol{\Lambda} \boldsymbol{V} \nonumber \\
    \boldsymbol{\Lambda}^{-1} \boldsymbol{A}^\top \boldsymbol{\Phi} &= \boldsymbol{V} \nonumber
\end{align}
Define $\boldsymbol{W}_{N \times Q} = \boldsymbol{A} \boldsymbol{\Lambda}^{-1}$. Then, 
\begin{equation*}
    \boldsymbol{v}_q^\top = \sum_{i=1}^N w_{i,q} \boldsymbol{\phi}(\boldsymbol{Y}^{(r)}(\tau_i))
\end{equation*}
and the principal component is given by
\begin{align}
    \alpha_{m,q} &= \boldsymbol{\phi}(\boldsymbol{Y}^{(r)}(\tau_m)) \boldsymbol{v}_q \nonumber \\
    &= \boldsymbol{\phi}(\boldsymbol{Y}^{(r)}(\tau_m)) \sum_{i=1}^N w_{i,q} \boldsymbol{\phi}(\boldsymbol{Y}^{(r)}(\tau_i))^\top \nonumber \\
    &= \sum_{i=1}^N w_{i,q} \boldsymbol{\phi}(\boldsymbol{Y}^{(r)}(\tau_m)) \boldsymbol{\phi}(\boldsymbol{Y}^{(r)}(\tau_i))^\top \nonumber \\
    &= \sum_{i=1}^N w_{i,q} k(\boldsymbol{Y}^{(r)}(\tau_m), \boldsymbol{Y}^{(r)}(\tau_i)). \nonumber
\end{align}
Therefore, given a new sample point $\boldsymbol{Y}(\tau^*)$, its projection can be represented only by the eigen-decomposition of $\boldsymbol{K}$ and the kernel function $k(\cdot, \cdot)$ as 
\begin{equation}
    \alpha_{*, q} = \sum_{i=1}^N w_{i,q} k(\boldsymbol{Y}(\tau^*), \boldsymbol{Y}(\tau_i)). 
\end{equation}

In this paper, we choose the radial basis function (RBF) kernel, which is of the form
\begin{equation}
    k(\boldsymbol{x},\boldsymbol{y}) = \exp{\left(-\frac{||\boldsymbol{x}-\boldsymbol{y}||^2}{2 \sigma^2}\right)}, 
    \label{eq:rbf}
\end{equation}
where $\sigma > 0$ is the hyperparameter. We will discuss the estimation of $\sigma$ in Section \ref{sec:est_kernel}. The validation of the RBF kernel is proved in \cite{shawe2004kernel}. Some other choices of kernel functions include polynomial, graph, and ANOVA kernels. 

\subsection{Transformation of Functional Regression}
\label{sec:trans_fr}

In this section, we transform the functional regression part in Equation \eqref{eq:DNS_FR_m} into a weighted sum of finite factors using the Karhunen-Loeve theorem:

\newtheorem{theorem}{Theorem}

\begin{theorem}[Karhunen-Loeve theorem]
    Suppose $X_t$ is a zero-mean stochastic process for $t \in [a, b]$. $K(s,t)$ is the continuous covariance function. Then $X_t$ can be expressed as 
    \begin{equation*}
        X_t = \sum_{j=1}^\infty Z_j e_j(t),
    \end{equation*}
    where $Z_j = \int_a^b X_t e_j(t) dt$, and $e_j(t)$ are orthonormal basis functions defined in \eqref{eq:orthogonal_basis}. 
\end{theorem}
The orthonormal functions are defined as follows:

\newtheorem{definition}{Definition}

\begin{definition}
    Two real-valued functions $f(x)$ and $g(x)$ are orthonormal over the interval $[a,b]$ if 
    \begin{enumerate}
        \item $\int_a^b f(x) g(x) dx = 0$ 
        \item $||f(x)||_2 = ||g(x)||_2 = \left[ \int_a^b |f(x)|^2 dx \right]^{1/2} = \left[ \int_a^b |g(x)|^2 dx \right]^{1/2} = 1$
    \end{enumerate}
\end{definition}


In this paper, we choose $e_q(\tau_i) = \alpha_{i,q} \boldsymbol{v}_q$ \footnote{In reality, we don't know the eigenvectors $\boldsymbol{v}_q$. Therefore, we choose the eigenvectors of $K$ as a proxy. This will guarantee the orthogonality of the basis functions. } as the orthogonal basis functions, so that $\boldsymbol{\phi} (\boldsymbol{Y}^{(r)}(\tau_i)) = \sum_{q=1}^Q e_q(\tau_i)$. Therefore, we have
\begin{equation}
    e_q(\tau_i) = \alpha_{i,q} \boldsymbol{v}_q = \sum_{j=1}^N w_{j,q} k(\boldsymbol{Y}^{(r)}(\tau_i), \boldsymbol{Y}^{(r)}(\tau_j)) \boldsymbol{v}_q. 
    \label{eq:orthogonal_basis}
\end{equation}

Using the Karhunen-Loeve theorem, we express $Y_t^{(r)}(s)$ and $\gamma_i(s)$ as follows: 
\begin{equation}
    Y_t^{(r)}(s) = \sum_{j=1}^\infty U_{tj} e_j(s)
\end{equation}
and
\begin{equation}
    \gamma_i(s) = \sum_{k=1}^\infty \gamma_{i,k} e_k(s),
\end{equation}
where $U_{tj} = \int_0^{\tau_N} Y_t^{(r)}(s) e_j(s) ds$ and $\gamma_{i,k} = \int_0^{\tau_N} \gamma_i(s) e_k(s) ds$. 
We then have:
\begin{align}
    \int_0^{\tau_N} \gamma_i(s) Y_t^{(r)}(s) ds =& \int_0^{\tau_N} \left( \sum_{k=1}^\infty \gamma_{i,k} e_k(s) \right) \left( \sum_{j=1}^\infty U_{tj} e_j(s) \right) ds \nonumber \\
    =& \sum_{j=k, j=1}^\infty \gamma_{i,j} U_{tj} \int_0^{\tau_N} \left( e_j(s) \right)^2 ds + \sum_{j \ne k} \gamma_{i,k} U_{tj} \int_0^{\tau_N} e_k(s) e_j(s) ds 
    \nonumber \\
    =& 
    \sum_{j=1}^\infty \gamma_{i,j} U_{tj} \approx \sum_{j=1}^Q \gamma_{i,j} U_{tj}.
\end{align}
Thus, the DNS-FR model given in Equations \eqref{eq:DNS_FR_m} and \eqref{eq:DNS_FR_s} can be rewritten as: 
\begin{align}
    &Y_t(\tau_i) = F_{1,t} + F_{2,t} \left( \frac{1 - e^{-\lambda \tau_i}}{\lambda \tau_i} \right) + F_{3,t} \left( \frac{1 - e^{-\lambda \tau_i}}{\lambda \tau_i} - e^{-\lambda \tau_i} \right) + \sum_{j=1}^Q \gamma_{i,j} U_{tj} + \epsilon_t(\tau_i), \label{eq:DNS_FR_m2} \\
    &F_{j,t} = \psi_{j,0} + \psi_{j,1} F_{j, t-1} + \eta_{j, t}, \;\; j \in \{ 1,2,3 \}. \label{eq:DNS_FR_s2}
\end{align}
In matrix notation, this becomes:
\begin{align}
    \boldsymbol{Y}_t &= \boldsymbol{\Lambda} \boldsymbol{F}_t + \boldsymbol{\Gamma} \boldsymbol{U}_t + \boldsymbol{\epsilon}_t, \;\; \boldsymbol{\epsilon}_t \sim N(0, \boldsymbol{\Sigma}_{\epsilon}), \label{eq:FR_m_mat} \\ 
    \boldsymbol{F}_t &= \boldsymbol{\Psi}_0 + \boldsymbol{\Psi}_1 \boldsymbol{F}_{t-1} + \boldsymbol{\eta}_t, \;\; \boldsymbol{\eta}_t \sim N(0, \boldsymbol{\Sigma}_{\eta}). 
    \label{eq:FR_s_mat}
\end{align}
Given the assumptions that the $F_{j,t}$ are uncorrelated, we have:
\begin{equation*}
    \boldsymbol{\Sigma}_{\eta} = \left[ \begin{matrix}
    \sigma_{\eta_1}^2 & 0 & 0 \\
    0 & \sigma_{\eta_2}^2 & 0 \\
    0 & 0 & \sigma_{\eta_3}^2 
    \end{matrix} \right].
\end{equation*}
However, $\boldsymbol{\Sigma}_{\epsilon}$ is not, generally, a diagonal matrix. That is because in the bond market, different bonds with different maturities are usually correlated to each other. Therefore, in this paper, we will consider the following three structures for $\boldsymbol{\Sigma}_{\epsilon}$, for both the DNS model and the DNS-FR model: 
\begin{enumerate}[Structure 1.]
    \item $\boldsymbol{\Sigma}_{\epsilon}$ is diagonal, and bonds with different maturities have different variances: 
    \begin{equation*}
        \boldsymbol{\Sigma}_{\epsilon} = \left[ \begin{matrix} 
            \sigma_{\epsilon_1}^2 & 0 & \cdots & 0 \\
            0 & \sigma_{\epsilon_2}^2 & \cdots & 0 \\
            \vdots & \vdots & \ddots & \vdots \\
            0 & 0 & \cdots & \sigma_{\epsilon_N}^2
        \end{matrix} \right].
    \end{equation*}
    \item $\boldsymbol{\Sigma}_{\epsilon}$ has a diagonal band. Each bond is only correlated to the two bonds (one with shorter maturity, and one with longer maturity) which are closest to it, but is not correlated to others. Furthermore, we assume that all pairs of two adjacent bonds have the same correlation coefficient: 
    \begin{equation*}
        \boldsymbol{\Sigma}_{\epsilon} = \left[ \begin{matrix}
            \sigma_{\epsilon_1}^2 & \rho \sigma_{\epsilon_1} \sigma_{\epsilon_2} & 0 & \cdots & 0 & 0 \\ 
            \rho \sigma_{\epsilon_1} \sigma_{\epsilon_2} & \sigma_{\epsilon_2}^2 & \rho \sigma_{\epsilon_2} \sigma_{\epsilon_3} & \cdots & 0 & 0 \\
            0 & \rho \sigma_{\epsilon_2} \sigma_{\epsilon_3} & \sigma_{\epsilon_3}^2 & \cdots & 0 & 0 \\ 
            \vdots & \vdots & \vdots & \ddots & \vdots \\
            0 & 0 & 0 & \cdots & \sigma_{\epsilon_{N-1}}^2 & \rho \sigma_{\epsilon_{N-1}} \sigma_{\epsilon_N} \\
            0 & 0 & 0 & \cdots & \rho \sigma_{\epsilon_{N-1}} \sigma_{\epsilon_N} & \sigma_{\epsilon_N}^2
        \end{matrix} \right]
    \end{equation*}
    However, it should be noted that this matrix is not always positive definite for all $\rho \in [-1, 1]$. Actually, it has been shown in \cite{johnson1996conditions} that only when $\rho$ ranges from $-\frac{1}{2} \sqrt{1 + \frac{\pi^2}{1 + 4N^2}}$ to $\frac{1}{2} \sqrt{1 + \frac{\pi^2}{1 + 4N^2}}$, a positive definite matrix is guaranteed. Therefore, assuming $\theta \in \mathbb{R}$ is the input parameter, we take a transformation 
    \begin{equation}
        \rho = \sqrt{1 + \frac{\pi^2}{1 + 4N^2}} \times \frac{1}{1 + e^{-\theta}} - \frac{1}{2} \sqrt{1 + \frac{\pi^2}{1 + 4N^2}}
    \end{equation}
    to make sure that $\rho$ is in the correct range.
    \item $\boldsymbol{\Sigma}_{\epsilon}$ is a full covariance matrix. To avoid a very high dimensionality of parameter space, we assume that this covariance is generated by two parameters in the following form: 
    \begin{equation*}
        \boldsymbol{\Sigma}_{\epsilon} = \left[ \begin{matrix}
            \sigma^2 & \sigma^2 \rho & \cdots & \sigma^2 \rho^{N-1} \\ 
            \sigma^2 \rho & \sigma^2 & \cdots & \sigma^2 \rho^{N-2} \\
            \vdots & \vdots & \ddots & \vdots \\
            \sigma^2 \rho^{N-1} & \sigma^2 \rho^{N-2} & \cdots & \sigma^2
        \end{matrix} \right]
    \end{equation*}
    In this case, we assume that all bonds have the same variance, and the covariance of two bonds decays as the difference in maturities increases. 
\end{enumerate}
In Section \ref{sec:empirical_analysis}, we will compare the in-sample estimation accuracy of the models using these three covariance structures, respectively. 

\section{Estimation Methodology}
\label{sec:estimation}

In this section, we present the estimation methods for the model. We first discuss the estimation of the hyperparameter $\gamma$ of the RBF kernel function in Section \ref{sec:est_kernel}. Then in Section \ref{sec:kf}, we introduce the Kalman Filter to estimate the hidden factors in a standard linear state-space model. All parameters of the state-space model are estimated by maximising the marginal likelihood function in Section \ref{sec:mle}.

\subsection{Estimation of Hyperparameters for the Kernel Function}
\label{sec:est_kernel}

The hyperparameter $\gamma$ of the RBF kernel is estimated through a grid search. We choose $\gamma$ from the range 0.001 to 1 with a step size of 0.001. At each grid point, the pre-image measurement error is calculated. The point on the grid with the minimum error is selected as the optimal $\gamma$.

\subsection{Kalman Filter}
\label{sec:kf}

The Kalman Filter is used to estimate all parameters of the DNS-FR model. We start by reparameterising Equations \eqref{eq:FR_m_mat} and \eqref{eq:FR_s_mat} into a standard state-space model:
\begin{align}
    &\boldsymbol{Y}_t - \boldsymbol{\Lambda \mu} - \boldsymbol{\Gamma} \boldsymbol{U}_t = \boldsymbol{\Lambda} (\boldsymbol{F}_t - \boldsymbol{\mu}) + \boldsymbol{B} \boldsymbol{v}_t, 
    \label{eq:FR_ssm_m} \\
    &\boldsymbol{F}_t - \boldsymbol{\mu} = \boldsymbol{\Psi}_1 (\boldsymbol{F}_{t-1} - \boldsymbol{\mu}) + \boldsymbol{D} \boldsymbol{w}_t,
    \label{eq:FR_ssm_s}
\end{align}
where $\boldsymbol{\mu}$ is the mean vector of the hidden state vector such that $\boldsymbol{\mu} - \boldsymbol{\Psi}_1 \boldsymbol{\mu} = \boldsymbol{\Psi}_0$, and $\boldsymbol{B}$ and $\boldsymbol{D}$ are Cholesky decomposition of $\boldsymbol{\Sigma}_{\epsilon}$ and $\boldsymbol{\Sigma}_{\eta}$ respectively. $\boldsymbol{v}_t$ and $\boldsymbol{w}_t$ are uncorrelated unit-variance white noise processes. We define $\boldsymbol{Z}_t := \boldsymbol{Y}_t - \boldsymbol{\Lambda \mu} - \boldsymbol{\Gamma} \boldsymbol{U}_t$ and $\boldsymbol{X}_t := \boldsymbol{F}_t - \boldsymbol{\mu}$ representing the measurement and state vectors. Therefore, Equations \eqref{eq:FR_ssm_m} and \eqref{eq:FR_ssm_s} can be rewritten as:
\begin{align}
    &\boldsymbol{Z}_t = \boldsymbol{\Lambda} \boldsymbol{X}_t + \boldsymbol{B} \boldsymbol{v}_t, 
    \label{eq:FR_ssm_m_deflated} \\
    &\boldsymbol{X}_t = \boldsymbol{\Psi}_1 \boldsymbol{X}_{t-1} + \boldsymbol{D} \boldsymbol{w}_t. 
    \label{eq:FR_ssm_s_deflated}
\end{align}
For the DNS model, we have similar notations but $\boldsymbol{\Gamma} \boldsymbol{U}_t = 0$. 

Moreover, we use the following notations to represent the expectation and covariance matrix of the state vector $\boldsymbol{X}_t$: 
\begin{align}
\boldsymbol{a}_{t|t-1} &:= \mathbb{E}(\boldsymbol{X}_t | \boldsymbol{Z}_{1:t-1}),& \boldsymbol{P}_{t|t-1} &:= Cov(\boldsymbol{X}_t | \boldsymbol{Z}_{1:t-1}), \nonumber \\
\boldsymbol{a}_t &:= \mathbb{E}(\boldsymbol{X}_t | \boldsymbol{Z}_{1:t}),& \boldsymbol{P}_t &:= Cov(\boldsymbol{X}_t | \boldsymbol{Z}_{1:t}) \nonumber 
\end{align} 
$\boldsymbol{Z}_{1:t}$ represents all vectors $\boldsymbol{Z}_{1}, \boldsymbol{Z}_{2}, \dots, \boldsymbol{Z}_{t}$. We assume that $\boldsymbol{X}_t$ is a Markov process, so that the distribution at time $t$ depends only on the state in the previous time $t-1$. 

Now, we present the algorithm of Kalman Filter. The system starts from an initial state $\boldsymbol{X}_0 \sim N(\boldsymbol{a}_0, \boldsymbol{P}_0)$, and all point estimates $\boldsymbol{a}_t$ and covariance $\boldsymbol{P}_t$ are calculated recursively in a two-stage process involving a prediction stage and a update stage. 

Firstly, given observations $\boldsymbol{Z}_{1:t-1}$ and state $\boldsymbol{X}_{t-1}$, the prediction of the distribution of $\boldsymbol{X}_t$ is calculated by 
\begin{equation}
    f(\boldsymbol{X}_t | \boldsymbol{Z}_{1:t-1}) = \int f(\boldsymbol{X}_{t} | \boldsymbol{X}_{t-1}, \boldsymbol{Z}_{1:t-1}) f(\boldsymbol{X}_{t-1} | \boldsymbol{Z}_{1:t-1}) d \boldsymbol{X}_{t-1},
\end{equation}
where $f(\cdot)$ represents the density function. As $\boldsymbol{X}_{t} | \boldsymbol{X}_{t-1}, \boldsymbol{Z}_{1:t-1} \sim N(\boldsymbol{\Psi}_1 \boldsymbol{X}_{t-1}, \boldsymbol{\Sigma}_{\eta})$ and $\boldsymbol{X}_{t-1} | \boldsymbol{Z}_{1:t-1} \sim N(\boldsymbol{a}_{t-1}, \boldsymbol{P}_{t-1})$, we have $\boldsymbol{X}_t | \boldsymbol{Z}_{1:t-1} \sim N(\boldsymbol{\Psi}_1 \boldsymbol{a}_{t-1}, \boldsymbol{\Psi}_1 \boldsymbol{P}_{t-1} \boldsymbol{\Psi}_1^\top + \boldsymbol{\Sigma}_{\eta})$. The point estimates and covariance matrix are 
\begin{equation}
    \boldsymbol{a}_{t|t-1} = \boldsymbol{\Psi}_1 \boldsymbol{a}_{t-1}
    \label{eq:kf_prediction_mean}
\end{equation}
and 
\begin{equation}
    \boldsymbol{P}_{t|t-1} = \boldsymbol{\Psi}_1 \boldsymbol{P}_{t-1} \boldsymbol{\Psi}_1^\top + \boldsymbol{\Sigma}_{\eta}. 
    \label{eq:kf_prediction_cov}
\end{equation}
Next, when a new observation $\boldsymbol{Z}_t$ is available, we update the distribution as 
\begin{equation}
    f(\boldsymbol{X}_t | \boldsymbol{Z}_{1:t}) \propto f(\boldsymbol{Z}_t | \boldsymbol{X}_{t}, \boldsymbol{Z}_{1:t-1}) f(\boldsymbol{X}_t | \boldsymbol{Z}_{1:t-1}), 
\end{equation}
which is a direct result of Bayes' Theorem. From the measurement equation, we have $\boldsymbol{Z}_t | \boldsymbol{X}_{t}, \boldsymbol{Z}_{1:t-1} \sim N(\boldsymbol{\Lambda} \boldsymbol{X}_t, \boldsymbol{\Sigma}_{\epsilon})$. Then, using some basic properties of normal distribution, we have $\boldsymbol{X}_t | \boldsymbol{Z}_{1:t} \sim N(\boldsymbol{a}_{t|t-1} + \boldsymbol{K}_t(\boldsymbol{Z}_t - \boldsymbol{\Lambda} \boldsymbol{a}_{t|t-1}), (\boldsymbol{I} - \boldsymbol{K}_t \boldsymbol{\Lambda}) \boldsymbol{P}_{t|t-1})$, where $\boldsymbol{I}$ is the identity matrix, and $\boldsymbol{K}_t = \boldsymbol{P}_{t|t-1} \boldsymbol{\Lambda}^\top (\boldsymbol{\Lambda} \boldsymbol{P}_{t|t-1} \boldsymbol{\Lambda}^\top + \boldsymbol{\Sigma}_{\epsilon})^{-1}$ is the Kalman gain matrix. The updated point estimates and covariance matrix are
\begin{equation}
    \boldsymbol{a}_{t} = \boldsymbol{a}_{t|t-1} + \boldsymbol{K}_t(\boldsymbol{y}_t - \boldsymbol{\Lambda} \boldsymbol{a}_{t|t-1})
    \label{eq:kf_update_mean}
\end{equation}
and 
\begin{equation}
    \boldsymbol{P}_{t} = (\boldsymbol{I} - \boldsymbol{K}_t \boldsymbol{\Lambda}) \boldsymbol{P}_{t|t-1}. 
    \label{eq:kf_update_cov}
\end{equation}
Finally, we repeat all the steps for $t \in \{ 1, \dots, N \}$. 

\subsection{Maximmum Marginal Likelihood Estimation}
\label{sec:mle}

In this section, we discuss the method for estimating unknown parameters, denoted by $\boldsymbol{\theta}$, by maximising the marginal likelihood. 

The marginal distribution of $\boldsymbol{Z}_t | \boldsymbol{Z}_{1:t-1}$ is given by 
\begin{equation}
    f(\boldsymbol{Z}_t | \boldsymbol{Z}_{1:t-1}) = \int f(\boldsymbol{Z}_t | \boldsymbol{X}_t, \boldsymbol{Z}_{1:t-1}) f(\boldsymbol{X}_t | \boldsymbol{Z}_{1:t-1}) d\boldsymbol{X}_t. 
\end{equation}
From the previous section, we know that $\boldsymbol{Z}_t | \boldsymbol{X}_{t}, \boldsymbol{Z}_{1:t-1} \sim N(\boldsymbol{\Lambda} \boldsymbol{X}_t, \boldsymbol{\Sigma}_{\epsilon})$ and $\boldsymbol{X}_t | \boldsymbol{Z}_{1:t-1} \sim N(\boldsymbol{a}_{t|t-1}, \boldsymbol{P}_{t|t-1})$. Therefore, we have $\boldsymbol{Z}_t | \boldsymbol{Z}_{1:t-1} \sim N(\boldsymbol{\Lambda} \boldsymbol{a}_{t|t-1}, \boldsymbol{\Lambda} \boldsymbol{P}_{t|t-1} \boldsymbol{\Lambda}^\top + \boldsymbol{\Sigma}_{\epsilon})$. 
Moreover, we define the estimation error of $\boldsymbol{Z}_t$ as 
\begin{equation}
    \boldsymbol{e}_t = \boldsymbol{Z}_t - \hat{\boldsymbol{Z}}_t = \boldsymbol{Z}_t - \boldsymbol{\Lambda} \boldsymbol{a}_{t|t-1}
\end{equation}
and the covariance matrix is 
\begin{equation}
    \boldsymbol{L}_t = Cov(\boldsymbol{e}_t) = Cov(\boldsymbol{Z}_t) = \boldsymbol{\Lambda} \boldsymbol{P}_{t|t-1} \boldsymbol{\Lambda}^\top + \boldsymbol{\Sigma}_{\epsilon}. 
\end{equation}
Ignoring the constant terms, the log-likelihood function is given by
\begin{equation}
    l(\boldsymbol{\theta}; \boldsymbol{Z}_{1:N}) = -\frac{1}{2} \sum_{t=1}^{N} \left( \boldsymbol{e}_t^\top \boldsymbol{L}_t^{-1} \boldsymbol{e}_t + \log{|\boldsymbol{L}_t|} \right). 
    \label{eq:ll}
\end{equation}
The maximum likelihood estimation (MLE) $\hat{\boldsymbol{\theta}}$ maximises the Equation \eqref{eq:ll}. 

\section{Forecasting Methodology}
\label{sec:forecast}

In this section, we discuss the procedures for forecasting the response country's yields over a $h$-month horizon. Firstly, we discuss the prediction of the measurement $\boldsymbol{Z}_{N+1:N+h}$. We start with the 1-step ahead forecasting and extend to $h$-step ahead forecasting.

The conditional density of the measurement is 
\begin{equation}
    f(\boldsymbol{Z}_{N+1} | \boldsymbol{Z}_{1:N}) = \int f(\boldsymbol{Z}_{N+1} | \boldsymbol{X}_{N+1}, \boldsymbol{Z}_{1:N}) f(\boldsymbol{X}_{N+1} | \boldsymbol{Z}_{1:N}) d\boldsymbol{X}_{N+1}.
\end{equation}
Using the results of Section \ref{sec:mle}, we have $\boldsymbol{Z}_{N+1} | \boldsymbol{Z}_{1:N} \sim N(\boldsymbol{\Lambda} \boldsymbol{a}_{N+1|N}, \boldsymbol{L}_{N+1})$. So our prediction is 
\begin{equation}
    \hat{\boldsymbol{Z}}_{N+1} = \boldsymbol{\Lambda} \boldsymbol{a}_{N+1|N} = \boldsymbol{\Lambda} \boldsymbol{\Psi}_1 \boldsymbol{a}_N. 
\end{equation}
Then, for a $h$-step ahead forecasting, we can consider it as a sequence of 1-step ahead forecasting. The conditional density is given by
\begin{equation}
    f(\boldsymbol{Z}_{N+h} | \boldsymbol{Z}_{1:N}, \hat{\boldsymbol{Z}}_{N+1:N+h-1}) = \int f(\boldsymbol{Z}_{N+h} | \boldsymbol{X}_{N+h}, \boldsymbol{Z}_{1:N}, \hat{\boldsymbol{Z}}_{N+1:N+h-1}) f(\boldsymbol{X}_{N+h} | \boldsymbol{Z}_{1:N}, \hat{\boldsymbol{Z}}_{N+1:N+h-1}) d\boldsymbol{X}_{N+h}. 
\end{equation}
Using a similar method, we have $\boldsymbol{Z}_{N+h} | \boldsymbol{Z}_{1:N}, \hat{\boldsymbol{Z}}_{N+1:N+h-1} \sim N(\boldsymbol{\Lambda} \boldsymbol{a}_{N+h|N+h-1}, \boldsymbol{L}_{N+h})$. As we do not have the real values for $\hat{\boldsymbol{Z}}_{N+1:N+h-1}$ but only the predicted values $\hat{\hat{\boldsymbol{Z}}}_{N+1:N+h-1}$, the prediction error $\boldsymbol{e}_{i} = 0$ for all $N+1 \le i \le N+h$ and therefore the point estimates of the state variable $\boldsymbol{a}_{N+h|N+h-1} = \boldsymbol{\Psi}_1 \boldsymbol{a}_{N+h-1|N+h-2} = \dots = \boldsymbol{\Psi}_1^h \boldsymbol{a}_{N}$. Our prediction is 
\begin{equation}
    \hat{\boldsymbol{Z}}_{N+h} = \boldsymbol{\Lambda} \boldsymbol{a}_{N+h|N+h-1} = \boldsymbol{\Lambda} \boldsymbol{\Psi}_1^h \boldsymbol{a}_{N}. 
    \label{eq:forecasting}
\end{equation}
The corresponding covariance matrix is 
\begin{equation}
    \boldsymbol{L}_{N+h} := Cov(\boldsymbol{Z}_{N+h}) = Cov(\boldsymbol{Y}_{N+h}) = \boldsymbol{\Lambda} Cov(\boldsymbol{X}_{N+h}) \boldsymbol{\Lambda} + \boldsymbol{\Sigma}_{\epsilon},
    \label{eq:forecasting_cov}
\end{equation}
where 
\begin{equation}
    Cov(\boldsymbol{X}_{N+h}) = \boldsymbol{\Psi}_1 Cov(\boldsymbol{X}_{N+h-1}) \boldsymbol{\Psi}_1^\top + \boldsymbol{\Sigma}_{\eta}
\end{equation}
is calculated recursively. 

Now, we discuss the prediction of response country's yields. We divide the forecasting into the following three steps: 
\begin{enumerate}[Step 1.]
    \item Forecast the US yields $\hat{\boldsymbol{Y}}_{N+1}^{(r)}, \hat{\boldsymbol{Y}}_{N+2}^{(r)}, \dots, \hat{\boldsymbol{Y}}_{N+h}^{(r)}$ over the horizon $h$ using the DNS model by Equation \eqref{eq:forecasting}, where $\hat{\boldsymbol{Z}}_{N+i} = \hat{\boldsymbol{Y}}_{N+i}^{(r)} - \boldsymbol{\Lambda} \boldsymbol{\mu}$. 
    \item Reconstruct factors $\hat{\boldsymbol{U}}_1, \dots, \hat{\boldsymbol{U}}_{N+h}$ using both in-sample data $\boldsymbol{Y}_1^{(r)}, \dots, \boldsymbol{Y}_t^{(r)}$ and predicted data $\hat{\boldsymbol{Y}}_{N+1}^{(r)}, \dots, \hat{\boldsymbol{Y}}_{N+h}^{(r)}$ through KPCA. Keep the hyperparameter constant. 
    \item Forecast the response country's yields $\hat{\boldsymbol{Y}}_{N+1}, \dots, \hat{\boldsymbol{Y}}_{N+h}$ by Equation \eqref{eq:forecasting} using DNS-FR model, where $\hat{\boldsymbol{Z}}_{N+i} = \hat{\boldsymbol{Y}}_{N+i} - \boldsymbol{\Lambda} \boldsymbol{\mu} - \boldsymbol{\Gamma} \hat{\boldsymbol{U}}_{N+i}$. The covariance matrix is given in Equation \eqref{eq:forecasting_cov} as the terms $\boldsymbol{\Lambda} \boldsymbol{\mu}$ and $\boldsymbol{\Gamma} \hat{\boldsymbol{U}}_{N+i}$ are constants. 
\end{enumerate} 
In Step 2, both in-sample data and predicted data are used for two reasons. Firstly, the forecast horizon is not too long compared to the in-sample horizon. If only predicted data is used, the extracted factors may not be accurate. Secondly, incorporating the true trajectories into the reconstruction produces a more stable representation.

\section{Data}
\label{sec:data}

In this paper, we focus on examining the relative spreads of seven countries/regions (the UK, Germany, France, Italy, Japan, Australia, and the European Union) with respect to US bond yields, which serve as the reference yield. The data used in this analysis consist of monthly data from January 2010 to December 2020.\footnote{Data was obtained from \url{https://www.tradingview.com/}. } We use the first ten years as in-sample data to estimate unknown parameters and hidden state variables, and the data for the last year for out-of-sample forecasting.

The original data contains the following maturities: 
\begin{itemize}
    \item United States (US): with maturities 1, 3, 6 months, and 1, 2, 3, 5, 7, 10, and 30 years. 
    \item United Kingdom (UK): with maturities 1, 3, 6 months, 1, 2, 3, 5, 10, 20, and 30 years. 
    \item Germany (DE): with maturities 3, 6, 9 months, and 1, 2, 3, 5, 10, 20, and 30 years. 
    \item France (FR): with maturities 1, 3, 6, 9 months, and 2, 3, 5, 10, 20, and 30 years.
    \item Italy (IT): with maturities 6, 9 months, and 2, 3, 5, 10, and 30 years. 
    \item Japan (JP): with maturities 6 months, and 1, 2, 3, 5, 10, 20, and 30 years. 
    \item Australia (AU): with maturities 1, 2, 3, 5, and 10 years.  
    \item European Union (EU): with maturities 1, 3, 6, 9 months, 1, 2, 3, 5, 7, 10, 20, and 30 years. 
\end{itemize}

\begin{figure}[h]
    \centering
    \begin{subfigure}{0.45\textwidth}
        \includegraphics[width=\textwidth]{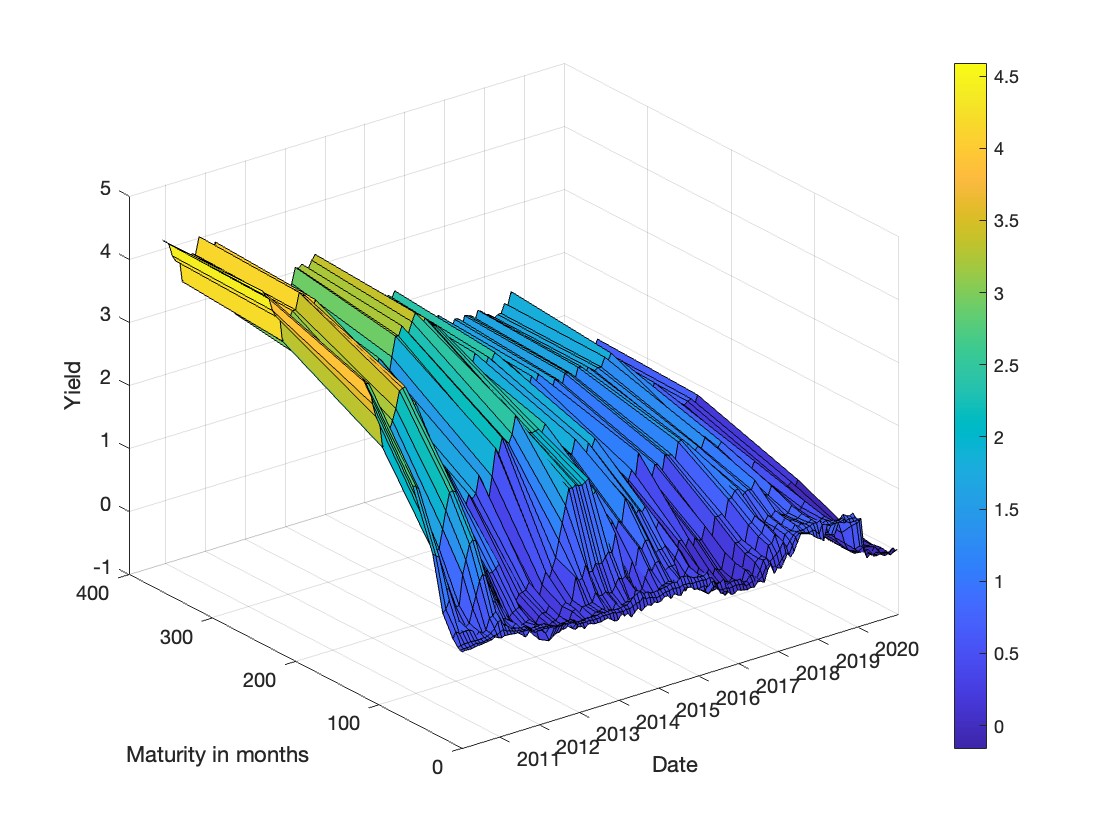}
        \caption{Original data. }
    \end{subfigure}
    \hfill
    \begin{subfigure}{0.45\textwidth}
        \includegraphics[width=\textwidth]{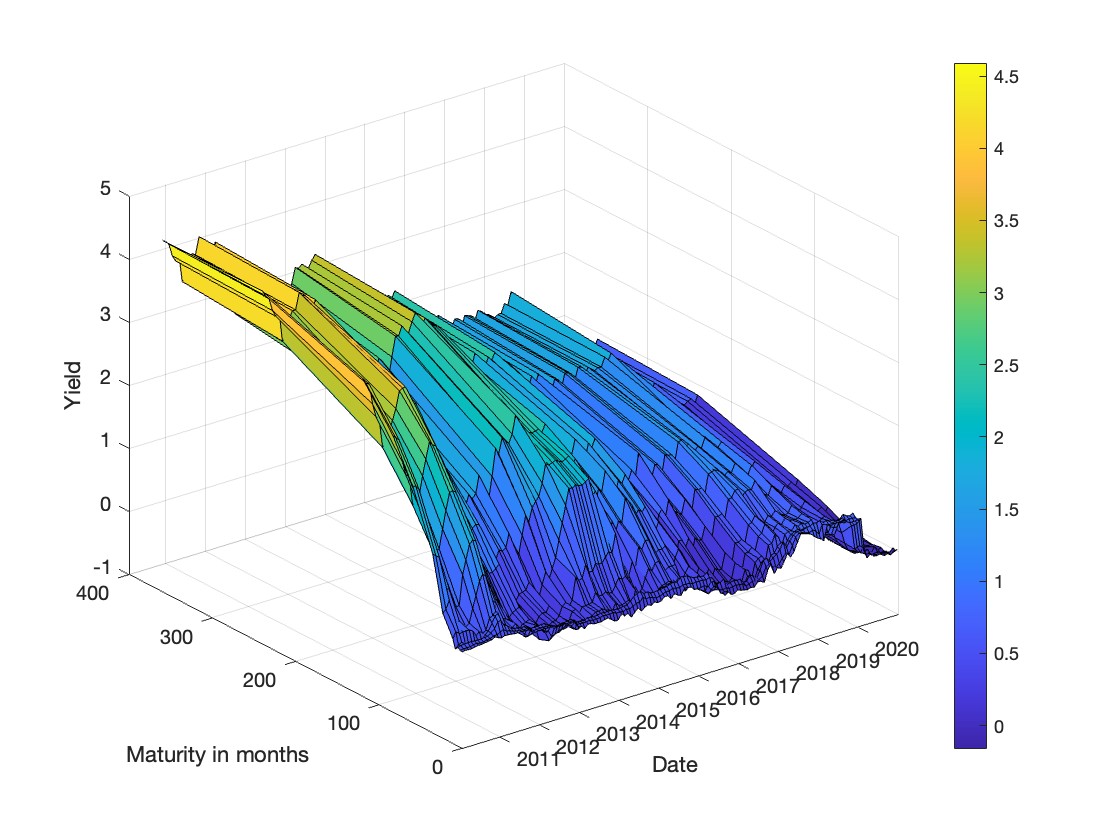}
        \caption{Data after matching maturity. }
    \end{subfigure}   
    \caption{UK yield curves from January 2010 to December 2020. The left figure is for the original data, and the right figure is for the data after maturity matching. 9-month yield and 7-year (84-month) yield are interpolated in the right figure. }
    \label{fig:data_UK}
\end{figure}

To provide a better interpretation, we match each country's maturities to 1, 3, 6, 9 months, 1, 2, 3, 5, 7, 10, 20, and 30 years. For countries/regions without all these maturities, we estimated the missing contracts using the static Nelson-Siegel model:
\begin{equation}
    Y_t(\tau_i) = F_{1,t} + F_{2,t} \left( \frac{1 - e^{-\lambda \tau_i}}{\lambda \tau_i} \right) + F_{3,t} \left( \frac{1 - e^{-\lambda \tau_i}}{\lambda \tau_i} - e^{-\lambda \tau_i} \right) + v_t(\tau_i), 
    \label{eq:SNS}
\end{equation}
where $F_{1,t}$, $F_{2,t}$, and $F_{3,t}$ are hidden factors representing the level, slope and curvature. The difference between Equation \eqref{eq:SNS} and the DNS model described in Section \ref{sec:DNS} is that for Equation \eqref{eq:SNS}, we do not assume any dynamics of the factors $F_{i,t}$, for $i \in \{ 1,2,3 \}$. Instead, they are estimated day by day using the least squares method. Figure \ref{fig:data_UK} shows the UK yield curves, for the original data and the data after maturity matching, respectively. The 9-month and 7-year bond yields are interpolated in the right sub-figure. Additionally, all missing values are replaced by interpolated values.

\section{Empirical Analysis}
\label{sec:empirical_analysis}

In this section, we present the empirical results. We first show the in-sample estimations using the DNS model and the DNS-FR model in Section \ref{sec:result_DNS} and Section \ref{sec:result_DNS_FR}, respectively. Section \ref{sec:result_forecasting} gives the out-of-sample forecasting. Then, we test the performance of DNS-FR model under a permanent shock and a temporary shock in Section \ref{sec:result_stress}. Finally, one application of the DNS-FR model in the construction of a bond ladder portfolio for risk management is given in Section \ref{sec:result_bond_ladder}. 

\subsection{Dynamic Nelson-Siegel Model}
\label{sec:result_DNS}

We first present the estimation results of the DNS model, which serves as a benchmark. Model parameters and hidden state variables are jointly estimated by maximising the marginal likelihood function described in Section \ref{sec:mle}. In this paper, $\lambda$ is fixed at 0.0609, as in \cite{diebold2006forecasting}.

The in-sample Root Mean Squared Error (RMSE) using covariance structure 2 is provided in Table \ref{tbl:DNS_estimation}. The in-sample RMSE for the DNS model using covariance structures 1 and 3 are given in Appendix \ref{app:DNS}. We first discuss the differences between these covariance structures, as outlined in Section \ref{sec:trans_fr}. Overall, the differences in mean RMSE are minimal. For long-end maturities, specifically 20-year and 30-year maturities, covariance structure 3 provides the most accurate estimation. However, for short-end and middle maturities, covariance structure 3 exhibits the highest RMSE. It also has the highest mean RMSE for most countries/regions, except Japan. Therefore, covariance structure 3 is not the best choice for this study. Covariance structures 1 and 2 are more evenly matched, with structure 2 outperforming structure 1 for some countries/regions and vice versa. Considering that covariance structure 2 shows some dependencies between yields with different maturities, which is more realistic in the bond market, while structure 1 assumes all yields are uncorrelated, we choose to use structure 2 for further analysis.

\begin{table}[width=.9\linewidth,cols=9,pos=h]
    \caption{In-sample RMSE for DNS model using covariance structure 2.}
    \label{tbl:DNS_estimation}
    \begin{tabular*}{\tblwidth}{@{} LLLLLLLLL @{}}
        \toprule
         Maturity & UK & FR & IT & DE & JP & AU & EU & US \\
         \midrule
         1 month & 0.1313 & 0.1139 & 0.0288 & 0.0047 & 2.26e-05 & 0.0048 & 0.1984 & 0.0645 \\
         3 months & 0.0212 & 0.0755 & 0.0007 & 0.0571 & 3.20e-15 & 0.0005 & 0.1307 & 0.0126 \\
         6 months & 0.1027 & 0.0449 & 0.0679 & 0.0677 & 0.0323 & 0.0026 & 0.0492 & 0.0688 \\
         9 months & 0.0511 & 0.0738 & 0.0722 & 0.0891 & 0.0071 & 0.0012 & 0.0590 & 0.0324 \\
         1 year & 0.0999 & 0.0142 & 3.64e-13 & 0.1009 & 0.0160 & 0.0106 & 0.0867 & 0.0752 \\
         2 years & 0.0102 & 0.0725 & 0.1096 & 0.0210 & 0.0144 & 0.0272 & 0.0145 & 0.0179 \\
         3 years & 0.1223 & 0.1770 & 0.2409 & 0.1216 & 0.0195 & 0.0589 & 0.1305 & 0.1082 \\
         5 years & 0.2390 & 0.2764 & 0.3754 & 0.1658 & 0.0268 & 0.0217 & 0.2097 & 0.1263 \\
         7 years & 0.1563 & 0.2023 & 0.3206 & 0.0950 & 0.1574 & 0.0957 & 0.2067 & 0.0787 \\
         10 years & 0.1374 & 0.1716 & 0.2214 & 0.1607 & 0.3325 & 0.2469 & 0.0904 & 0.1598 \\
         20 years & 0.4039 & 0.3410 & 0.0532 & 0.4377 & 0.9466 & 0.6271 & 0.2843 & 0.3872 \\
         30 years & 0.4706 & 0.4769 & 0.1233 & 0.4846 & 1.1036 & 0.8269 & 0.3049 & 0.5018 \\
         Mean & 0.1622 & 0.1700 & 0.1345 & 0.1505 & 0.2214 & 0.1603 & 0.1471 & 0.1361 \\
         \bottomrule
    \end{tabular*}
\end{table}

Focusing on covariance structure 2, most countries/regions exhibit a high RMSE for long-end maturities, with the exception of Italy. Italy has a low RMSE for long-end maturities but a high RMSE for middle maturities, between 3 years and 10 years. Similar patterns are observed for the UK, France, and the European Union, which also show high RMSEs for middle maturities. Conversely, the DNS model fits the short-end maturities well across all countries/regions. 

These results suggest that while the DNS model generally captures the dynamics well for short-end maturities, it struggles with accuracy at longer maturities for most countries/regions. This discrepancy is likely due to the higher volatility and risk associated with longer-term bonds, which are not fully captured by the DNS model. 

\FloatBarrier 

\subsection{DNS-FR Model}
\label{sec:result_DNS_FR}

In this section, we present the estimation results of the DNS-FR model. To estimate the yields of response countries, we first extract factors from US Treasury yields using kPCA as discussed in Section \ref{sec:kpca}. The estimated hyperparameter for the RBF kernel is $\gamma = 0.083$. After extracting these factors, we estimate the yields using the same methodology as the DNS model.

We select 3 kPCA factors for this study. Compared to the RMSE using 2 factors, the in-sample RMSE with 3 factors shows significant improvement, particularly for UK yields. Increasing the number of factors to 4 or 5 further reduces the RMSE for most countries/regions, but the improvement is marginal. Given the substantial increase in parameter space and associated computational challenges, we determine that 3 factors strike the optimal balance. Similar to the DNS model, the choice among the three covariance structures shows minimal difference. For consistency, we use covariance structure 2 for the remainder of this paper.

Table \ref{tbl:DNS_FR_estimation} presents the in-sample RMSE for the DNS-FR model with 3 factors and covariance structure 2. The in-sample RMSE for other covariance structures and factor counts are provided in Appendix \ref{app:DNS_FR}. The DNS-FR model significantly reduces the mean RMSE compared to the DNS model across all countries/regions. Specifically, for the UK, France, Germany, and European Union yields, the DNS-FR model outperforms the DNS model at all maturities. For Italy yields, while the DNS model shows lower RMSE at 3-month, 1-year, and 30-year maturities, the differences are minimal. For other maturities, the DNS-FR model performs better. For Japan yields, the DNS model has a slight advantage at 1-month and 3-month maturities, whereas for Australia yields, it is more accurate at 9-month, 1-year, and 2-year maturities. Otherwise, the DNS-FR model provides superior yield estimates.

As discussed in Section \ref{sec:result_DNS}, the DNS model struggles with accurately estimating the yield curve for 20-year and 30-year maturities. The DNS-FR model significantly improves this issue. For UK and European Union yields, the RMSE magnitude for long-end maturities aligns closely with other maturities. The DNS model already provides accurate long-end maturity estimates for Italy yields, so the DNS-FR model does not offer much improvement here. For France, Germany, Japan, and Australia yields, the RMSE for long-end maturities remains high, but the DNS-FR model still achieves significant reductions.

For the remainder of this paper, we focus on UK yields, as the DNS-FR model performs well for both short and long maturities.

\begin{table}[width=.9\linewidth,cols=8,pos=h]
    \caption{In-sample RMSE for DNS-FR model with 3 factors using covariance structure 2.}
    \label{tbl:DNS_FR_estimation}
    \begin{tabular*}{\tblwidth}{@{} LLLLLLLL @{}}
        \toprule
         Maturity & UK & FR & IT & DE & JP & AU & EU \\
         \midrule
         1 month & 0.0585 & 0.0474 & 0.0234 & 3.98e-15 & 0.0001 & 0.0021 & 0.0911 \\
         3 months & 0.0115 & 0.0544 & 0.0030 & 0.0458 & 0.0002 & 0.0002 & 0.0580 \\
         6 months & 0.0447 & 0.0336 & 0.0638 & 0.0451 & 0.0156 & 0.0002 & 0.0455 \\
         9 months & 0.0193 & 0.0404 & 0.0682 & 0.0483 & 0.0006 & 0.0034 & 0.0357 \\
         1 year & 0.0684 & 0.0098 & 2.87e-15 & 0.0496 & 0.0145 & 0.0289 & 0.0456 \\
         2 years & 5.51e-15 & 0.0551 & 0.1007 & 0.0032 & 0.0131 & 0.0380 & 1.41e-14 \\
         3 years & 0.0802 & 0.0578 & 0.1536 & 0.0443 & 0.0084 & 0.0457 & 0.0424 \\
         5 years & 0.1639 & 0.1332 & 0.2277 & 0.0915 & 0.0063 & 0.0147 & 0.0910 \\
         7 years & 0.1059 & 0.0811 & 0.2088 & 0.0588 & 0.0403 & 0.0566 & 0.1152 \\
         10 years & 0.1203 & 0.0830 & 0.1591 & 0.0741 & 0.0870 & 0.1473 & 0.0696 \\
         20 years & 0.0669 & 0.1982 & 0.0426 & 0.1534 & 0.2508 & 0.4285 & 0.1042 \\
         30 years & 0.0775 & 0.3181 & 0.1251 & 0.2709 & 0.3515 & 0.5899 & 0.1679 \\
         Mean & 0.0681 & 0.0927 & 0.0980 & 0.0738 & 0.0657 & 0.1130 & 0.0722 \\
         \bottomrule
    \end{tabular*}
\end{table}

We close this subsection by discussing the structure of the yield curve. Typically, in the bond market, the yield curve is in a contango structure, where short-term interest rates are lower than long-term rates due to the higher risk associated with long-term debt. However, under certain conditions, this relationship can invert, leading to a backwardation structure where long-term bonds have lower yields than short-term bonds. An inverted yield curve is rare and noteworthy as it often indicates an unusual economic environment. Therefore, it is essential to test if models can accurately estimate yield curves in both contango and backwardation structures.

\begin{figure}[h]
    \centering
    \begin{subfigure}{0.32\textwidth}
        \includegraphics[width=\textwidth]{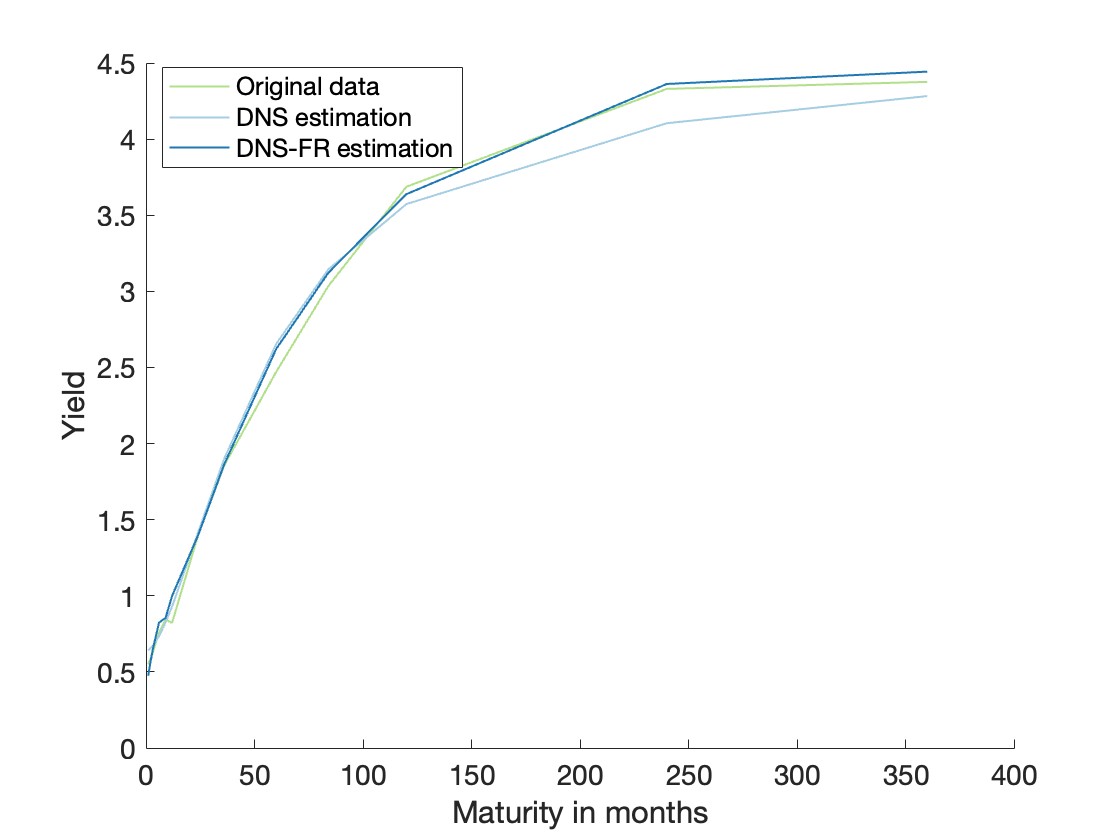}
        \caption{January 2011}
    \end{subfigure}
    \hfill
    \begin{subfigure}{0.32\textwidth}
        \includegraphics[width=\textwidth]{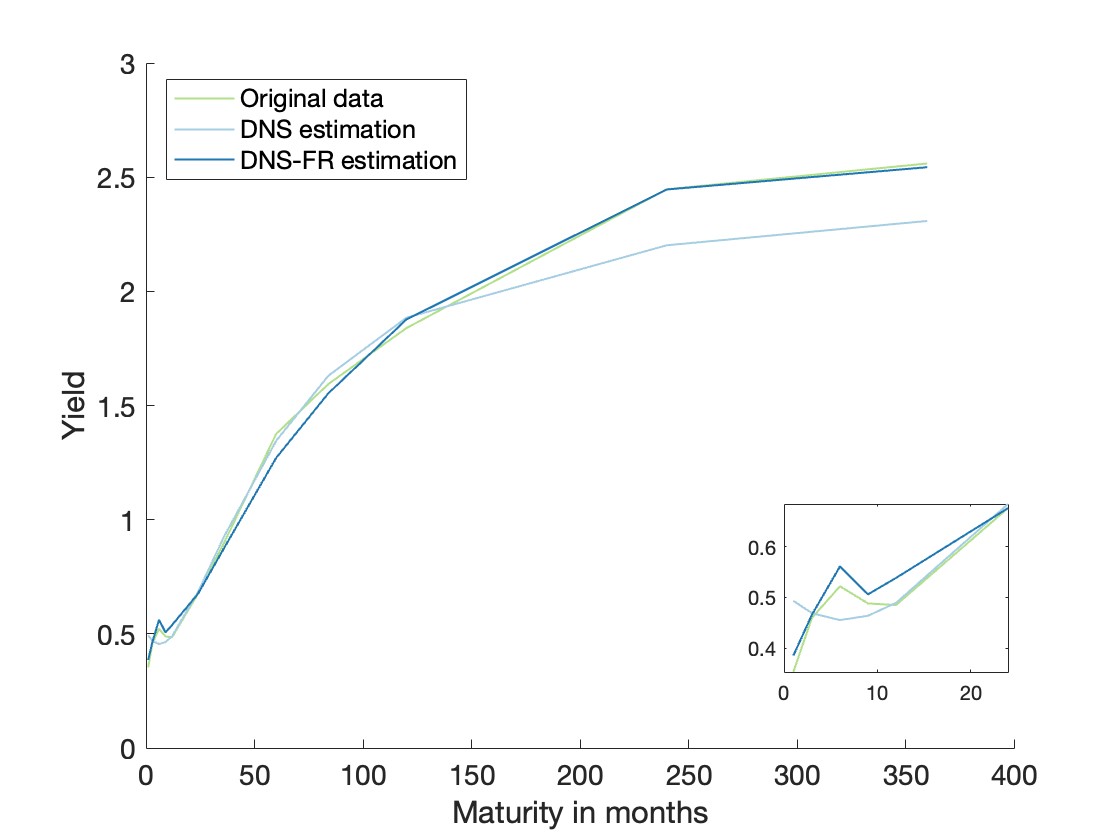}
        \caption{August 2015}
    \end{subfigure}
    \hfill
    \begin{subfigure}{0.32\textwidth}
        \includegraphics[width=\textwidth]{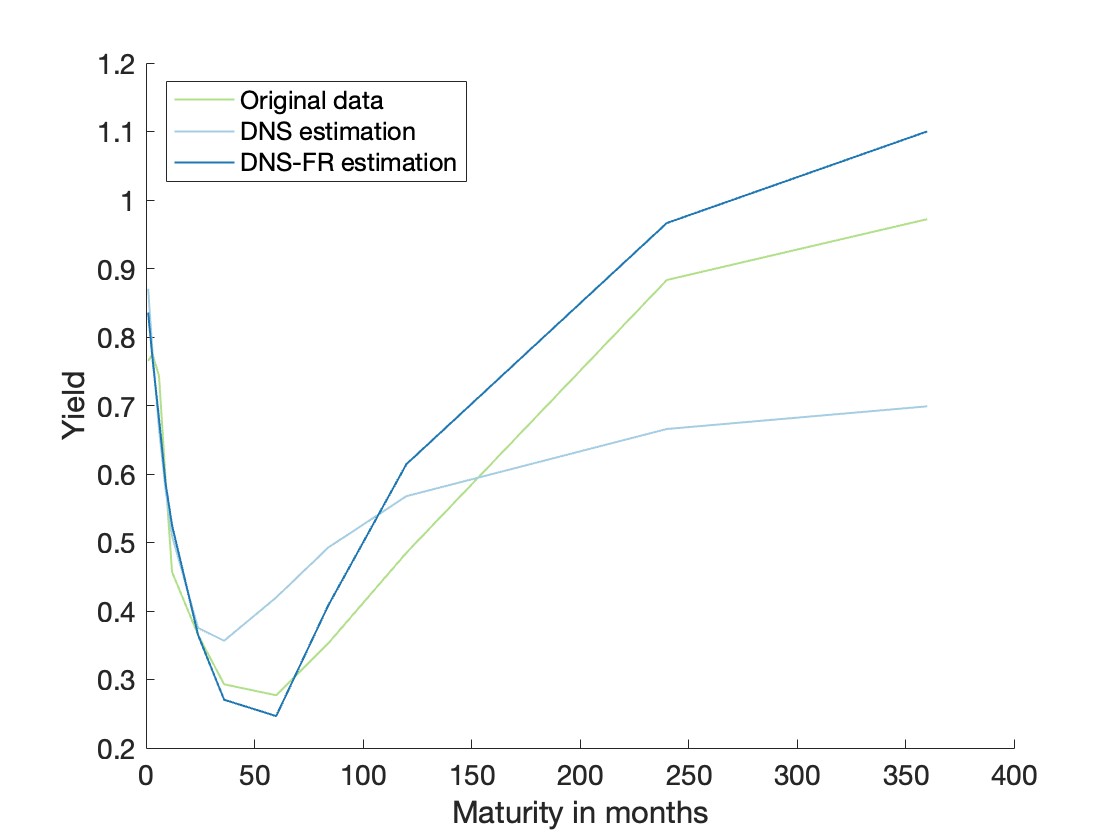}
        \caption{September 2019}
    \end{subfigure}    
    \caption{Estimated UK yield curves by DNS model and DNS-FR model on three different months. }
    \label{fig:estimation}
\end{figure}

Figure \ref{fig:estimation} illustrates the estimated UK yield curves by the DNS and DNS-FR models for January 2011, August 2015, and September 2019. In January 2011, the yield curve is in contango. For bonds with maturities less than 120 months, both models perform similarly. However, for longer maturities, the DNS-FR model provides better estimates. In August 2015, also in contango, the DNS-FR model continues to outperform, especially for long-term bonds. Moreover, the DNS-FR model captures short-term fluctuations accurately, whereas the DNS model offers a smoothed estimation. In September 2019, the yield curve is in backwardation for maturities less than 60 months, reverting to contango thereafter. The DNS-FR model again demonstrates superior performance in both backwardation and contango periods.

\FloatBarrier 

\subsection{Forecasting}
\label{sec:result_forecasting}

In this section, we compare the performance of the DNS model and the DNS-FR model in out-of-sample forecasting, using the methods discussed in Section \ref{sec:forecast}. Forecast accuracy is evaluated using the RMSE for 12-step ahead ($h=12$) forecasts.

Tables \ref{tbl:forecast_DNS} and \ref{tbl:forecast_DNS_FR} present the RMSE for the DNS model and the DNS-FR model, respectively. Interestingly, for all countries/regions, the DNS model provides a lower mean RMSE than the DNS-FR model. However, it is important to recall that, in the DNS-FR model forecasting process described in Section \ref{sec:forecast}, we first forecast the yields of the reference country (US Treasury) using the DNS model. In Table \ref{tbl:forecast_DNS}, the mean RMSE for the US Treasury yield prediction is notably high, at least double that of the mean RMSE for other countries/regions. Consequently, the reconstructed factors $\hat{\boldsymbol{U}}_{N+1}, \dots, \hat{\boldsymbol{U}}_{N+h}$ may not accurately reflect the US Treasury market during this period.

\begin{table}[width=.9\linewidth,cols=9,pos=h]
    \caption{12-step ahead forecasting RMSE for the DNS model.}
    \label{tbl:forecast_DNS}
    \begin{tabular*}{\tblwidth}{@{} LLLLLLLLL @{}}
        \toprule
         Maturity & UK & FR & IT & DE & JP & AU & EU & US \\
         \midrule
         1 month & 0.6645 & 0.1348 & 0.2305 & 0.0698 & 0.0639 & 0.5642 & 0.1164 & 1.5016 \\
         3 months & 0.6158 & 0.0543 & 0.2390 & 0.0687 & 0.0585 & 0.5202 & 0.0574 & 1.4229 \\
         6 months & 0.5456 & 0.0724 & 0.2657 & 0.1169 & 0.0510 & 0.4678 & 0.0796 & 1.3378 \\
         9 months & 0.5257 & 0.0969 & 0.2817 & 0.1718 & 0.0435 & 0.4285 & 0.1203 & 1.2959 \\
         1 year & 0.4797 & 0.0662 & 0.3215 & 0.2097 & 0.0346 & 0.4075 & 0.1537 & 1.2309 \\
         2 years & 0.4659 & 0.0541 & 0.4272 & 0.1702 & 0.0317 & 0.3174 & 0.0994 & 1.1570 \\
         3 years & 0.4744 & 0.1127 & 0.5100 & 0.1073 & 0.0350 & 0.3259 & 0.0652 & 1.1464 \\
         5 years & 0.5096 & 0.2305 & 0.5819 & 0.0721 & 0.0356 & 0.2357 & 0.1444 & 1.1109 \\
         7 years & 0.4605 & 0.2481 & 0.6054 & 0.0970 & 0.0558 & 0.1184 & 0.2012 & 0.9822 \\
         10 years & 0.3946 & 0.1886 & 0.5505 & 0.0783 & 0.1137 & 0.1576 & 0.1798 & 0.8818 \\
         20 years & 0.1243 & 0.1984 & 0.4190 & 0.1188 & 0.4541 & 0.8816 & 0.1665 & 0.5196 \\
         30 years & 0.1224 & 0.3753 & 0.4180 & 0.2515 & 0.6326 & 1.3228 & 0.1251 & 0.3364 \\
         Mean & 0.4486 & 0.1527 & 0.4042 & 0.1277 & 0.1342 & 0.4790 & 0.1258 & 1.0770 \\
         \bottomrule
    \end{tabular*}
\end{table}

\begin{table}[width=.9\linewidth,cols=8,pos=h]
    \caption{12-step ahead forecasting RMSE for the DNS-FR model.}
    \label{tbl:forecast_DNS_FR}
    \begin{tabular*}{\tblwidth}{@{} LLLLLLLL @{}}
        \toprule
         Maturity & UK & FR & IT & DE & JP & AU & EU \\
         \midrule
         1 month & 0.7252 & 0.1168 & 0.5875 & 0.1389 & 0.5606 & 0.9644 & 0.1422 \\
         3 months & 0.6624 & 0.0722 & 0.6008 & 0.1475 & 0.5646 & 0.9418 & 0.0718 \\
         6 months & 0.5646 & 0.1130 & 0.6271 & 0.1464 & 0.5688 & 0.9130 & 0.0846 \\
         9 months & 0.5482 & 0.1336 & 0.6608 & 0.1254 & 0.5684 & 0.8896 & 0.1264 \\
         1 year & 0.5093 & 0.1150 & 0.6893 & 0.1323 & 0.5734 & 0.8992 & 0.1586 \\
         2 years & 0.4720 & 0.0862 & 0.7953 & 0.1639 & 0.5442 & 0.7814 & 0.1042 \\
         3 years & 0.4860 & 0.0787 & 0.8922 & 0.2051 & 0.5148 & 0.7346 & 0.0527 \\
         5 years & 0.5358 & 0.1996 & 0.8578 & 0.2234 & 0.4452 & 0.5133 & 0.1345 \\
         7 years & 0.5288 & 0.1967 & 0.8045 & 0.2245 & 0.4137 & 0.2970 & 0.2128 \\
         10 years & 0.5026 & 0.1437 & 0.6503 & 0.1679 & 0.3622 & 0.0862 & 0.1681 \\
         20 years & 0.1794 & 0.3798 & 0.4682 & 0.1488 & 0.1159 & 0.7989 & 0.1601  \\
         30 years & 0.1464 & 0.6761 & 0.5102 & 0.2830 & 0.2720 & 1.2561 & 0.2321 \\
         Mean & 0.4884 & 0.1926 & 0.6787 & 0.1756 & 0.4587 & 0.7563 & 0.1373 \\
         \bottomrule
    \end{tabular*}
\end{table}

To further compare the performance of the DNS model and the DNS-FR model, we conduct a moving window analysis. The moving window is set as follows: we start with a 5-year window from January 2010 to December 2014, treating this period as in-sample data to estimate model parameters and hidden state variables. We then estimate the yields and calculate the mean RMSE over all maturities and data points within this window. Next, we forecast the yields 12 steps ahead, treating the next 12 months (i.e., from January 2015 to December 2015) as out-of-sample data, and calculate the mean RMSE. We then shift the window forward by one month (i.e., one data point) and repeat the calculations, continuing this process until the last available data point.

\begin{figure}[h]
    \centering
    \begin{subfigure}{0.45\textwidth}
        \includegraphics[width=\textwidth]{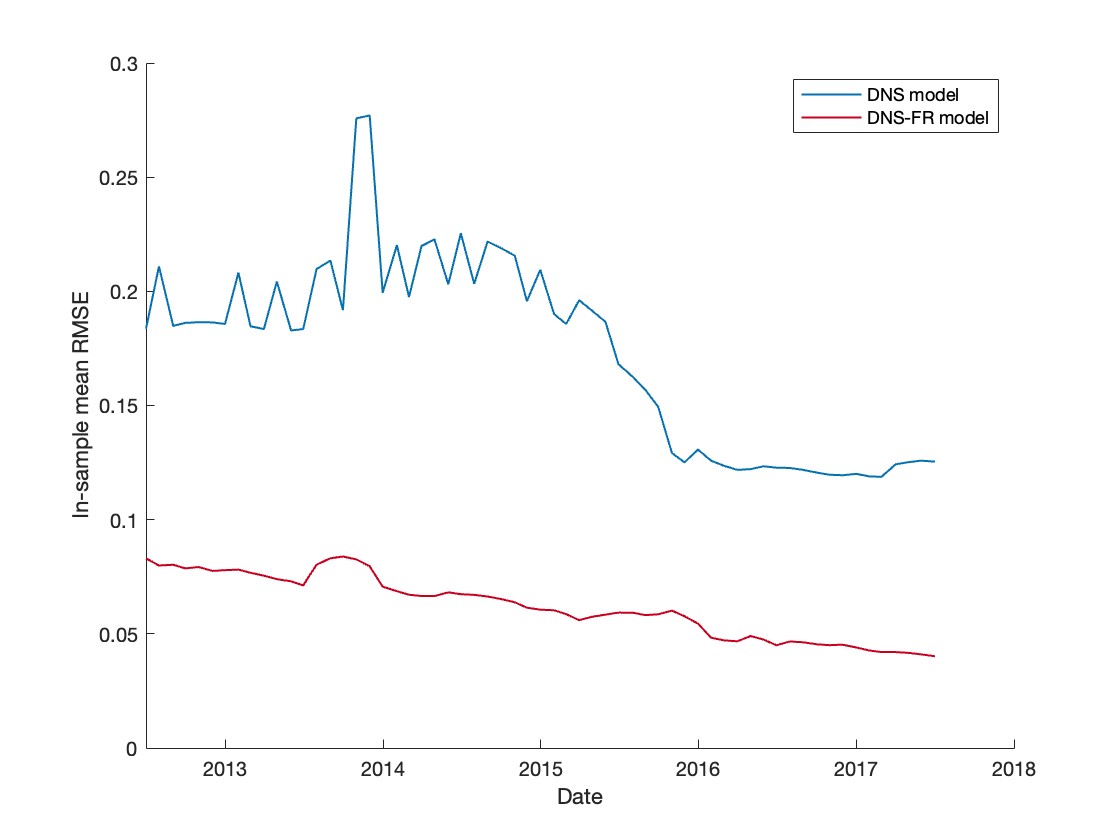}
        \caption{In-sample mean RMSE.}
    \end{subfigure}
    \hfill
    \begin{subfigure}{0.45\textwidth}
        \includegraphics[width=\textwidth]{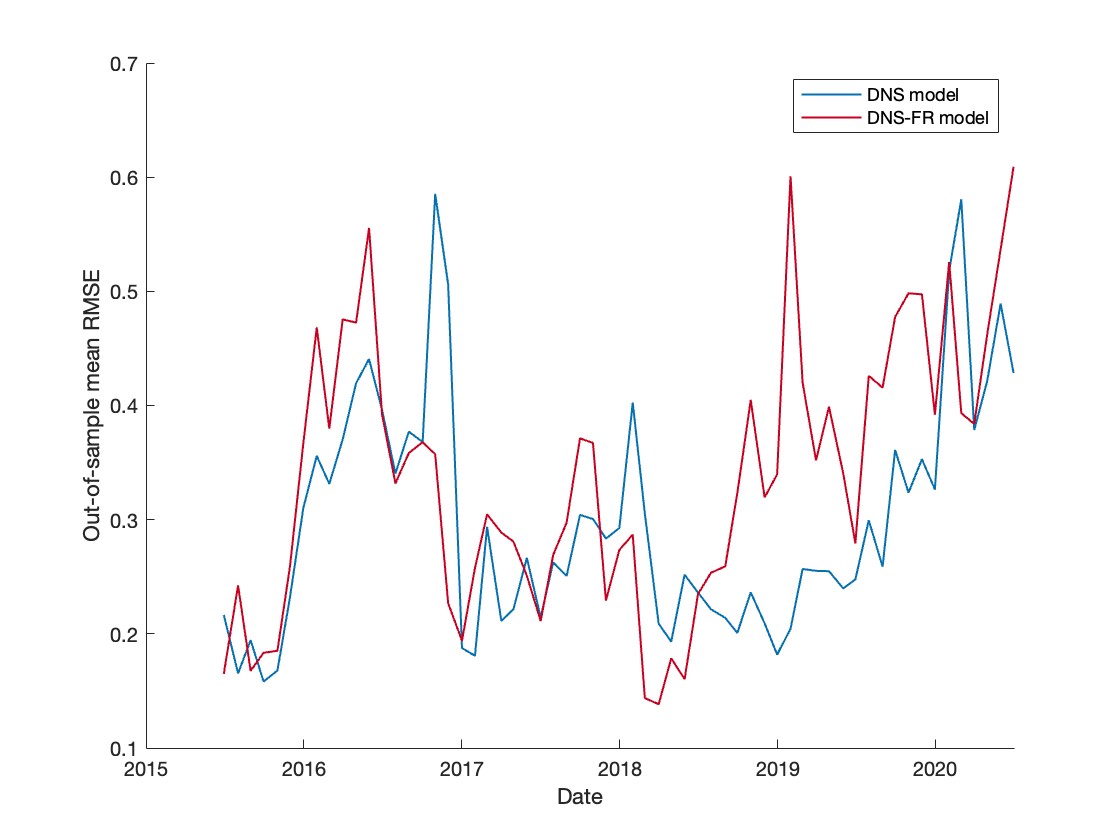}
        \caption{Out-of-sample mean RMSE.}
    \end{subfigure}
    \caption{In-sample and out-of-sample mean RMSE for UK yields using a 5-year moving window, move forward for 1 month each time. The date of each point represents the middle date of the moving window.}
    \label{fig:mv}
\end{figure}

Figure \ref{fig:mv} shows the in-sample (left) and out-of-sample (right) mean RMSE for UK yields using a moving window. The black curve represents the mean RMSE for the DNS model, while the red curve represents the mean RMSE for the DNS-FR model. The date of each point corresponds to the midpoint of the window. From the in-sample estimation, it is evident that the DNS-FR model significantly outperforms the DNS model, with the mean RMSE being only half of that of the DNS model before 2016. Even after 2016, the gap between the two curves remains substantial. In contrast, for the out-of-sample forecasting, the two curves intertwine. For some periods, such as before mid-2016, the DNS-FR model has a lower mean RMSE, while for other periods, such as from mid-2016 to mid-2017, the DNS model has a lower mean RMSE. Additionally, the differences between the two curves are limited, indicating that both models have similar forecasting performance. The same figures for other countries/regions are provided in Appendix \ref{app:moving_window}, and similar conclusions are drawn for all other countries/regions.

\FloatBarrier 

\subsection{Stress Testing}
\label{sec:result_stress}

In this section, we conduct a stress testing analysis to answer the following question: If different shocks are applied to the US Treasury market, how do the bond markets of other countries/regions respond? We define two types of stress testing scenarios, each with four cases:
\begin{itemize}
    \item Scenario 1: Temporary shocks (January 2015 to December 2015). 
    \begin{itemize}
        \item Case 1.1: Short-end maturities (1, 3, 6, 9 months, and 1, 2, 3, 5 years) yields double.
        \item Case 1.2: Middle maturities (7 and 10 years) yields double.
        \item Case 1.3: Long-end maturities (20 and 30 years) yields double.
        \item Case 1.4: Entire yield curve doubles. 
    \end{itemize}
    \item Scenario 2: Permanent shocks (beginning January 2015).
    \begin{itemize}
        \item Case 2.1: Short-end maturities (1, 3, 6, 9 months, and 1, 2, 3, 5 years) yields double.
        \item Case 2.2: Middle maturities (7 and 10 years) yields double.
        \item Case 2.3: Long-end maturities (20 and 30 years) yields double.
        \item Case 2.4: Entire yield curve doubles. 
    \end{itemize}
\end{itemize}

To estimate the yields of the response countries, we first apply each stress testing scenario to the US Treasury yield curve. The time series of the US Treasury yields under different stress testing scenarios are provided in Appendix \ref{app:time_series_st}. Using the amended US Treasury data, we extract kPCA factors. For consistency, we use three factors in this study, with the estimated hyperparameter $\gamma$ for each scenario listed in Table \ref{tbl:gamma}. Finally, we estimate the yields of the response countries.

\begin{table}[width=.4\linewidth,cols=2,pos=h]
    \caption{Estimated hyperparameter $\gamma$ using 3 factors for different stress testing cases. }
    \label{tbl:gamma}
    \begin{tabular*}{\tblwidth}{@{} LL @{}}
        \toprule
         Stress testing cases & Estimated $\gamma$ \\
         \midrule
         Original data & 0.083 \\
         Stress testing 1, case 1.1 & 0.091 \\
         Stress testing 1, case 1.2 & 0.080 \\
         Stress testing 1, case 1.3 & 0.115 \\
         Stress testing 1, case 1.4 & 0.081 \\
         Stress testing 2, case 2.1 & 0.059 \\
         Stress testing 2, case 2.2 & 0.096 \\
         Stress testing 2, case 2.3 & 0.183 \\
         Stress testing 2, case 2.4 & 0.037 \\
         \bottomrule
    \end{tabular*}
\end{table}

\begin{figure}[h]
    \centering
    \begin{subfigure}{0.45\textwidth}
        \includegraphics[width=\textwidth]{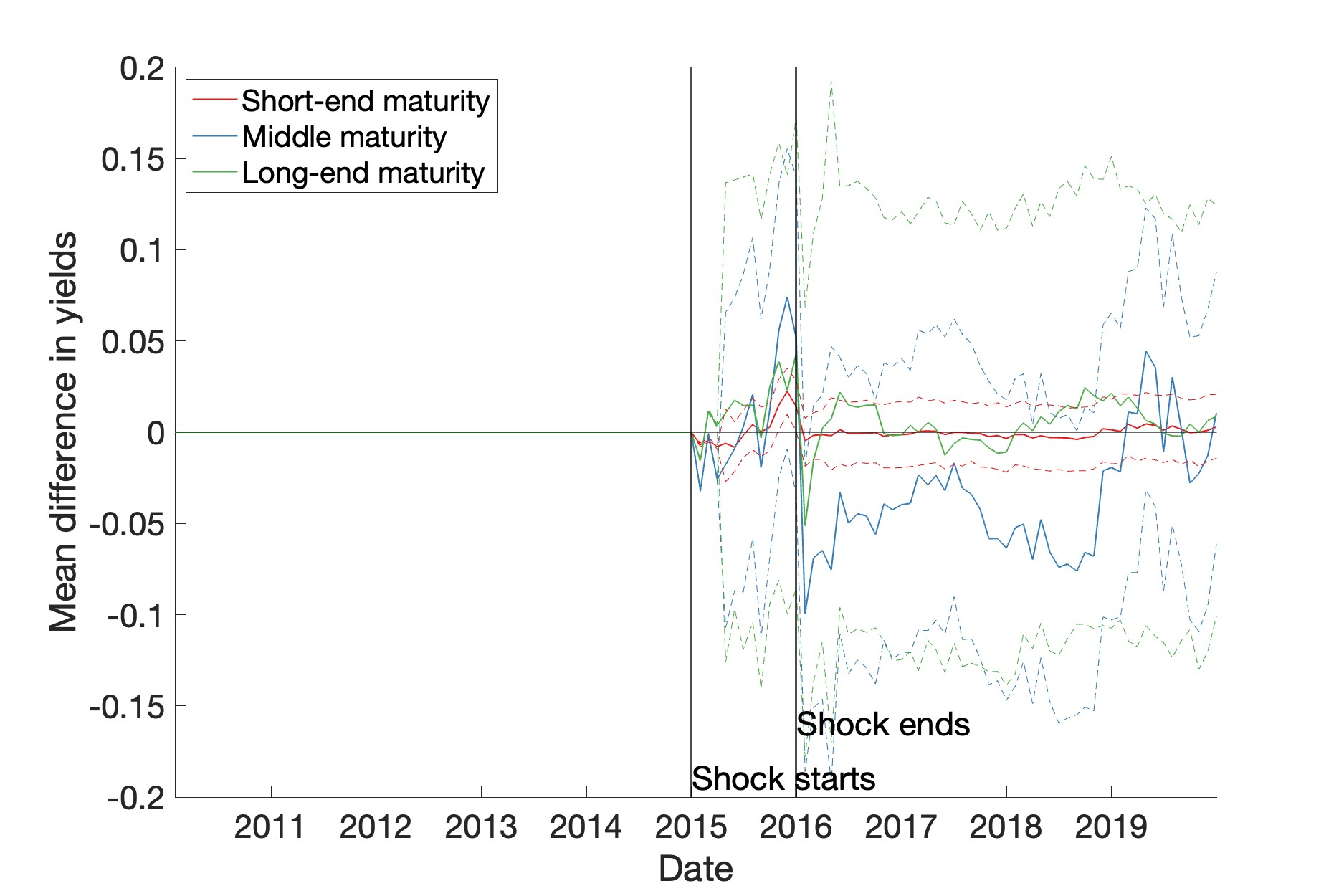}
        \caption{Case 1.1: the shock is applied to the short-end maturities of US Treasury.}
    \end{subfigure}
    \hfill
    \begin{subfigure}{0.45\textwidth}
        \includegraphics[width=\textwidth]{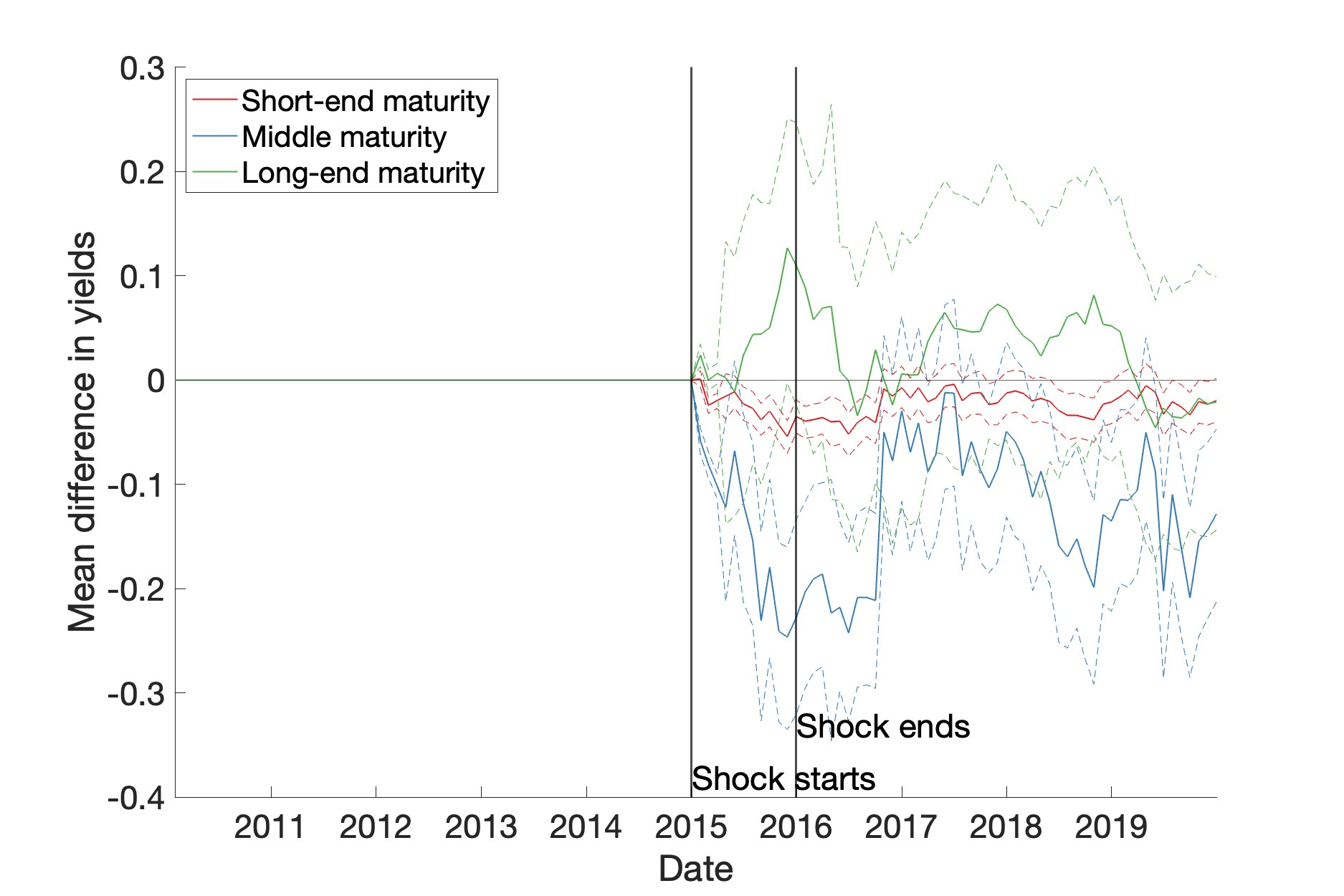}
        \caption{Case 1.2: the shock is applied to the middle maturities of US Treasury.}
    \end{subfigure}
    \hfill
    \begin{subfigure}{0.45\textwidth}
        \includegraphics[width=\textwidth]{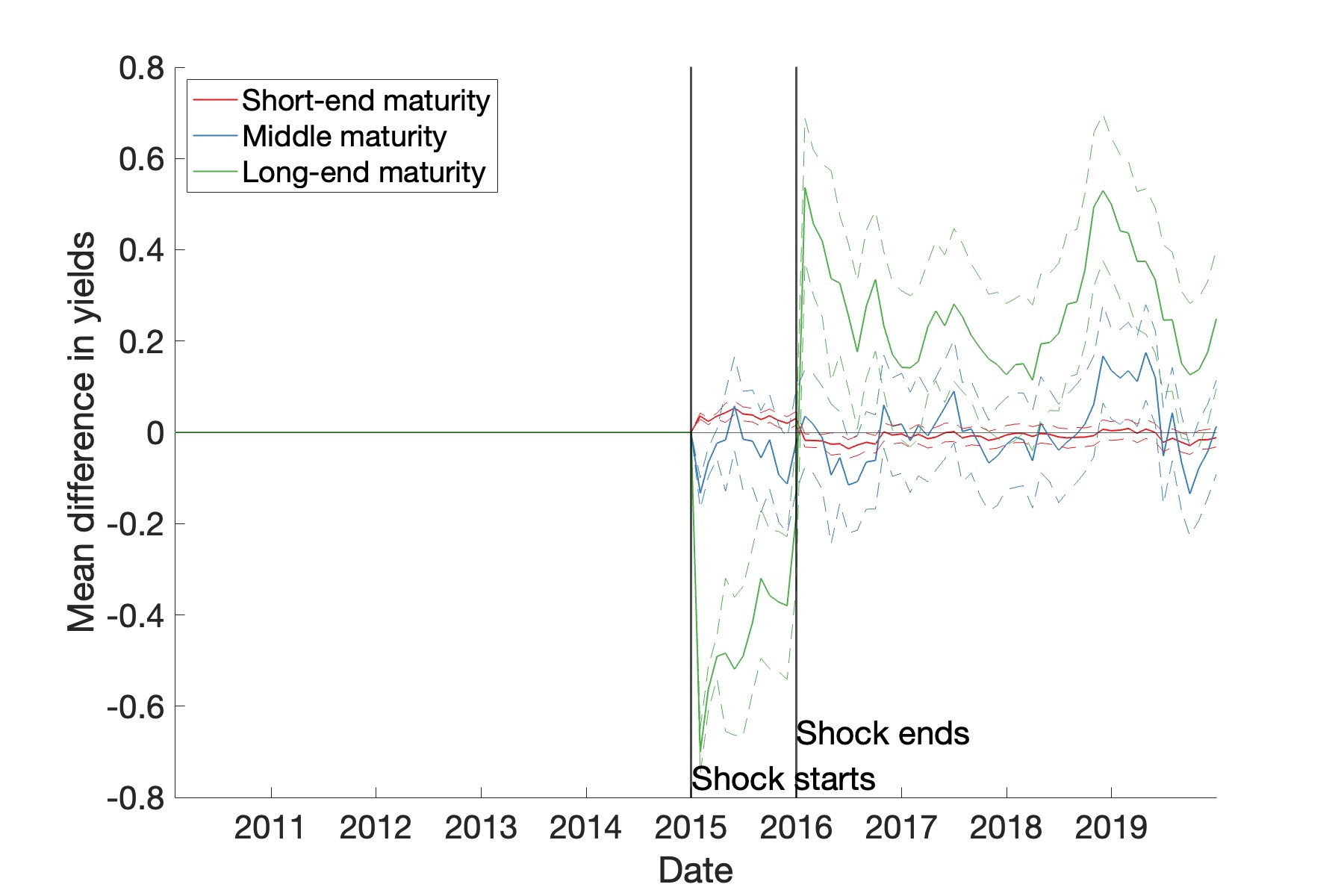}
        \caption{Case 1.3: the shock is applied to the long-end maturities of US Treasury.}
    \end{subfigure} 
    \hfill
    \begin{subfigure}{0.45\textwidth}
        \includegraphics[width=\textwidth]{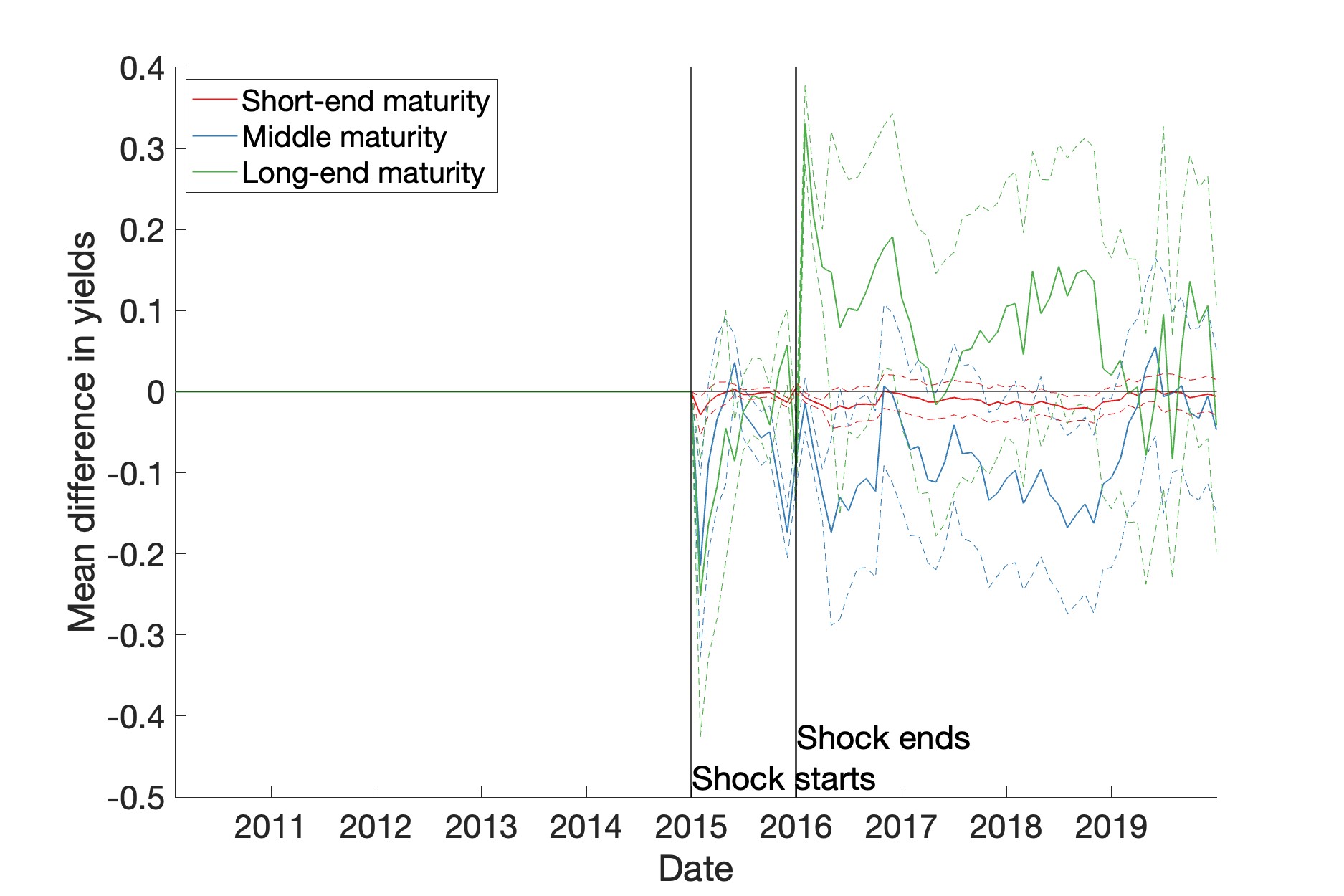}
        \caption{Case 1.4: the shock is applied to all maturities of US Treasury.}
    \end{subfigure}  
    \caption{Mean difference in percentage points of in-sample estimations between stress testing scenario 1 and original data for UK yields. The mean values are taken over short-end ($(0, 5]$ years), middle ($(5, 10]$ years), and long-end ( $(10, 30]$ years) maturities. Dashed lines represent the lower and upper bounds of 95\% confidence interval for each curve. }
    \label{fig:in_sample_diff_st1}
\end{figure}

\begin{figure}[h]
    \centering
    \begin{subfigure}{0.45\textwidth}
        \includegraphics[width=\textwidth]{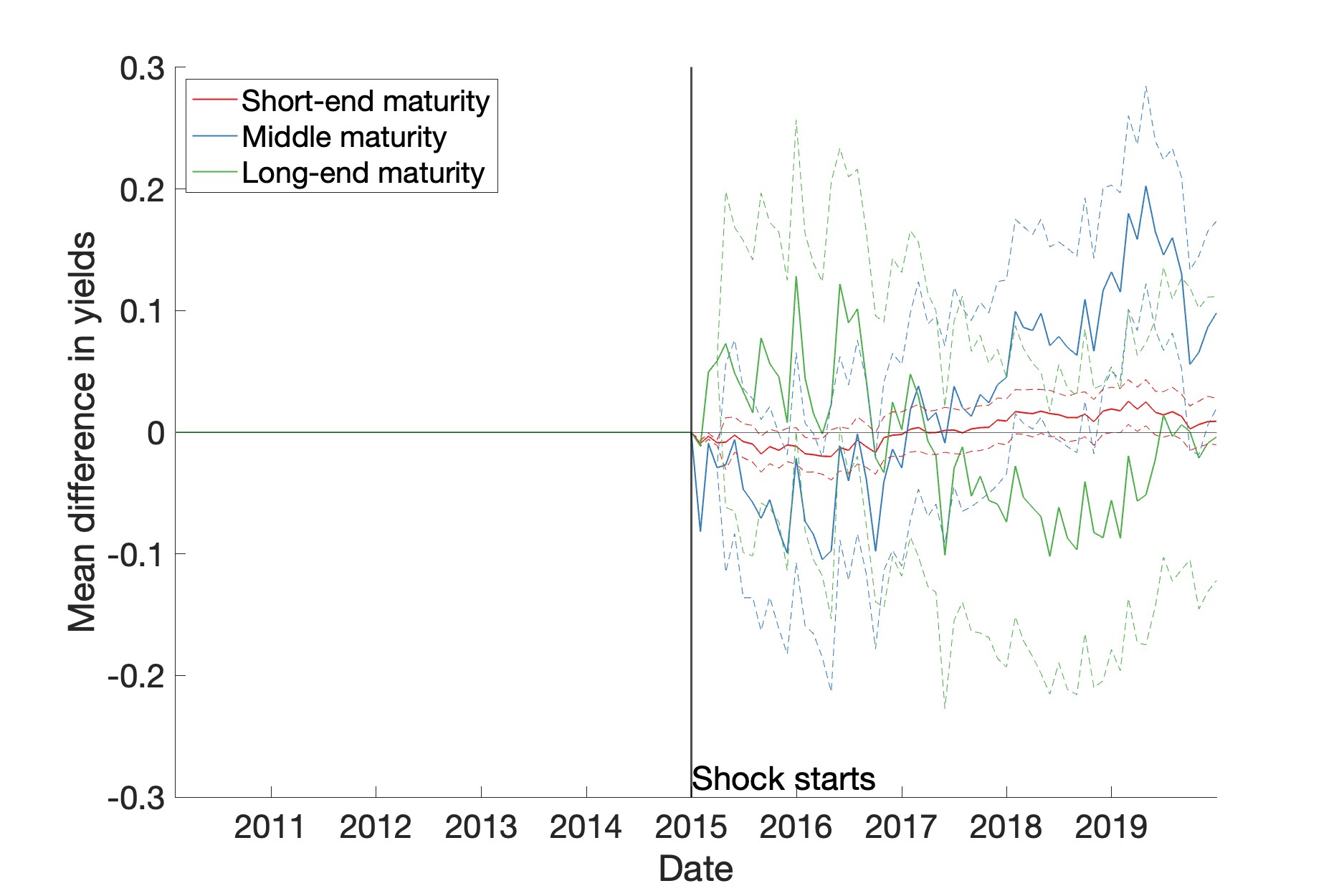}
        \caption{Case 2.1: the shock is applied to the short-end maturities of US Treasury.}
    \end{subfigure}
    \hfill
    \begin{subfigure}{0.45\textwidth}
        \includegraphics[width=\textwidth]{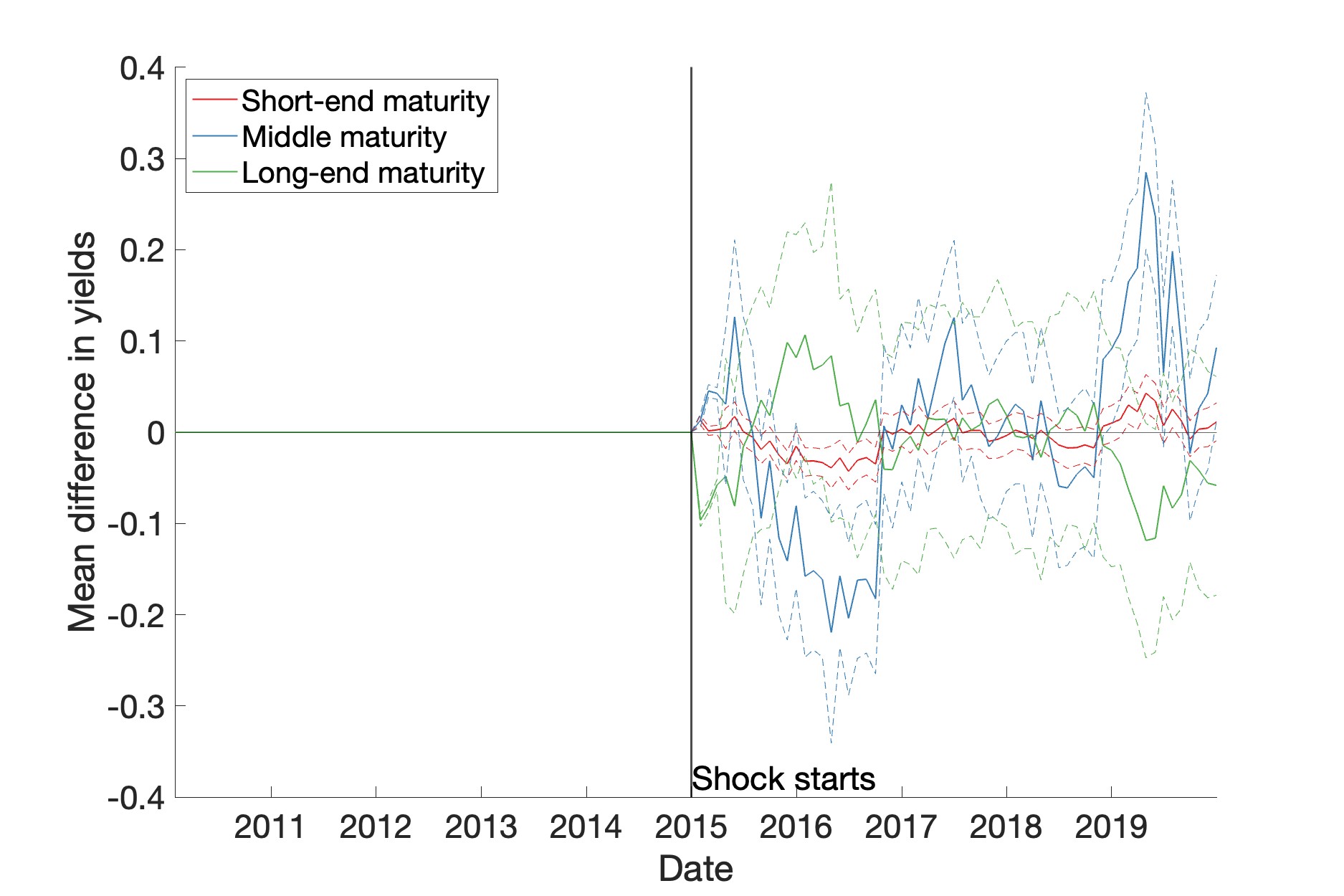}
        \caption{Case 2.2: the shock is applied to the middle maturities of US Treasury.}
    \end{subfigure}
    \hfill
    \begin{subfigure}{0.45\textwidth}
        \includegraphics[width=\textwidth]{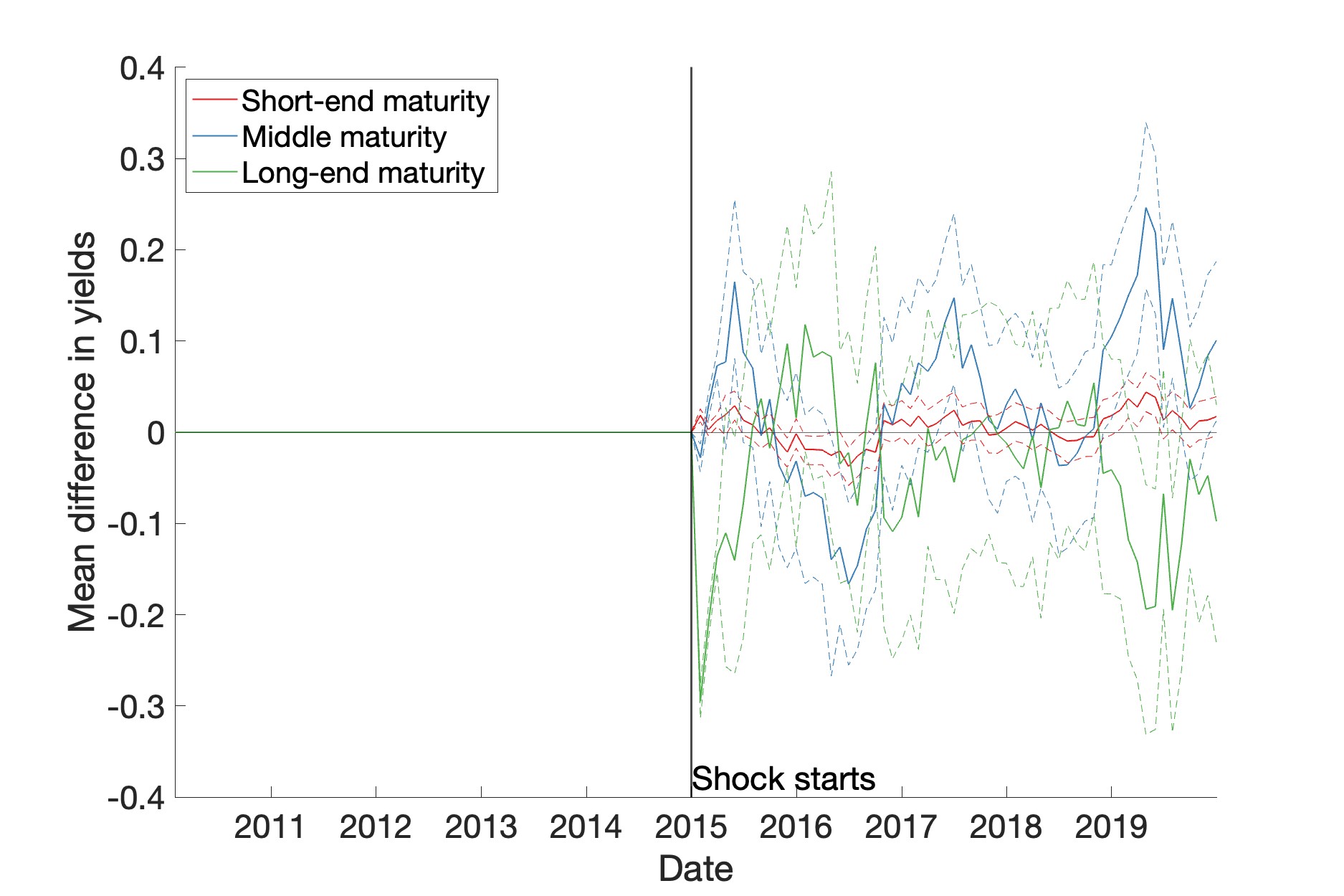}
        \caption{Case 2.3: the shock is applied to the long-end maturities of US Treasury.}
    \end{subfigure} 
    \hfill
    \begin{subfigure}{0.45\textwidth}
        \includegraphics[width=\textwidth]{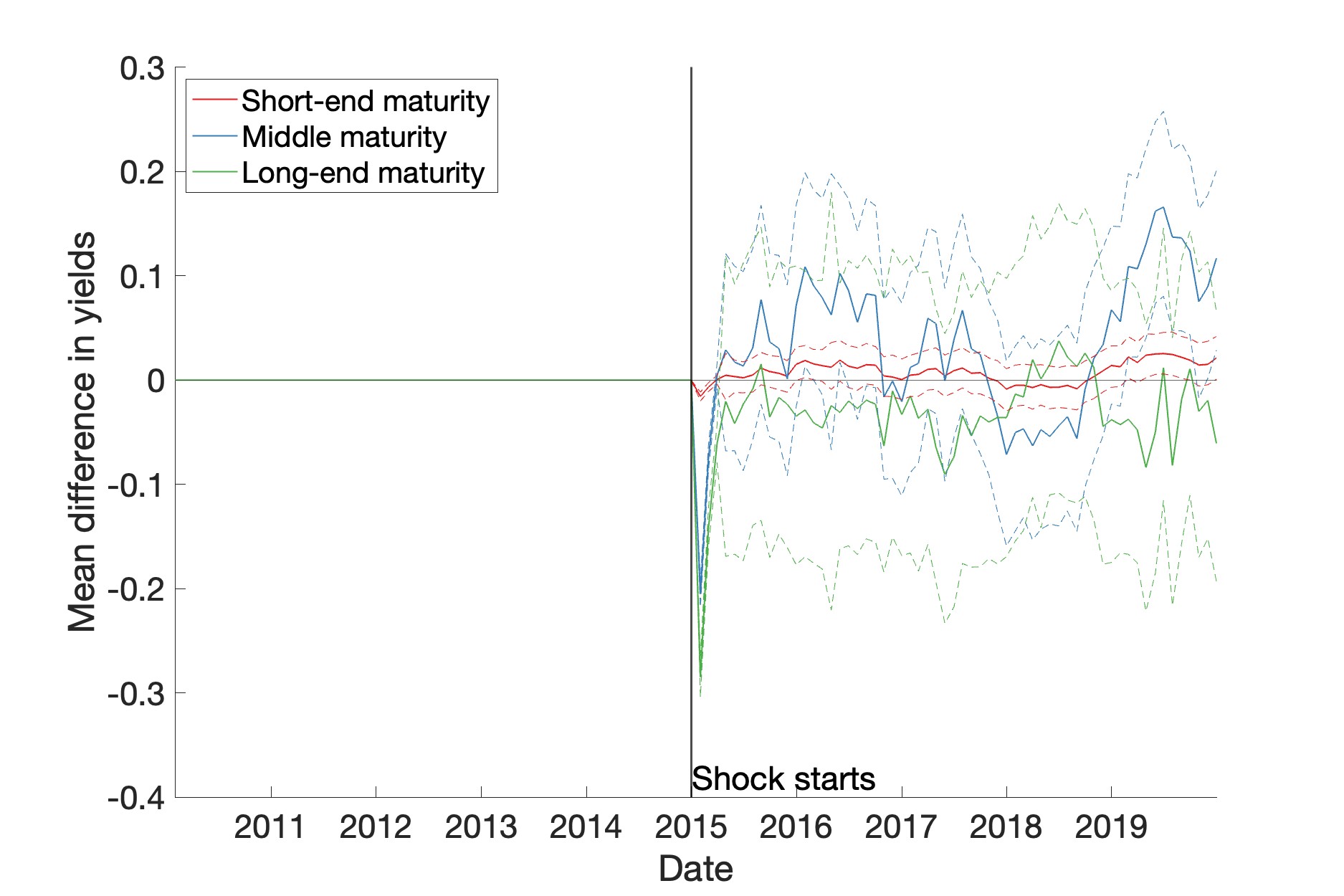}
        \caption{Case 2.4: the shock is applied to all maturities of US Treasury.}
    \end{subfigure}  
    \caption{Mean difference in percentage points of in-sample estimations between stress testing scenario 2 and original data for UK yields. The mean values are taken over short-end ($(0, 5]$ years), middle ($(5, 10]$ years), and long-end ( $(10, 30]$ years) maturities. Dashed lines represent the lower and upper bounds of 95\% confidence interval for each curve. }
    \label{fig:in_sample_diff_st2}
\end{figure}

Figures \ref{fig:in_sample_diff_st1} and \ref{fig:in_sample_diff_st2} show the mean difference in in-sample estimation for UK yields under temporary and permanent shock scenarios, respectively, compared to the estimation under original US Treasury data. Each sub-figure corresponds to cases 1.1 to 1.4 and 2.1 to 2.4. We categorise bonds into three classes: short-end ($(0, 5]$ years), middle ($(5, 10]$ years), and long-end ($(10, 30]$ years). The mean value is taken over all maturities in each class at each time point. The dashed lines represent the lower and upper bounds of the 95\% confidence interval \footnote{The confidence interval is calculated numerically. We first simulate $n=1000$ samples from the distribution $\boldsymbol{X}_t | \boldsymbol{Z}_{1:t} \sim N(\boldsymbol{a}_t, \boldsymbol{P}_t)$ at each time point $t$ using the original US yields and the yields in each stress testing scenario, where $\boldsymbol{a}_t$ and $\boldsymbol{P}_t$ are given in Equations \eqref{eq:kf_update_mean} and \eqref{eq:kf_update_cov}. Then, we calculate the estimated yields by $\boldsymbol{Y}_t = \boldsymbol{\Lambda} \boldsymbol{X}_t + \boldsymbol{\Lambda} \boldsymbol{\mu} + \boldsymbol{\Gamma} \boldsymbol{U}_t$. The lower and upper bounds of the confidence interval are the 2.5th and 97.5th percentiles of the mean difference when using the original US data and each stress testing data.}, while the vertical black lines indicate the start and end of the shock period for the temporary shocks.

We first consider the effects of the temporary shocks in Figure \ref{fig:in_sample_diff_st1}. When the shock is applied to the short-end maturities of the US Treasury yield curve (case 1.1), the effect on UK yields is limited. The confidence intervals cover 0 at almost all time points for all three classes. The short-end maturities of UK yields may have some changes during the shock but will return to normal levels once the shock ends. When the shock is applied to the middle maturities of the US Treasury (case 1.2), there is a statistically significant effect on the short-end and middle maturities of UK yields. The long-end maturities of UK yields are not affected. Interestingly, even though the shock ended in December 2015, the effects persist in the long run. In case 1.3, when the shock is applied to the long-end maturities of the US Treasury, the entire yield curve is affected, but the long-end maturities are most impacted. Even after the shock, the mean differences in the estimation of long-end maturities remain significantly away from 0. Finally, in case 1.4, when the shock is applied to the entire US Treasury yield curve, the mean differences in the estimation of middle maturities of UK yields are affected in some intervals, for example, from July 2015 to September 2016, and from September 2017 to December 2019, while the short-end and long-end maturities are not significantly affected.

One should note that for the temporary shock, even though it ended in December 2015, some maturities of UK yields are still affected in the long run in some cases, such as the middle maturities in case 1.2 and the long-end maturities in case 1.3. This can be explained as follows. In a regression model, the change in the covariate at time $t$ only affects the response variable at time $t$. However, if the covariate is autoregressive, the change in the covariate at time $t$ will be accumulated over a long period.

Now we move on to the permanent shock in Figure \ref{fig:in_sample_diff_st2}. Unlike the temporary shock, a permanent shock usually does not affect the UK yields in the long run. During the first few months after the shock starts, all the short-end, middle, and long-end maturities of UK yields react to this shock. However, after that, the yield curve returns to normal levels and fluctuates in the short run. The permanent shock affects the middle and long-end maturities more than the short-end maturities in all four cases. 

Another notable feature is the green peaks almost immediately after the shocks start in cases 1.3/2.3 and 1.4/2.4 in Figures \ref{fig:in_sample_diff_st1} and \ref{fig:in_sample_diff_st2}. This indicates that whenever a temporary or permanent shock is applied to the long-end maturities or the entire US Treasury yield curve, the long-end maturities of UK bonds react immediately. However, the UK bond market may overreact to this shock. Subsequently, the effects of this shock decrease, and the yields return to normal levels in a few months. If the shock is temporary, when it ends, the UK bond market, especially the long-end maturities, will overreact again. This is because, in our stress testing setup, when a temporary shock ends, the US Treasury yields are halved, causing another dramatic change in the US Treasury yield curve.

\begin{figure}[h]
    \centering
    \begin{subfigure}{0.3\textwidth}
        \includegraphics[width=\textwidth]{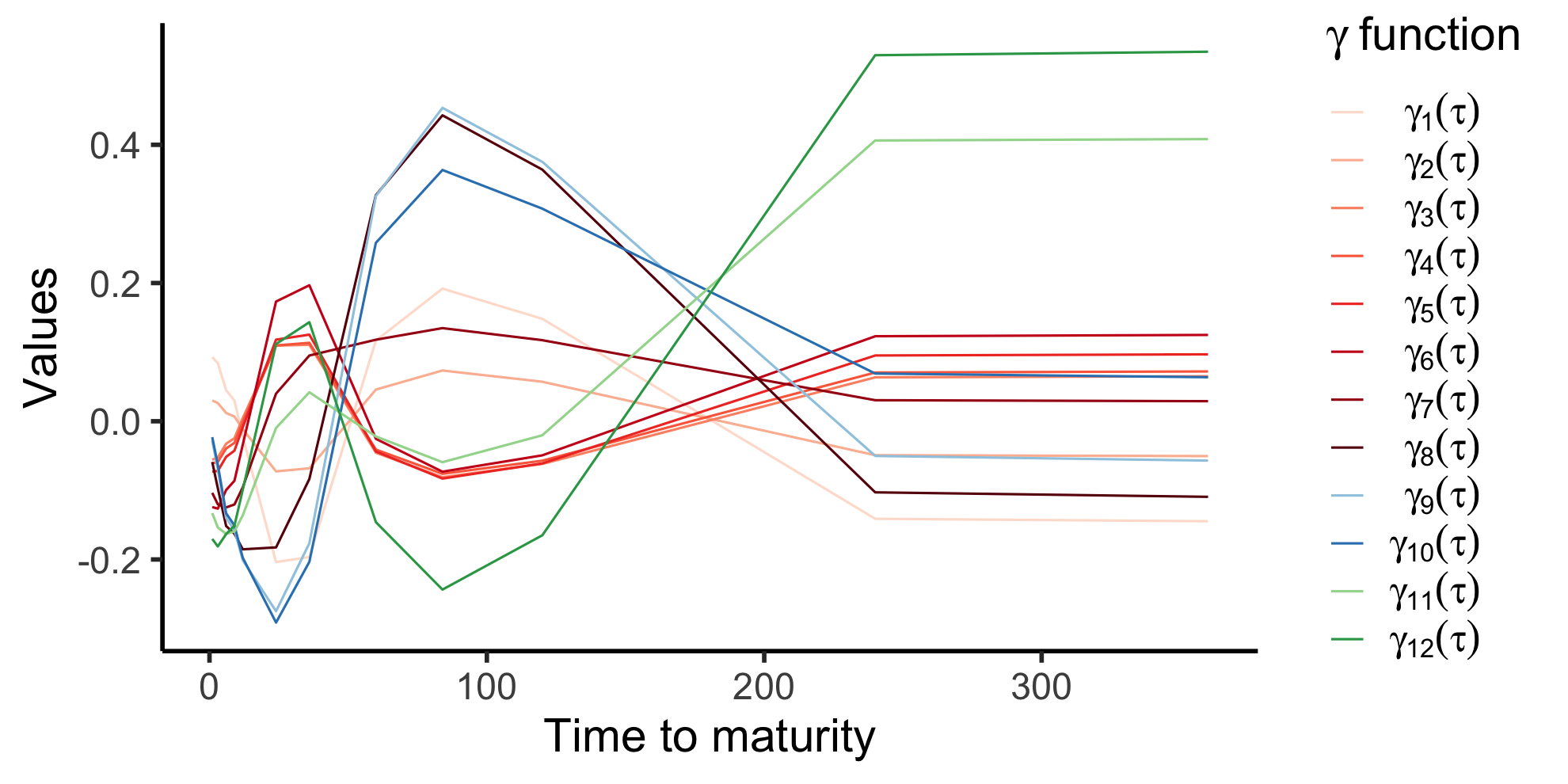}
        \caption{Functional coefficient for original data.}
    \end{subfigure}
    \hfill
    \begin{subfigure}{0.3\textwidth}
        \includegraphics[width=\textwidth]{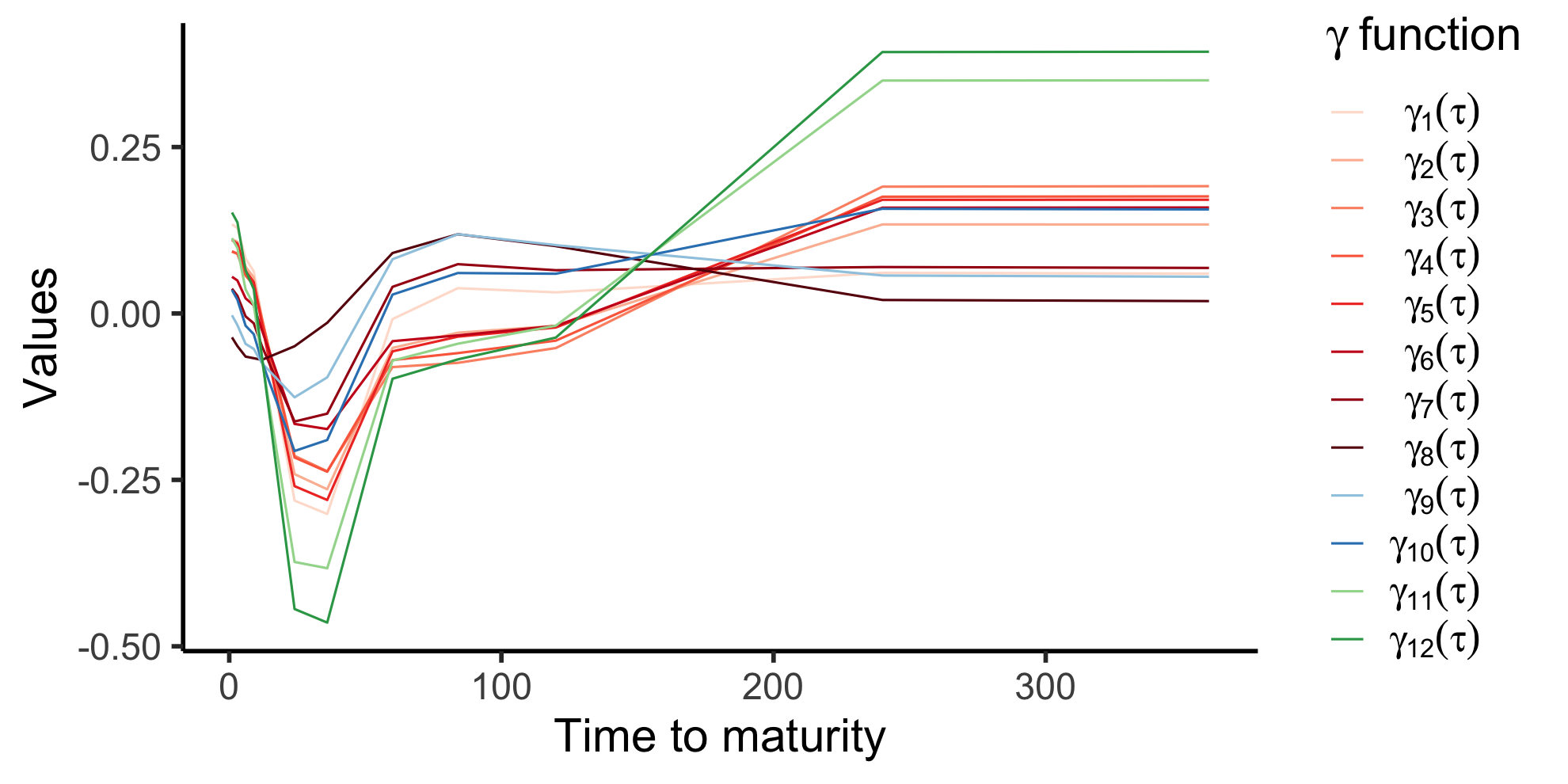}
        \caption{Functional coefficient for stress testing 1.}
    \end{subfigure}
    \hfill
    \begin{subfigure}{0.3\textwidth}
        \includegraphics[width=\textwidth]{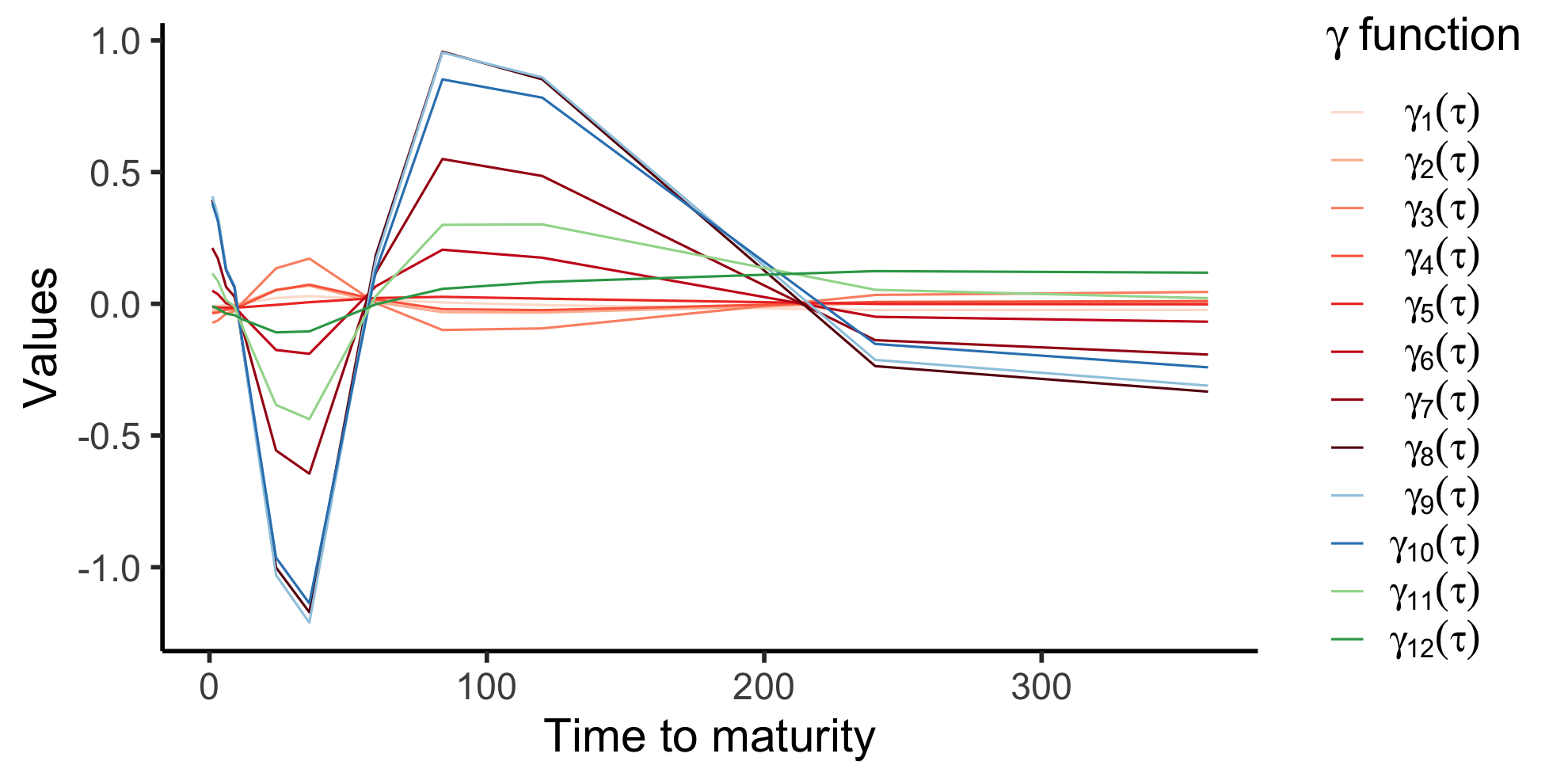}
        \caption{Functional coefficient for stress testing 2.}
    \end{subfigure}    
    \caption{Functional coefficients for UK yields.}
    \label{fig:func_coe}
\end{figure}

Finally, Figure \ref{fig:func_coe} provides the estimated functional coefficients for the UK yields for the original US Treasury data, stress testing scenario 1 data, and stress testing scenario 2 data. For the stress testing data, we only consider cases 1.4 and 2.4, i.e., shocks applied to the entire yield curve. In each sub-figure, different colours represent different functional coefficients $\gamma_i(\tau)$ for $i \in \{ 1, \dots, 12 \}$, in Equation \eqref{eq:DNS_FR_m}.

For the original US Treasury data, the long-end maturities of UK bonds are mostly affected by the long-end maturities of the US Treasury ($\gamma_{11}(\tau), \gamma_{12}(\tau)$), while the middle maturities of UK bonds are mostly affected by the middle maturities of the US Treasury ($\gamma_{9}(\tau), \gamma_{10}(\tau)$). All other maturities are evenly affected by the entire US Treasury yield curve. When a temporary shock is applied to the US Treasury, these relationships change slightly. The long-end maturities of UK bonds are still affected by the long-end maturities of the US Treasury, but they are also affected by the 1-year, 2-year, and 3-year US Treasury. The effects of the middle maturities of the US Treasury decrease. All other UK bonds with maturities less than or equal to 10 years are evenly affected by the entire US Treasury yield curve. As for the permanent shock, the effects of the long-end maturities of the US Treasury disappear. Instead, the middle maturities of the US Treasury contribute more to the UK bonds, especially the middle maturities of UK bonds ($\gamma_{8}(\tau), \gamma_{9}(\tau), \gamma_{10}(\tau)$).

\FloatBarrier

\subsection{Case Study: Bond Ladder Portfolio}
\label{sec:result_bond_ladder}

\begin{figure}[h]
    \centering
    \includegraphics[width=0.9\textwidth]{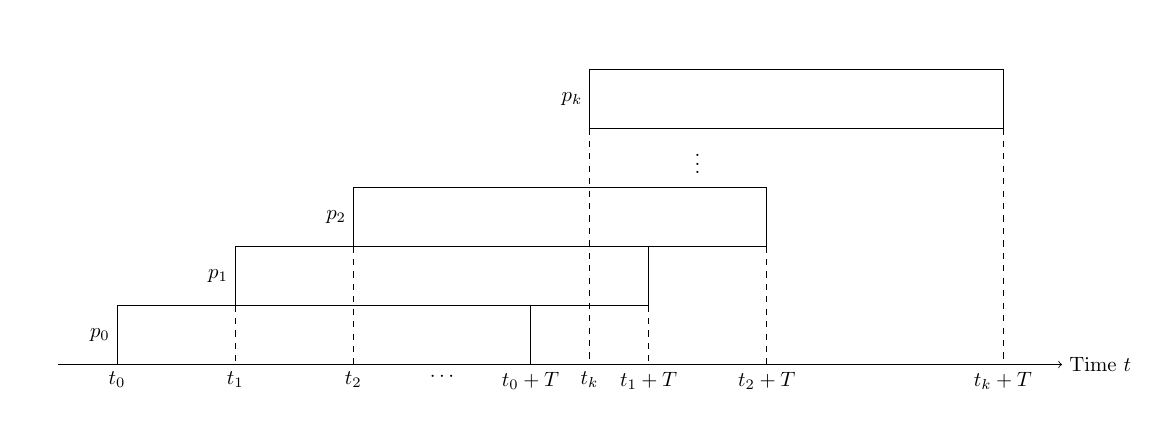}
    \caption{Construction of bond ladder portfolio. }
    \label{fig:bond_ladder}
\end{figure}

In this section, we present a case study of a bond ladder portfolio for risk management purposes. A bond ladder portfolio can be described as follows. Assume a US investor wants to construct an investment strategy consisting of cash and a UK bond with a maturity of $T$ months. At time $t_i$, the investor spends $p_i$ dollars to purchase bonds and deposits the remaining amount in a bank account. We denote the number of bonds as $N_i$ and the value of the portfolio at time $t_i$ as $W_i$ for $i \in \{ 0, 1, \dots, k \}$. The initial wealth $W_0$ is 12 million USD. For simplicity, we assume $p_1 = p_2 = \dots = p$, meaning the investor spends the same amount on bonds each month. The value of the portfolio $W_t$ can be calculated as follows, assuming the interest rate is compounded monthly and the investment interval is one month:
\begin{equation}
    W_i = \sum_{j=0}^k N_j F_i e^{-\tau_j Y_i(\tau_j)} \boldsymbol{1}_{\{ t_j \le t_i, \tau_j \ge 0\}} + \left( 1+\frac{r_i}{12} \right) C_{i-1}
\end{equation}
where $F_i$ is the face value in USD at time $t_i$ \footnote{The face value in GBP is usually a constant. However, as the GBP/USD exchange rate changes over time, the face value in USD also changes.}, $r_i$ is the risk-free interest rate \footnote{In this paper, we used the effective federal funds rate (EFFR) as a proxy for the risk-free interest rate. The EFFR data was sourced from the website of the Federal Reserve Bank of New York. Although the original data is provided on a daily basis, we converted it to monthly data for consistency.} for the period $[t_{i-1}, t_{i}]$, $\boldsymbol{1}_{\{ \cdot \}}$ is the indicator function, and $C_{i-1}$ is the amount in the cash account at time $t_{i-1}$. The number of bonds $N_j$ is calculated as:
\begin{equation}
    N_j = \frac{p}{F_j e^{-T Y_j(T)}}. 
\end{equation}
Figure \ref{fig:bond_ladder} illustrates the entire process of constructing the bond ladder portfolio.

In this study, we assume the investor makes 13 investments ($k+1=13$) over one year, from December 2019 to December 2020. The EFFR data and the GBP/USD exchange rate \footnote{Data was obtained from \url{https://www.tradingview.com/}.} for this period are given in Table \ref{tbl:EFFR}. Furthermore, we assume the bond yield for the first investment (made in December 2019) is known, while the yields for the subsequent 12 investments are predicted using the DNS-FR model \footnote{In this case study, the time to maturity $\tau_i$ can take any integer between 1 and $T$. However, we only know the coefficients $\gamma_{i,j}$ at $\tau_i=$ 1, 3, 6, 9 months, and 1, 2, 3, 5, 7, 10, 20, and 30 years. In this case, we first forecast the yields at those points, then take a linear interpolation to fill all the values between any two points.}. We consider three main questions:
\begin{enumerate}
    \item How does the value of the portfolio change over time?
    \item How do different stress tests affect the value of the portfolio?
    \item How does the value of the portfolio change with different maturities of the underlying bond?
\end{enumerate}
For the second question, we consider the stress testing scenarios 1.4 and 2.4 described in Section \ref{sec:result_stress}, i.e., the shocks are applied to the entire yield curve. For the third question, we consider three different bonds with maturities of 6 months, 1 year, and 30 years, respectively.

\begin{table}[width=.6\linewidth,cols=3,pos=h]
    \caption{Monthly EFFR data and the GBP/USD exchange rate for the year 2020.}
    \label{tbl:EFFR}
    \begin{tabular*}{\tblwidth}{@{} LLL @{}}
        \toprule
         Date & EFFR & GBP/USD exchange rate \\
         \midrule
         December 2019 & - & 1.32018 \\
         January 2020 & 1.59\% & 1.28231 \\
         February 2020 & 1.59\% & 1.24086 \\
         March 2020 & 0.08\% & 1.25907 \\
         April 2020 & 0.05\% & 1.23455 \\
         May 2020 & 0.05\% & 1.23992 \\
         June 2020 & 0.08\% & 1.30770 \\
         July 2020 & 0.10\% & 1.33690 \\
         August 2020 & 0.09\% & 1.29115 \\
         September 2020 & 0.09\% & 1.29457 \\
         October 2020 & 0.09\% & 1.33173 \\
         November 2020 & 0.09\% & 1.36561 \\
         December 2020 & 0.09\% & 1.36893 \\
         \bottomrule
    \end{tabular*}
\end{table}

\begin{figure}[h]
    \centering
    \includegraphics[width=0.6\textwidth]{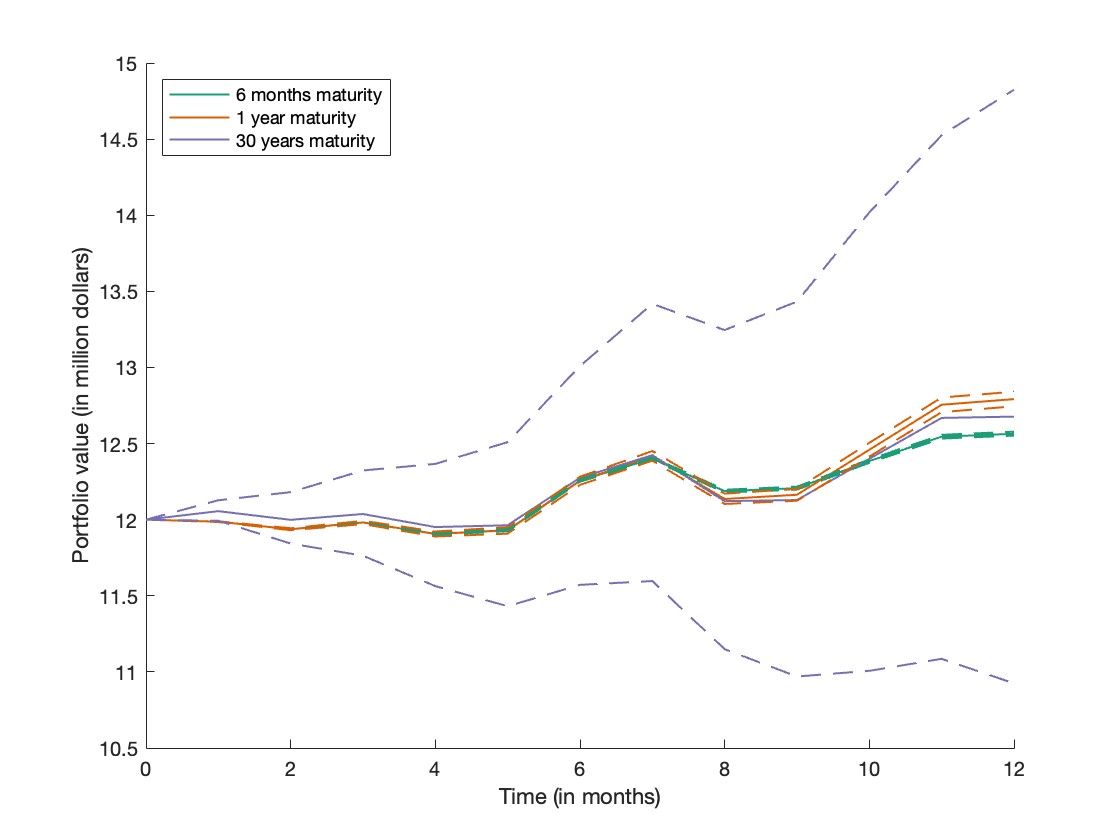}
    \caption{Predicted portfolio values.}
    \label{fig:portfolio_value}
\end{figure}

Figure \ref{fig:portfolio_value} shows the predicted portfolio values for the next 12 months for underlying bonds with maturities of 6 months, 1 year, and 30 years. The monthly investment amount is \$1 million for all maturities. For the first five months, the portfolio values of all bonds are lower than the initial wealth of \$12 million, mainly due to the decrease in the exchange rate. This also explains the decreases in portfolio values in the 8th and 9th months. At the end of 12 months, the 1-year bond has the maximum predicted portfolio value, followed by the 30-year bond, and the 6-month bond has the minimum predicted portfolio value. However, the differences among the three bonds are small. Another notable observation is that as maturities increase, so does risk. For example, at the end of 12 months, the portfolio value of the 6-month bond is roughly \$12.5 million with no uncertainty, while the portfolio value of the 30-year bond could be as high as \$14.5 million or as low as \$11 million, reflecting the common understanding that long-term debt usually carries higher risk.

\begin{figure}[h]
    \centering
    \begin{subfigure}{0.3\textwidth}
        \includegraphics[width=\textwidth]{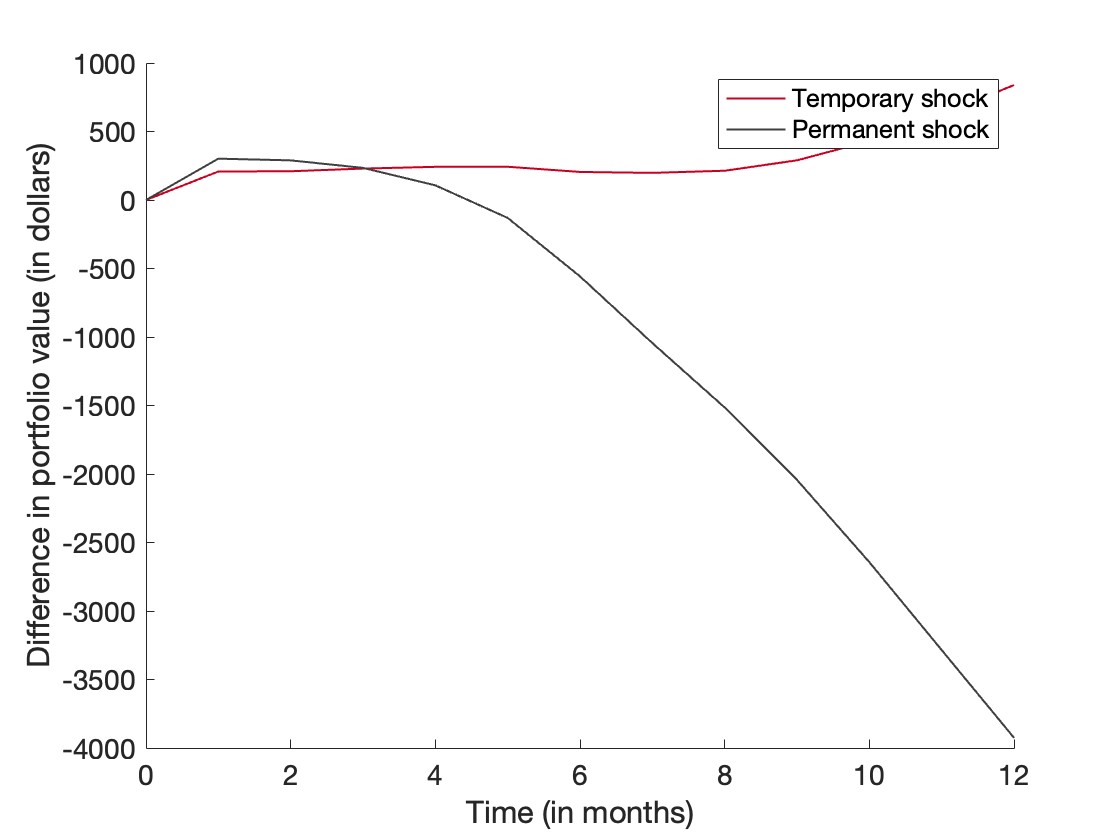}
        \caption{6-month maturity.}
    \end{subfigure}
    \hfill
    \begin{subfigure}{0.3\textwidth}
        \includegraphics[width=\textwidth]{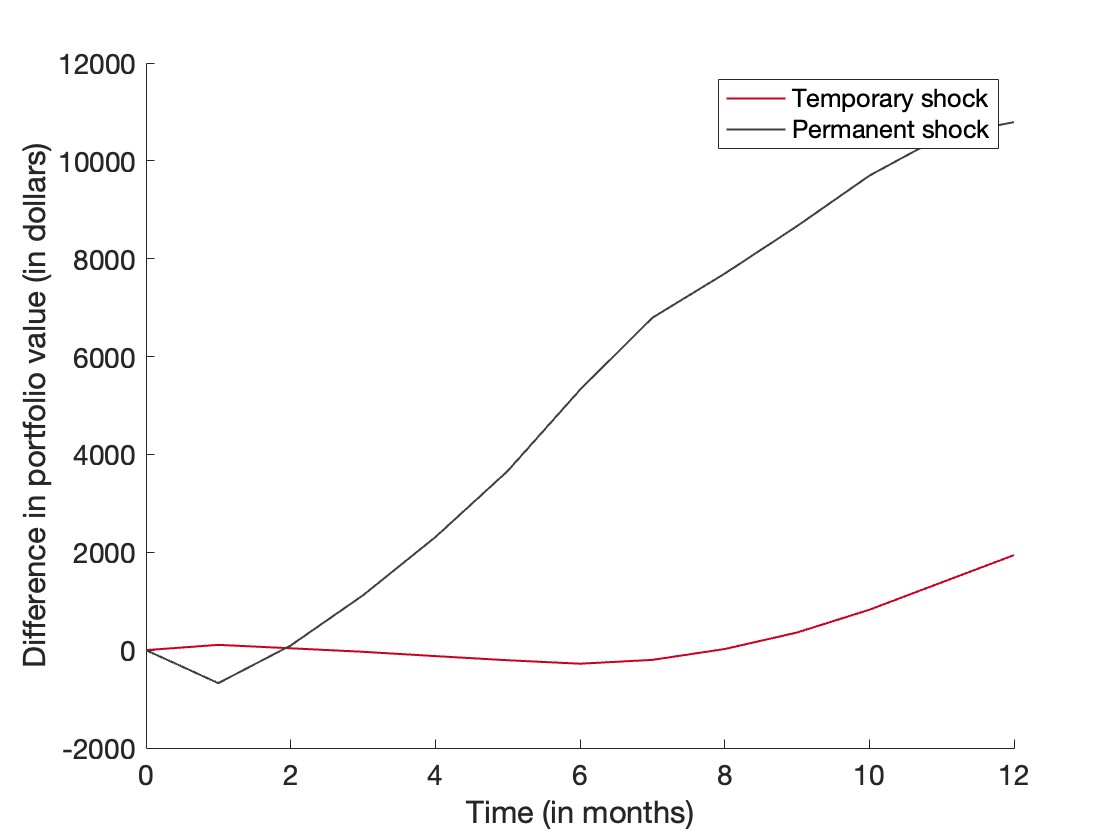}
        \caption{1-year maturity.}
    \end{subfigure}
    \hfill
    \begin{subfigure}{0.3\textwidth}
        \includegraphics[width=\textwidth]{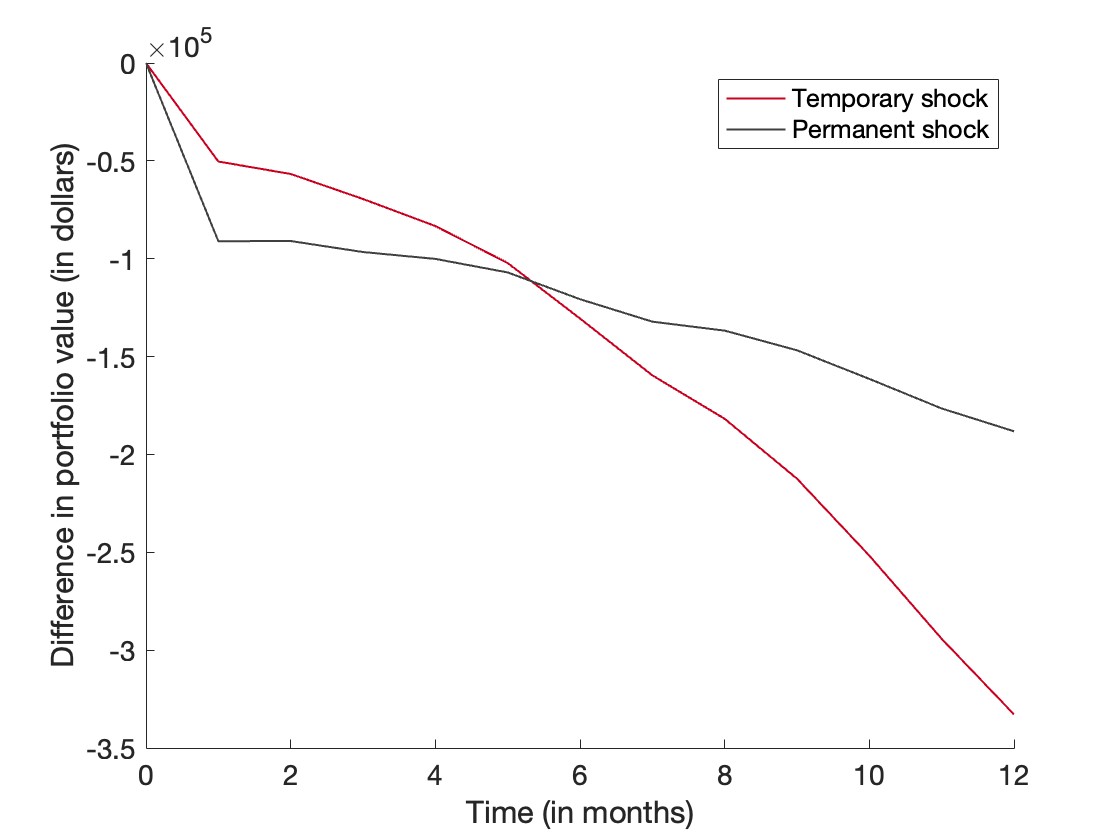}
        \caption{30-year maturity.}
    \end{subfigure}
    \caption{Differences of portfolio values between each stress testing scenario and original data, for different maturity of the underlying bond.}
    \label{fig:diff_st_bl}
\end{figure}

Figure \ref{fig:diff_st_bl} illustrates the differences in portfolio values between different shocks and the original data for the 6-month, 1-year, and 30-year bonds, respectively. A positive value indicates that the portfolio is worth more under the shock, while a negative value indicates that it is worth less. For short-end maturities (6-month and 1-year), the temporary shock has limited effects, but the permanent shock significantly impacts portfolio values. However, for the long-end maturity (30-year), the opposite is true. The permanent shock in 2015 has a larger effect on the portfolio values. Although this shock dissipated by the end of 2015, it continues to affect the long-end curve over an extended period.

\begin{figure}[h]
    \centering
    \includegraphics[width=0.6\textwidth]{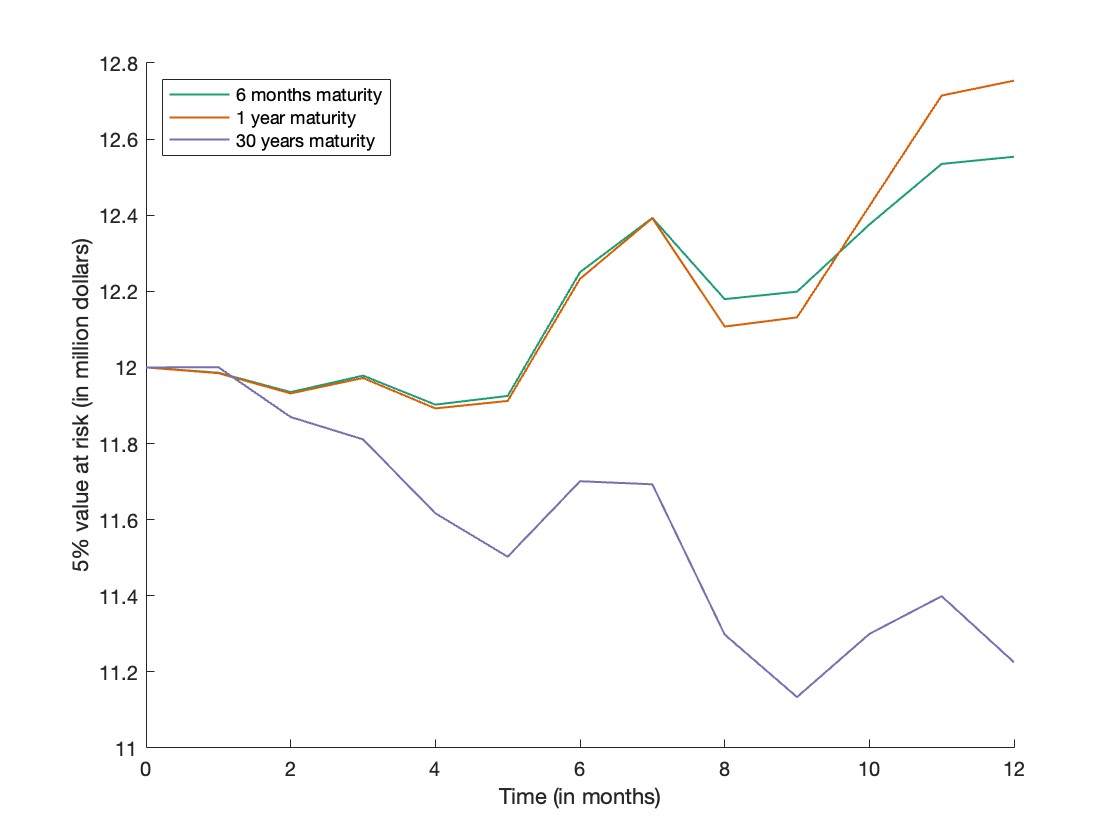}
    \caption{5\% value-at-risk (VaR) for bond ladder portfolios with different maturities.}
    \label{fig:VaR}
\end{figure}

In reality, VaR also attracts considerable attention. People are particularly concerned with the maximum potential loss under normal market conditions after excluding the worst outcomes with a total probability of $p\%$. Figure \ref{fig:VaR} shows the 5\% VaR for the portfolio value of bonds with 6-month, 1-year, and 30-year maturities, which is calculated numerically. At the end of 12 months, the 5\% VaR of the portfolio values of the 6-month and 1-year bonds are roughly \$12.5 and \$12.7 million, respectively, indicating a minimal chance of loss with these bonds even in the worst cases. In contrast, the 5\% VaR of the portfolio value of the 30-year bond is only about \$11.2 million, suggesting that investing in a 30-year bond could lead to a loss over a 1-year period in the worst case.

\begin{figure}[h]
    \centering
    \begin{subfigure}{0.3\textwidth}
        \includegraphics[width=\textwidth]{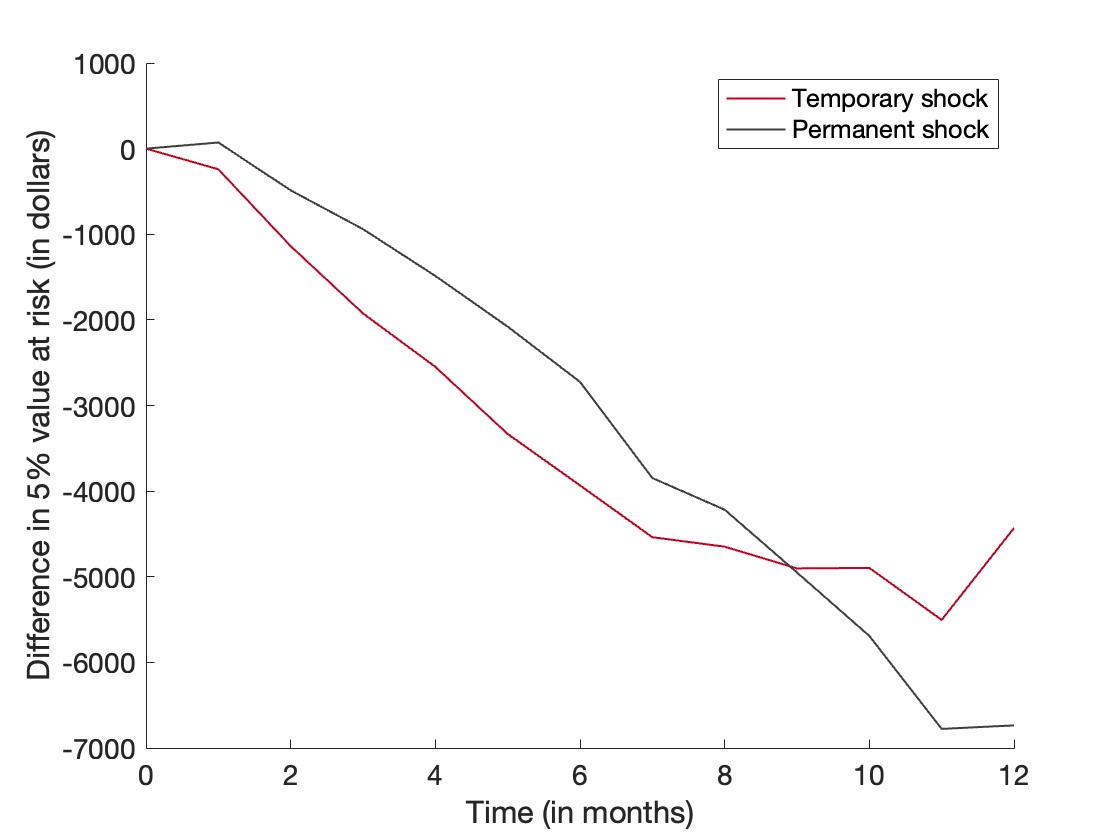}
        \caption{6-month maturity.}
    \end{subfigure}
    \hfill
    \begin{subfigure}{0.3\textwidth}
        \includegraphics[width=\textwidth]{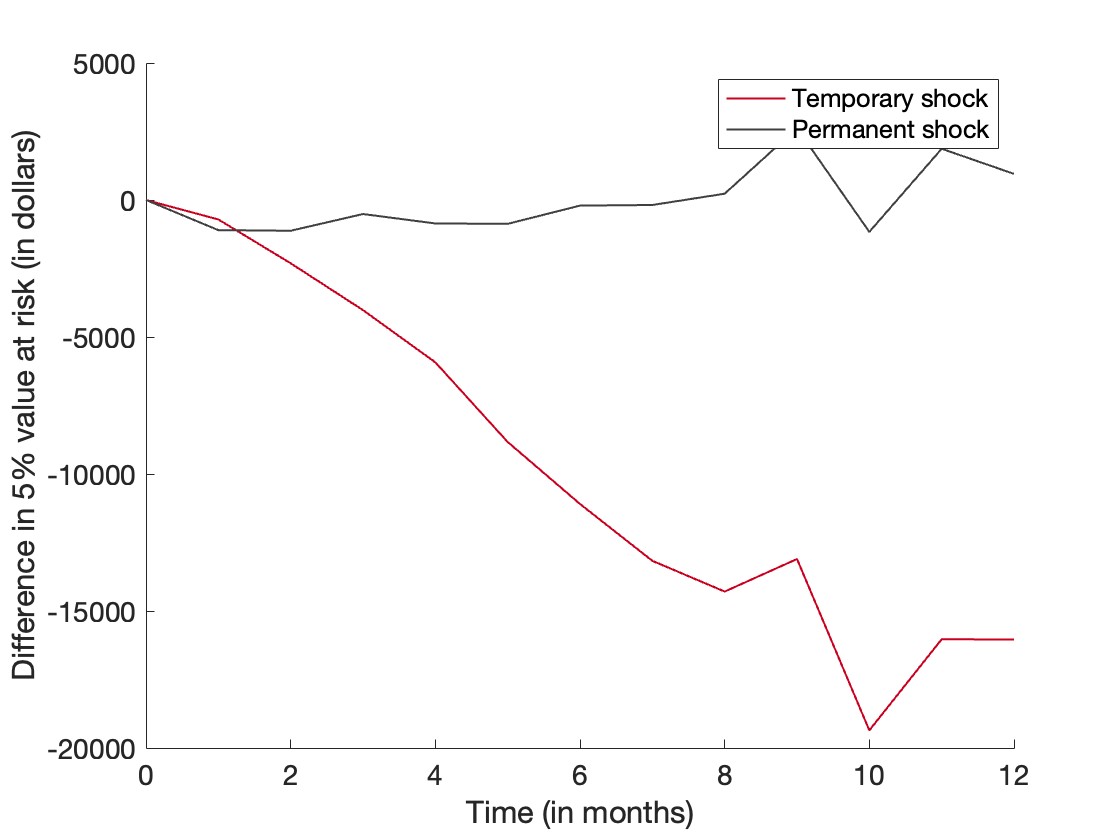}
        \caption{1-year maturity.}
    \end{subfigure}
    \hfill
    \begin{subfigure}{0.3\textwidth}
        \includegraphics[width=\textwidth]{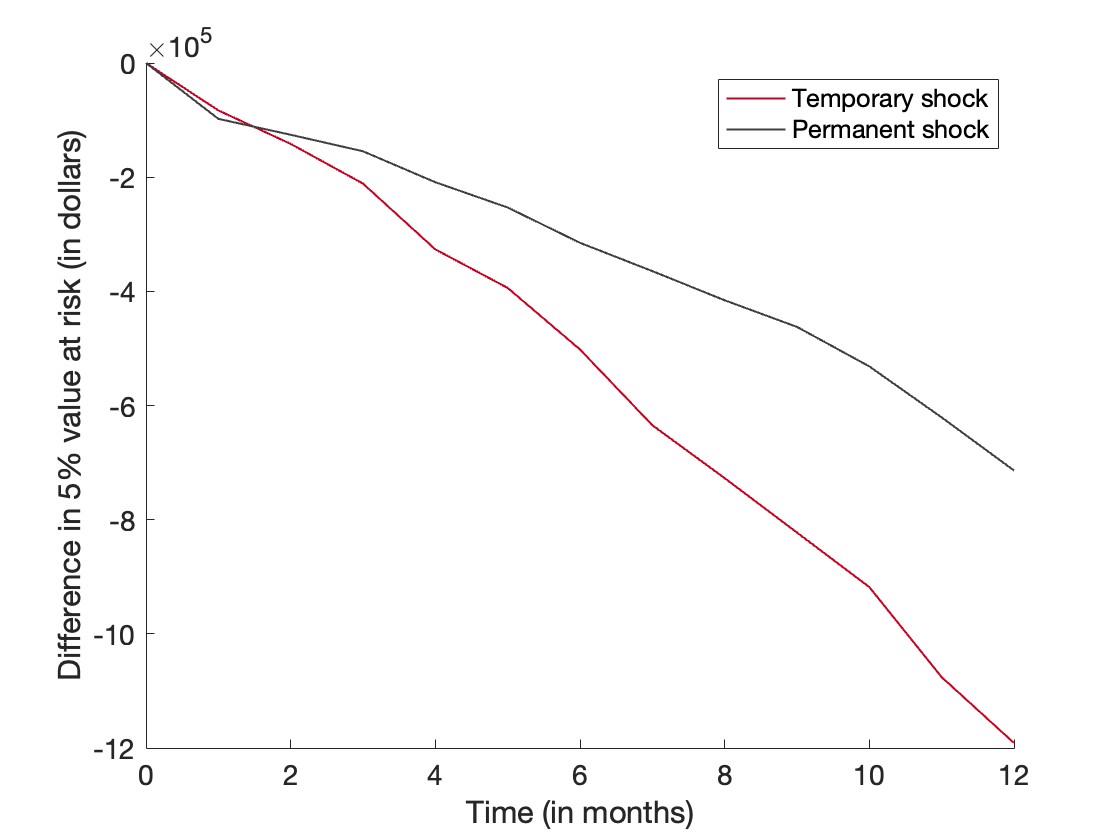}
        \caption{30-year maturity.}
    \end{subfigure}
    \caption{Differences of 5\% value-at-risk (VaR) between each stress testing scenario and original data, for different maturity of the underlying bond.}
    \label{fig:diff_VaR}
\end{figure}

Figure \ref{fig:diff_VaR} shows the difference in 5\% VaR between each stress testing scenario and the original US Treasury data for different maturities of UK bonds. Overall, if a shock is applied to the US Treasury, the 5\% VaR of the portfolio value of UK bonds will decrease. For the 6-month bond, both the temporary shock and the permanent shock have a similar effect on 5\% VaR. For the 1-year bond, the permanent shock has a limited effect on 5\% VaR, while the temporary shock has a significant influence. For the 30-year bond, both temporary and permanent shocks have a substantial influence on 5\% VaR, with the influence of the temporary shock being stronger. Another notable feature is the magnitude of the difference in 5\% VaR. For the 6-month bond, a shock can cause a decrease in 5\% VaR up to \$7,000 at the end of 12 months, while for the 30-year bond, a temporary shock causes a decrease of \$1,200,000 in 5\% VaR. Long-term bonds are more sensitive to the shock.

\FloatBarrier

\section{Conclusion}
\label{sec:conclusion}

The dynamic Nelson-Siegel (DNS) model has been pivotal in yield curve estimation over the past two decades. However, it falls short in capturing the relative spread between two economies. In this paper, we introduce a novel dynamic Nelson-Siegel functional regression (DNS-FR) model, which extends the DNS model by incorporating the relative spread between a reference country and a response country. To address estimation challenges, we use kernel principal component analysis (kPCA) to transform the functional regression into a finite-dimensional estimation problem. The Kalman Filter method is then employed to estimate unknown parameters and hidden state variables by maximising the marginal likelihood function.

Our empirical analysis, which includes eight countries/regions with the US as the reference country and seven response countries/regions, demonstrates that the DNS-FR model outperforms the DNS model in in-sample estimation, particularly for long-term bonds with maturities of 20 and 30 years, across all seven response countries/regions. Additionally, the DNS-FR model is superior in capturing unusual yield curve structures, such as local fluctuations and backwardation.

Furthermore, we conducted stress tests to analyse the effects of both temporary and permanent shocks applied to the US Treasury yield curve on the estimation of UK bond yields. The study reveals that middle and long-end maturities of UK bonds are mostly affected, regardless of whether the shock is applied to the short-end, middle, or long-end maturities of the US Treasury yield curve. Interestingly, for temporary shocks, their effects persist in the long run even after the shocks end.

Finally, we conducted a case study of a bond ladder portfolio, involving regular investments in UK bonds for risk management purposes. We forecasted the portfolio values 12 steps ahead with 6-month, 1-year, and 30-year bonds. The predicted portfolio values for the three bonds are similar, but the bond with a longer maturity carries higher risk, reflected in a wider confidence interval. The 5\% VaR suggests that investing in the 6-month and 1-year bonds is unlikely to result in a loss over one year, whereas investing in the 30-year bond may lead to a loss in the worst-case scenario. Furthermore, we analysed the influence of different types of shocks on the predicted portfolio values and VaR. The permanent shock has larger effects on short-end maturities, while the temporary shock, even though it ended a few years ago, has a more significant effect on long-end maturities.

\newpage

\appendix

\section{DNS Model Estimations}
\label{app:DNS}

\begin{table}[width=.9\linewidth,cols=9,pos=h]
    \caption{In-sample RMSE for DNS model using covariance structure 1.}
    \begin{tabular*}{\tblwidth}{@{} LLLLLLLLL @{}}
        \toprule
         Maturity & UK & FR & IT & DE & JP & AU & EU & US \\
         \midrule
         1 month & 0.2098 & 0.1217 & 0.0247 & 1.99e-14 & 1.08e-13 & 0.0068 & 0.1979 & 0.1214 \\
         3 months & 0.0890 & 0.0749 & 2.22e-15 & 0.0556 & 0.0025 & 1.08e-14 & 0.1256 & 0.0650 \\
         6 months & 0.0520 & 0.0318 & 0.0676 & 0.0647 & 0.0278 & 0.0031 & 0.0153 & 0.0378 \\
         9 months & 1.28e-15 & 0.0612 & 0.0719 & 0.0881 & 1.25e-14 & 5.43e-15 & 0.0621 & 2.83e-15 \\
         1 year & 0.0671 & 7.36e-16 & 4.18e-15 & 0.0937 & 0.0158 & 0.0110 & 0.0839 & 0.0537 \\
         2 years & 0.0130 & 0.0517 & 0.0953 & 4.59e-15 & 0.0140 & 0.0447 & 6.78e-16 & 1.08e-14 \\
         3 years & 0.0886 & 0.1237 & 0.1890 & 0.1081 & 0.0107 & 0.0826 & 0.1278 & 0.0922 \\
         5 years & 0.1687 & 0.1694 & 0.2604 & 0.1316 & 0.0144 & 0.0691 & 0.2068 & 0.0980 \\
         7 years & 0.0518 & 0.0668 & 0.1598 & 0.0389 & 0.1444 & 3.92e-15 & 0.1926 & 0.0268 \\
         10 years & 0.1781 & 0.2298 & 0.0221 & 0.1844 & 0.3146 & 0.1369 & 0.0493 & 0.1441 \\
         20 years & 0.5222 & 0.5228 & 0.2726 & 0.4993 & 0.9248 & 0.5017 & 0.2713 & 0.3931 \\
         30 years & 0.6001 & 0.6804 & 0.3863 & 0.5541 & 1.0806 & 0.6979 & 0.3156 & 0.5129 \\
         Mean & 0.1700 & 0.1778 & 0.1291 & 0.1516 & 0.2125 & 0.1295 & 0.1374 & 0.1287 \\
         \bottomrule
    \end{tabular*}
\end{table}

\begin{table}[width=.9\linewidth,cols=9,pos=h]
    \caption{In-sample RMSE for DNS model using covariance structure 3.}
    \begin{tabular*}{\tblwidth}{@{} LLLLLLLLL @{}}
        \toprule
         Maturity & UK & FR & IT & DE & JP & AU & EU & US \\
         \midrule
         1 month & 0.1211 & 0.1107 & 0.1117 & 0.0953 & 0.0919 & 0.1404 & 0.1213 & 0.0936 \\
         3 months & 0.0666 & 0.0762 & 0.0832 & 0.0836 & 0.0272 & 0.0914 & 0.0730 & 0.0462 \\
         6 months & 0.1791 & 0.1248 & 0.1322 & 0.0970 & 0.0662 & 0.0635 & 0.1240 & 0.1114 \\
         9 months & 0.1457 & 0.1761 & 0.1492 & 0.1554 & 0.1432 & 0.0748 & 0.1753 & 0.1030 \\
         1 year & 0.1833 & 0.1546 & 0.1539 & 0.1819 & 0.1813 & 0.0868 & 0.1978 & 0.1528 \\
         2 years & 0.1310 & 0.1378 & 0.1747 & 0.1231 & 0.1747 & 0.1056 & 0.1249 & 0.1034 \\
         3 years & 0.1172 & 0.1304 & 0.1491 & 0.1008 & 0.0609 & 0.1324 & 0.1099 & 0.0997 \\
         5 years & 0.2944 & 0.3120 & 0.2710 & 0.2409 & 0.2387 & 0.2397 & 0.2465 & 0.2169 \\
         7 years & 0.2955 & 0.3117 & 0.2548 & 0.2448 & 0.3152 & 0.2864 & 0.2933 & 0.2280 \\
         10 years & 0.2640 & 0.2540 & 0.2054 & 0.1932 & 0.3623 & 0.2661 & 0.1955 & 0.2127 \\
         20 years & 0.1606 & 0.1685 & 0.1227 & 0.1659 & 0.1294 & 0.1769 & 0.1754 & 0.1341 \\
         30 years & 0.1721 & 0.1871 & 0.1386 & 0.1521 & 0.1732 & 0.2646 & 0.1664 & 0.1581 \\
         Mean & 0.1775 & 0.1787 & 0.1622 & 0.1528 & 0.1637 & 0.1607 & 0.1669 & 0.1383 \\
         \bottomrule
    \end{tabular*}
\end{table}

\FloatBarrier
\newpage

\section{DNS-FR Model Estimations}
\label{app:DNS_FR}

\begin{table}[width=.9\linewidth,cols=8,pos=h]
    \caption{In-sample RMSE for DNS-FR model with 3 factors using covariance structure 1.}
    \begin{tabular*}{\tblwidth}{@{} LLLLLLLL @{}}
        \toprule
         Maturity & UK & FR & IT & DE & JP & AU & EU \\
         \midrule
         1 month & 0.0523 & 0.0554 & 0.0180 & 0.0131 & 0.0004 & 0.0045 & 0.0967 \\
         3 months & 0.0096 & 0.0387 & 3.25e-12 & 0.0412 & 0.0001 & 2.43e-13 & 0.0553 \\
         6 months & 0.0368 & 0.0316 & 0.0612 & 0.0397 & 0.0155 & 0.0020 & 0.0300 \\
         9 months & 0.0054 & 0.0334 & 0.0675 & 0.0430 & 1.11e-06 & 1.29e-10 & 0.0305 \\
         1 year & 0.0559 & 0.0043 & 1.51e-12 & 0.0432 & 0.0146 & 0.0076 & 0.0393 \\
         2 years & 0.0285 & 0.0458 & 0.0816 & 0.0105 & 0.0131 & 0.0304 & 0.0104 \\
         3 years & 0.0827 & 0.0495 & 0.0925 & 0.0373 & 0.0086 & 0.0547 & 0.0330 \\
         5 years & 0.1603 & 0.1104 & 0.0878 & 0.0664 & 0.0054 & 0.0381 & 0.0734 \\
         7 years & 0.1023 & 0.0438 & 7.21e-11 & 0.0340 & 0.0368 & 1.26e-09 & 0.0879 \\
         10 years & 0.1109 & 0.0770 & 0.1322 & 0.0406 & 0.0823 & 0.0885 & 0.0399 \\
         20 years & 0.0442 & 0.2360 & 0.3148 & 0.1275 & 0.2494 & 0.3725 & 0.1048 \\
         30 years & 0.0672 & 0.3620 & 0.4110 & 0.2222 & 0.3526 & 0.5353 & 0.1900 \\
         Mean & 0.0630 & 0.0907 & 0.1056 & 0.0599 & 0.0649 & 0.0945 & 0.0659 \\
         \bottomrule
    \end{tabular*}
\end{table}

\begin{table}[width=.9\linewidth,cols=8,pos=h]
    \caption{In-sample RMSE for DNS-FR model with 3 factors using covariance structure 3.}
    \begin{tabular*}{\tblwidth}{@{} LLLLLLLL @{}}
        \toprule
         Maturity & UK & FR & IT & DE & JP & AU & EU \\
         \midrule
         1 month & 0.0491 & 0.0577 & 0.0699 & 0.0480 & 0.0258 & 0.1142 & 0.0638 \\
         3 months & 0.0419 & 0.0646 & 0.0499 & 0.0495 & 0.0180 & 0.0829 & 0.0604 \\
         6 months & 0.0620 & 0.0679 & 0.1049 & 0.0510 & 0.0257 & 0.0598 & 0.0651 \\
         9 months & 0.0401 & 0.0734 & 0.0967 & 0.0607 & 0.0321 & 0.0579 & 0.0647 \\
         1 year & 0.0674 & 0.0830 & 0.0982 & 0.0614 & 0.0411 & 0.0615 & 0.0677 \\
         2 years & 0.0600 & 0.0943 & 0.1008 & 0.0496 & 0.0414 & 0.0710 & 0.0527 \\
         3 years & 0.0676 & 0.0728 & 0.0983 & 0.0522 & 0.0268 & 0.0846 & 0.0565 \\
         5 years & 0.1358 & 0.1775 & 0.1697 & 0.1115 & 0.0543 & 0.1395 & 0.1117 \\
         7 years & 0.0902 & 0.1783 & 0.1776 & 0.1192 & 0.0892 & 0.1758 & 0.1519 \\
         10 years & 0.1208 & 0.1846 & 0.1645 & 0.1219 & 0.1298 & 0.1694 & 0.1166 \\
         20 years & 0.0767 & 0.1163 & 0.0796 & 0.0979 & 0.0594 & 0.1299 & 0.0979 \\
         30 years & 0.0812 & 0.1204 & 0.0933 & 0.0974 & 0.0914 & 0.2160 & 0.1040 \\
         Mean & 0.0744 & 0.1076 & 0.1086 & 0.0767 & 0.0529 & 0.1135 & 0.0844 \\
         \bottomrule
    \end{tabular*}
\end{table}

\begin{table}[width=.9\linewidth,cols=8,pos=h]
    \caption{In-sample RMSE for DNS-FR model with 2 factors using covariance structure 2.}
    \begin{tabular*}{\tblwidth}{@{} LLLLLLLL @{}}
        \toprule
         Maturity & UK & FR & IT & DE & JP & AU & EU \\
         \midrule
         1 month & 0.0616 & 0.0520 & 0.0234 & 7.70e-16 & 0.0008 & 0.0035 & 0.1039 \\
         3 months & 0.0146 & 0.0568 & 0.0034 & 0.0461 & 7.03e-15 & 0.0001 & 0.0609 \\
         6 months & 0.0540 & 0.0353 & 0.0654 & 0.0479 & 0.0169 & 0.0007 & 0.0451 \\
         9 months & 0.0267 & 0.0416 & 0.0669 & 0.0489 & 0.0007 & 0.0043 & 0.0337 \\
         1 year & 0.0735 & 0.0090 & 1.11e-14 & 0.0548 & 0.0150 & 0.0122 & 0.0494 \\
         2 years & 0.0108 & 0.0562 & 0.1026 & 0.0066 & 0.0128 & 0.0416 & 0.0049 \\
         3 years & 0.1108 & 0.0705 & 0.1520 & 0.0550 & 0.0144 & 0.0672 & 0.0551 \\
         5 years & 0.2338 & 0.1571 & 0.2226 & 0.1075 & 0.0154 & 0.0495 & 0.1147 \\
         7 years & 0.1488 & 0.0963 & 0.2067 & 0.0708 & 0.0595 & 0.0132 & 0.1276 \\
         10 years & 0.1178 & 0.0987 & 0.1579 & 0.0744 & 0.1200 & 0.1035 & 0.0678 \\
         20 years & 0.2732 & 0.2377 & 0.0415 & 0.1587 & 0.2959 & 0.3900 & 0.1342 \\
         30 years & 0.3587 & 0.3541 & 0.1196 & 0.2407 & 0.3780 & 0.5537 & 0.1980 \\
         Mean & 0.1237 & 0.1055 & 0.0968 & 0.0760 & 0.0774 & 0.1033 & 0.0829 \\
         \bottomrule
    \end{tabular*}
\end{table}

\begin{table}[width=.9\linewidth,cols=8,pos=h]
    \caption{In-sample RMSE for DNS-FR model with 4 factors using covariance structure 2.}
    \begin{tabular*}{\tblwidth}{@{} LLLLLLLL @{}}
        \toprule
         Maturity & UK & FR & IT & DE & JP & AU & EU \\
         \midrule
         1 month & 0.0534 & 4.93e-16 & 0.0214 & 4.49e-16 & 0.0001 & 0.0024 & 0.0854 \\
         3 months & 0.0121 & 0.0622 & 0.0027 & 0.0446 & 2.30e-05 & 3.91e-05 & 0.0550 \\
         6 months & 0.0409 & 0.0451 & 0.0644 & 0.0462 & 0.0158 & 0.0001 & 0.0398 \\
         9 months & 0.0186 & 0.0425 & 0.0681 & 0.0476 & 0.0008 & 0.0026 & 0.0349 \\
         1 year & 0.0669 & 0.0090 & 1.11e-14 & 0.0499 & 0.0143 & 0.0095 & 0.0451 \\
         2 years & 1.54e-15 & 0.0612 & 0.0943 & 0.0047 & 0.0132 & 0.0254 & 0.0016 \\
         3 years & 0.0814 & 0.0753 & 0.1279 & 0.0439 & 0.0076 & 0.0426 & 0.0414 \\
         5 years & 0.1624 & 0.1478 & 0.1989 & 0.0739 & 0.0088 & 0.0114 & 0.0815 \\
         7 years & 0.1015 & 0.0971 & 0.1788 & 0.0464 & 0.0382 & 0.0421 & 0.0952 \\
         10 years & 0.1124 & 0.0898 & 0.1430 & 0.0585 & 0.0818 & 0.1441 & 0.0538 \\
         20 years & 0.0476 & 0.2126 & 0.0381 & 0.1168 & 0.2271 & 0.3782 & 0.0873 \\
         30 years & 0.0725 & 0.3117 & 0.1156 & 0.2054 & 0.3360 & 0.5343 & 0.1379 \\
         Mean & 0.0641 & 0.0962 & 0.0877 & 0.0615 & 0.0620 & 0.0994 & 0.0632 \\
         \bottomrule
    \end{tabular*}
\end{table}

\begin{table}[width=.9\linewidth,cols=8,pos=h]
    \caption{In-sample RMSE for DNS-FR model with 5 factors using covariance structure 2.}
    \begin{tabular*}{\tblwidth}{@{} LLLLLLLL @{}}
        \toprule
         Maturity & UK & FR & IT & DE & JP & AU & EU \\
         \midrule
         1 month & 0.0505 & 0.0682 & 0.0208 & 2.39e-15 & 0.0005 & 0.0023 & 0.0832 \\
         3 months & 0.0105 & 0.0676 & 0.0027 & 0.0451 & 1.58e-05 & 3.07e-05 & 0.0518 \\
         6 months & 0.0370 & 0.0234 & 0.0633 & 0.0446 & 0.0151 & 0.0002 & 0.0398 \\
         9 months & 0.0192 & 0.0461 & 0.0673 & 0.0416 & 0.0002 & 0.0026 & 0.0335 \\
         1 year & 0.0629 & 0.0303 & 1.37e-14 & 0.0473 & 0.0143 & 0.0088 & 0.0441 \\
         2 years & 1.48e-15 & 0.0772 & 0.0910 & 0.0055 & 0.0135 & 0.0260 & 0.0013 \\
         3 years & 0.0743 & 8.47e-16 & 0.1122 & 0.0451 & 0.0082 & 0.0449 & 0.0409 \\
         5 years & 0.1618 & 0.1097 & 0.1614 & 0.0677 & 0.0057 & 0.0117 & 0.0824 \\
         7 years & 0.1037 & 0.0400 & 0.1379 & 0.0339 & 0.0311 & 0.0518 & 0.0958 \\
         10 years & 0.1038 & 0.0756 & 0.1230 & 0.0424 & 0.0749 & 0.1256 & 0.0533 \\
         20 years & 0.0490 & 0.1942 & 0.0424 & 0.1068 & 0.2200 & 0.3684 & 0.0810 \\
         30 years & 0.0715 & 0.3599 & 0.0757 & 0.1890 & 0.3343 & 0.5233 & 0.1164 \\
         Mean & 0.0620 & 0.0910 & 0.0748 & 0.0557 & 0.0598 & 0.0971 & 0.0603 \\
         \bottomrule
    \end{tabular*}
\end{table}

\FloatBarrier

\section{In-Sample and Out-of-Sample Moving Window Results for Other Countries/Regions}
\label{app:moving_window}

\begin{figure}[h]
    \centering
    \begin{subfigure}{0.45\textwidth}
        \includegraphics[width=\textwidth]{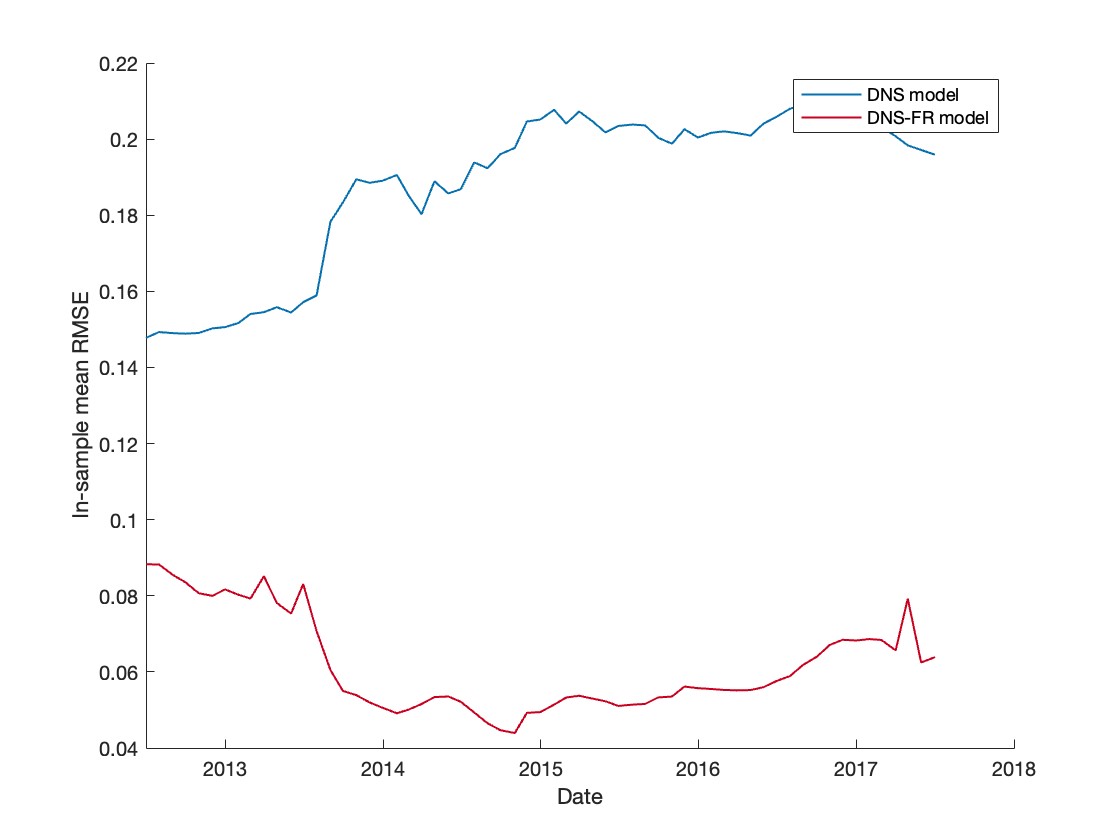}
        \caption{In-sample mean RMSE.}
    \end{subfigure}
    \hfill
    \begin{subfigure}{0.45\textwidth}
        \includegraphics[width=\textwidth]{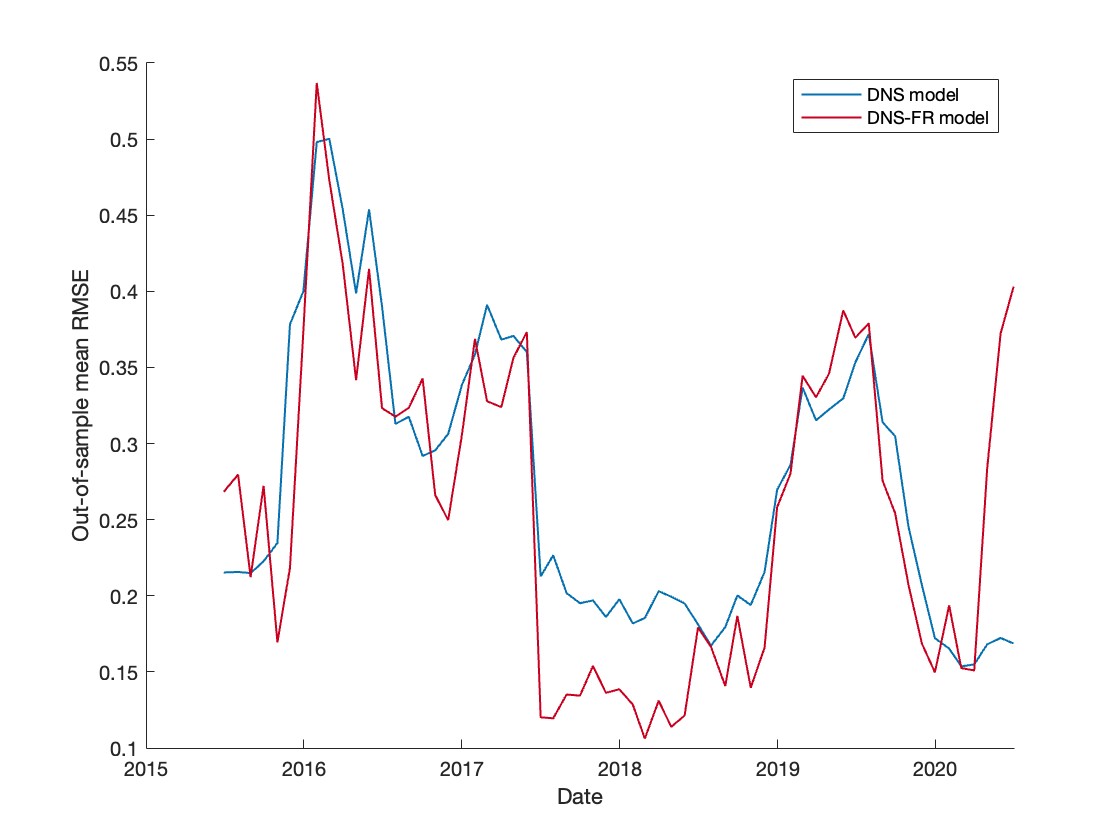}
        \caption{Out-of-sample mean RMSE.}
    \end{subfigure}
    \caption{In-sample and out-of-sample mean RMSE for France yields using a 5-year moving window, move forward for 1 month each time.}
\end{figure}

\begin{figure}[h]
    \centering
    \begin{subfigure}{0.45\textwidth}
        \includegraphics[width=\textwidth]{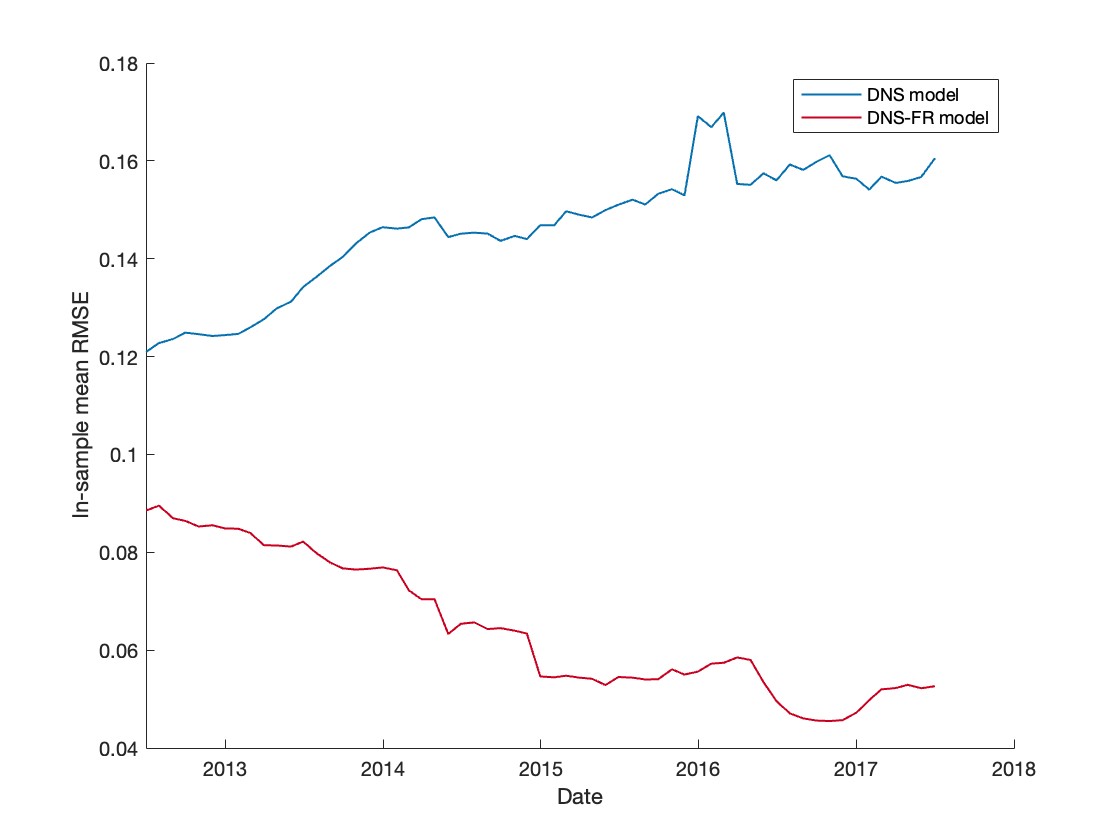}
        \caption{In-sample mean RMSE.}
    \end{subfigure}
    \hfill
    \begin{subfigure}{0.45\textwidth}
        \includegraphics[width=\textwidth]{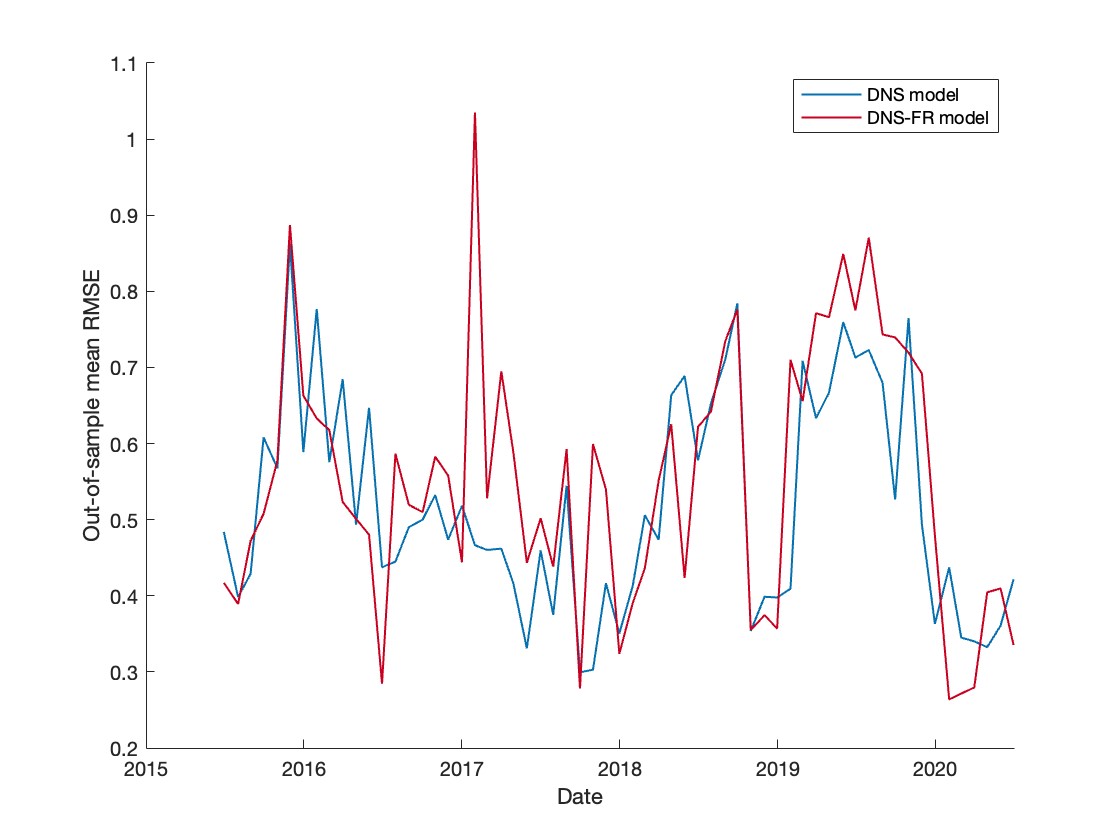}
        \caption{Out-of-sample mean RMSE.}
    \end{subfigure}
    \caption{In-sample and out-of-sample mean RMSE for Italy yields using a 5-year moving window, move forward for 1 month each time.}
\end{figure}

\begin{figure}[h]
    \centering
    \begin{subfigure}{0.45\textwidth}
        \includegraphics[width=\textwidth]{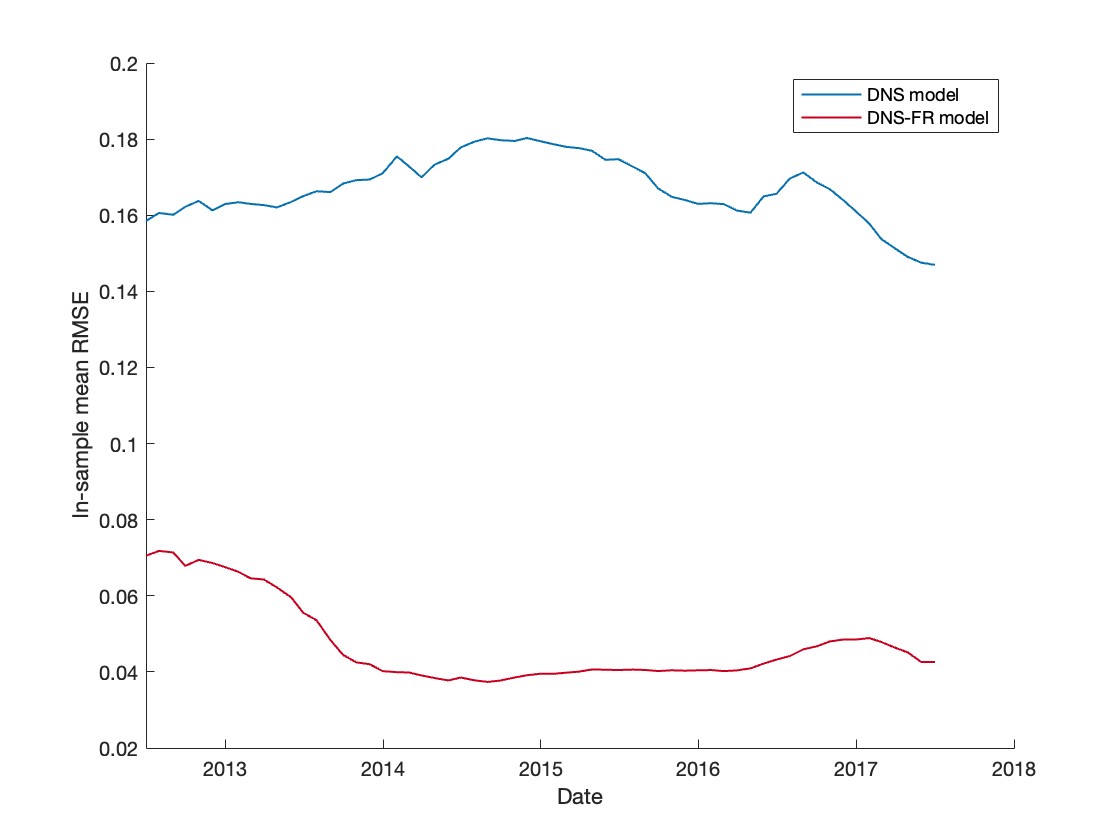}
        \caption{In-sample mean RMSE.}
    \end{subfigure}
    \hfill
    \begin{subfigure}{0.45\textwidth}
        \includegraphics[width=\textwidth]{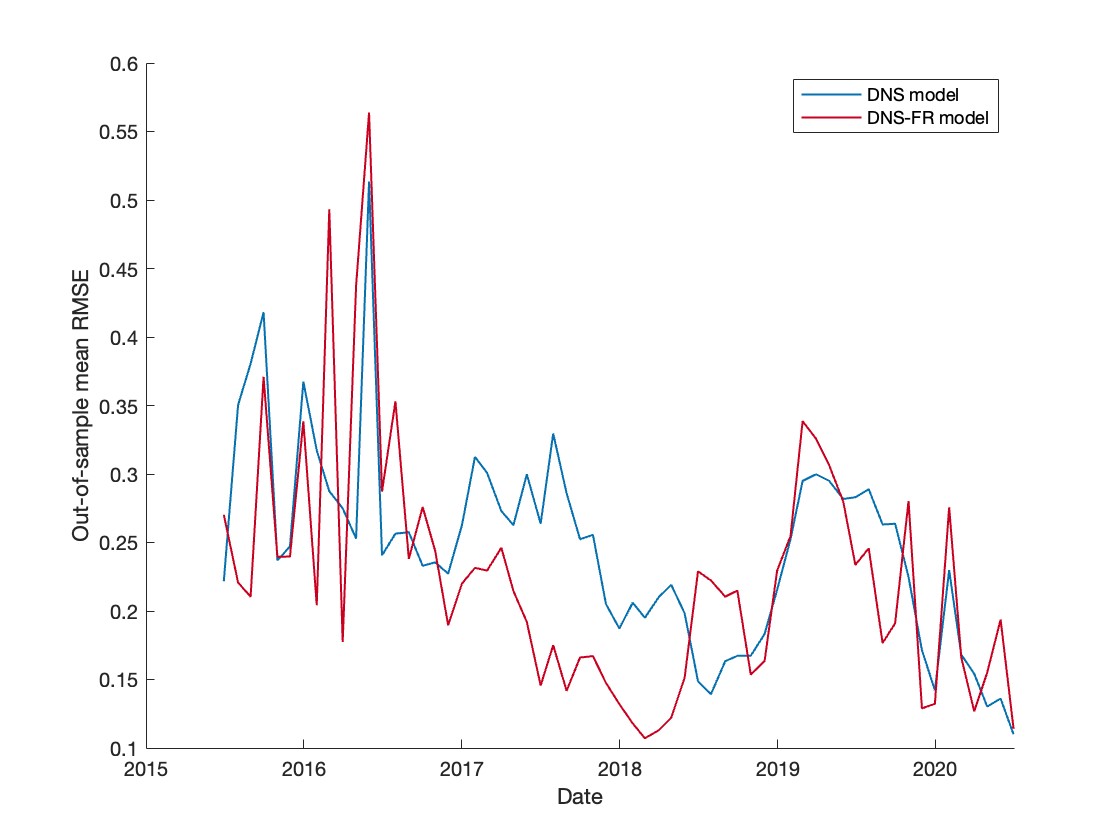}
        \caption{Out-of-sample mean RMSE.}
    \end{subfigure}
    \caption{In-sample and out-of-sample mean RMSE for Germany yields using a 5-year moving window, move forward for 1 month each time.}
\end{figure}

\begin{figure}[h]
    \centering
    \begin{subfigure}{0.45\textwidth}
        \includegraphics[width=\textwidth]{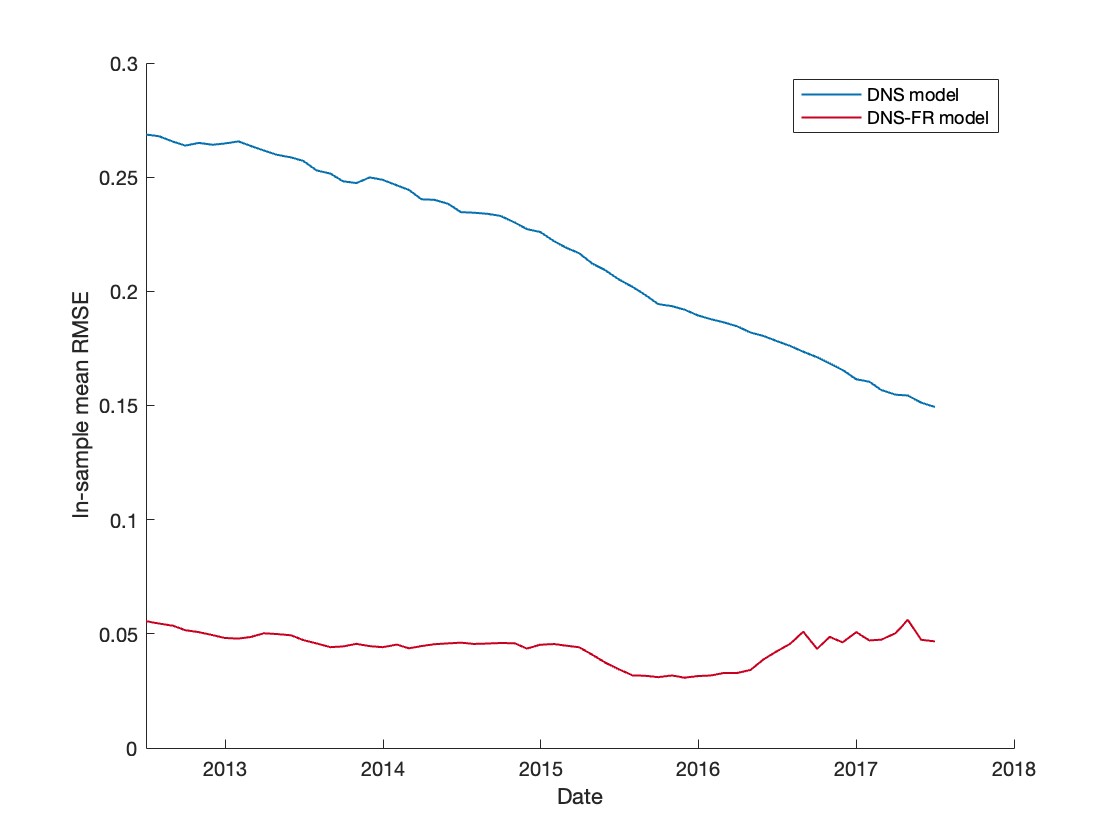}
        \caption{In-sample mean RMSE.}
    \end{subfigure}
    \hfill
    \begin{subfigure}{0.45\textwidth}
        \includegraphics[width=\textwidth]{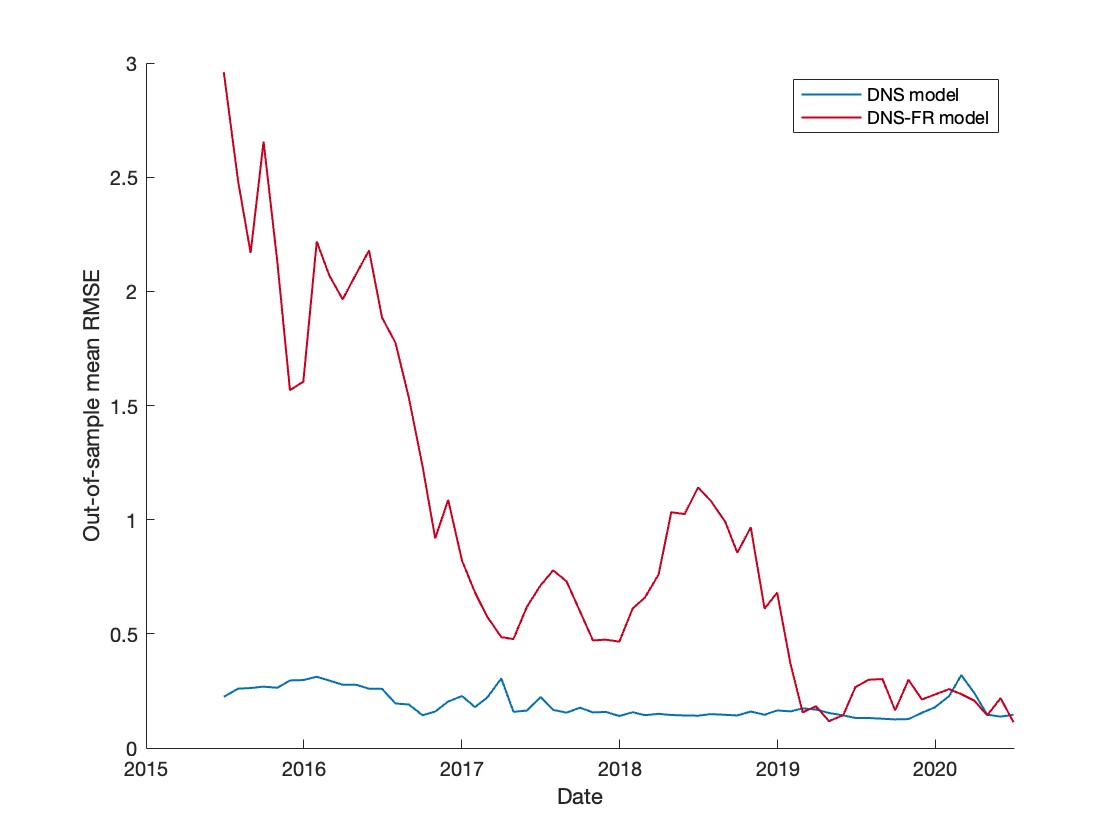}
        \caption{Out-of-sample mean RMSE.}
    \end{subfigure}
    \caption{In-sample and out-of-sample mean RMSE for Japan yields using a 5-year moving window, move forward for 1 month each time.}
\end{figure}

\begin{figure}[h]
    \centering
    \begin{subfigure}{0.45\textwidth}
        \includegraphics[width=\textwidth]{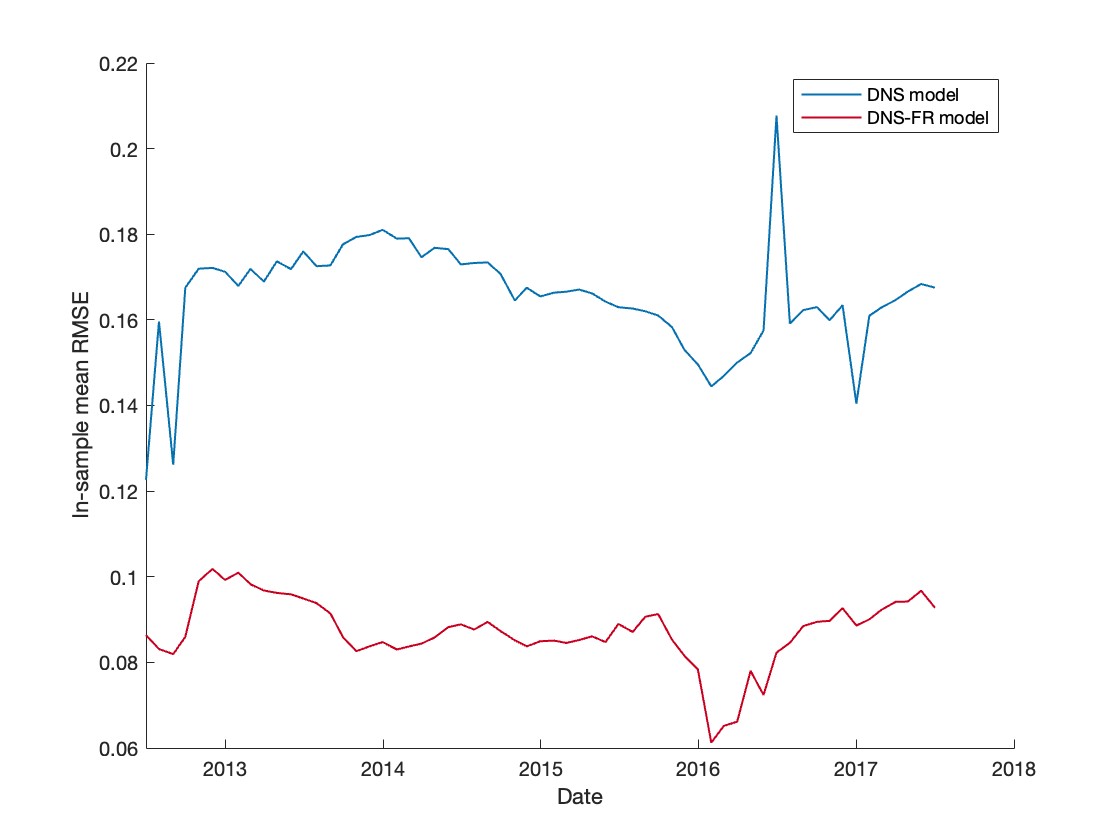}
        \caption{In-sample mean RMSE.}
    \end{subfigure}
    \hfill
    \begin{subfigure}{0.45\textwidth}
        \includegraphics[width=\textwidth]{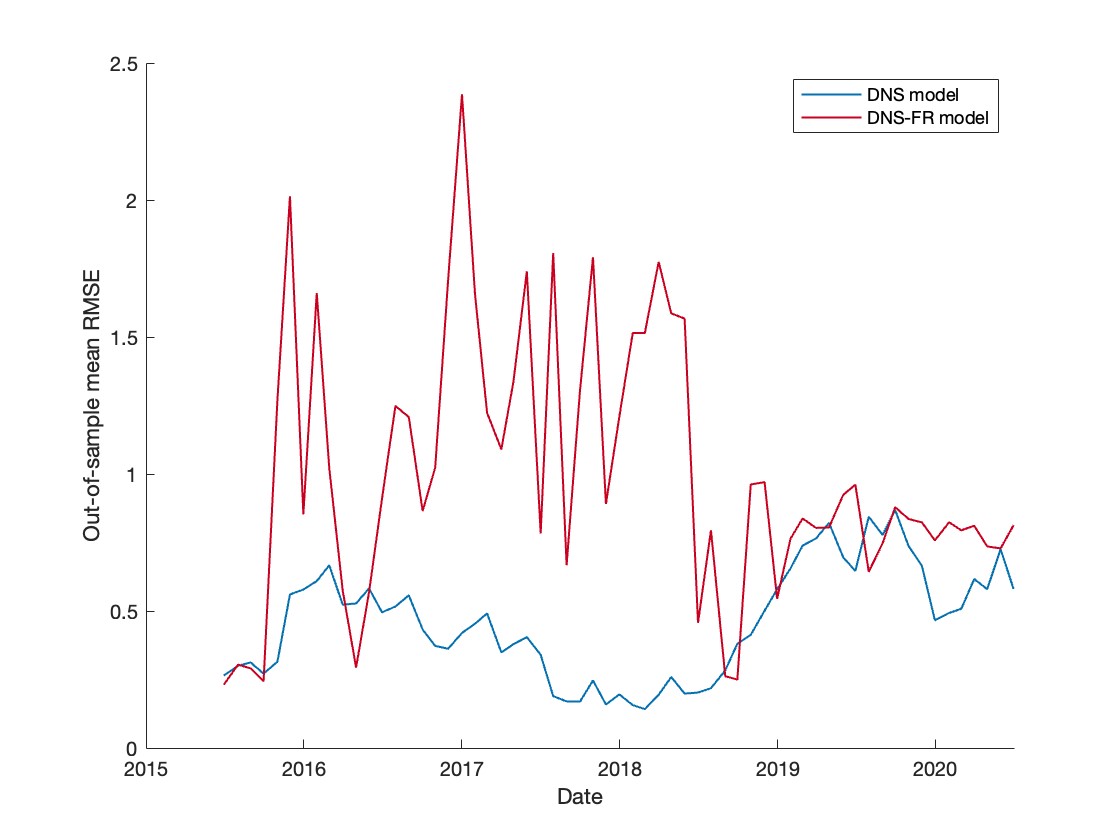}
        \caption{Out-of-sample mean RMSE.}
    \end{subfigure}
    \caption{In-sample and out-of-sample mean RMSE for Australia yields using a 5-year moving window, move forward for 1 month each time.}
\end{figure}

\begin{figure}[h]
    \centering
    \begin{subfigure}{0.45\textwidth}
        \includegraphics[width=\textwidth]{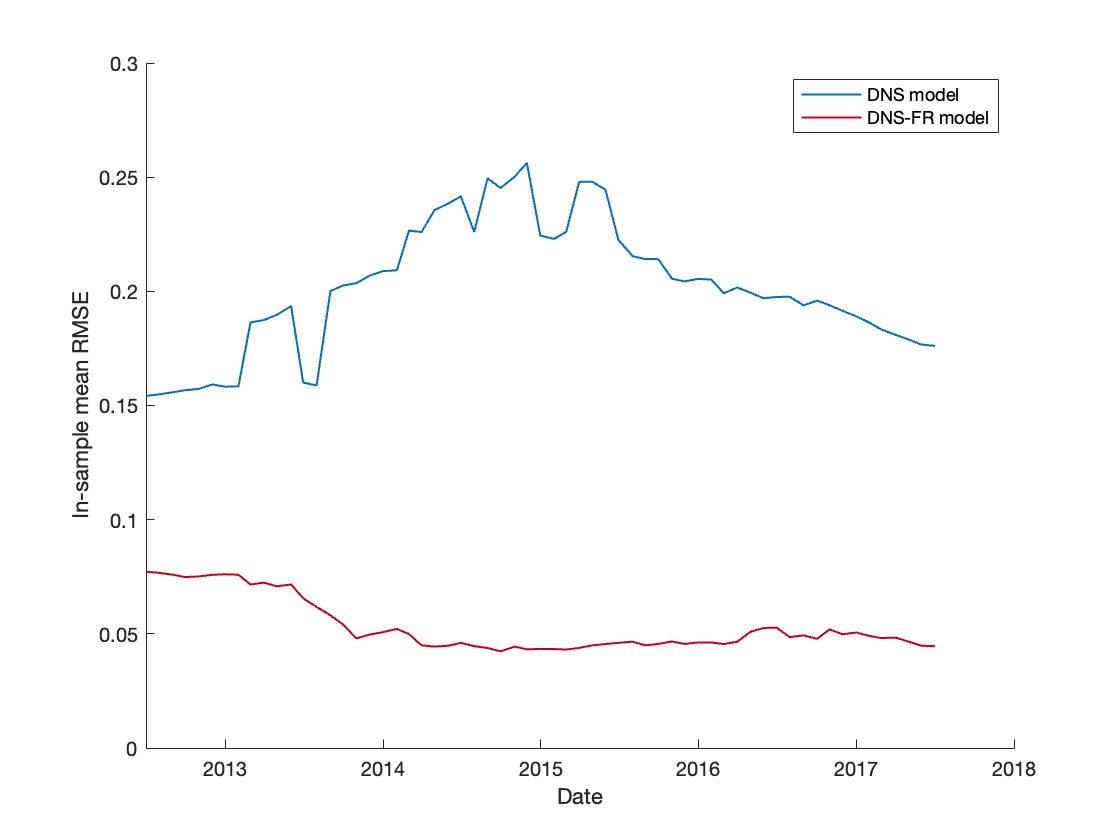}
        \caption{In-sample mean RMSE.}
    \end{subfigure}
    \hfill
    \begin{subfigure}{0.45\textwidth}
        \includegraphics[width=\textwidth]{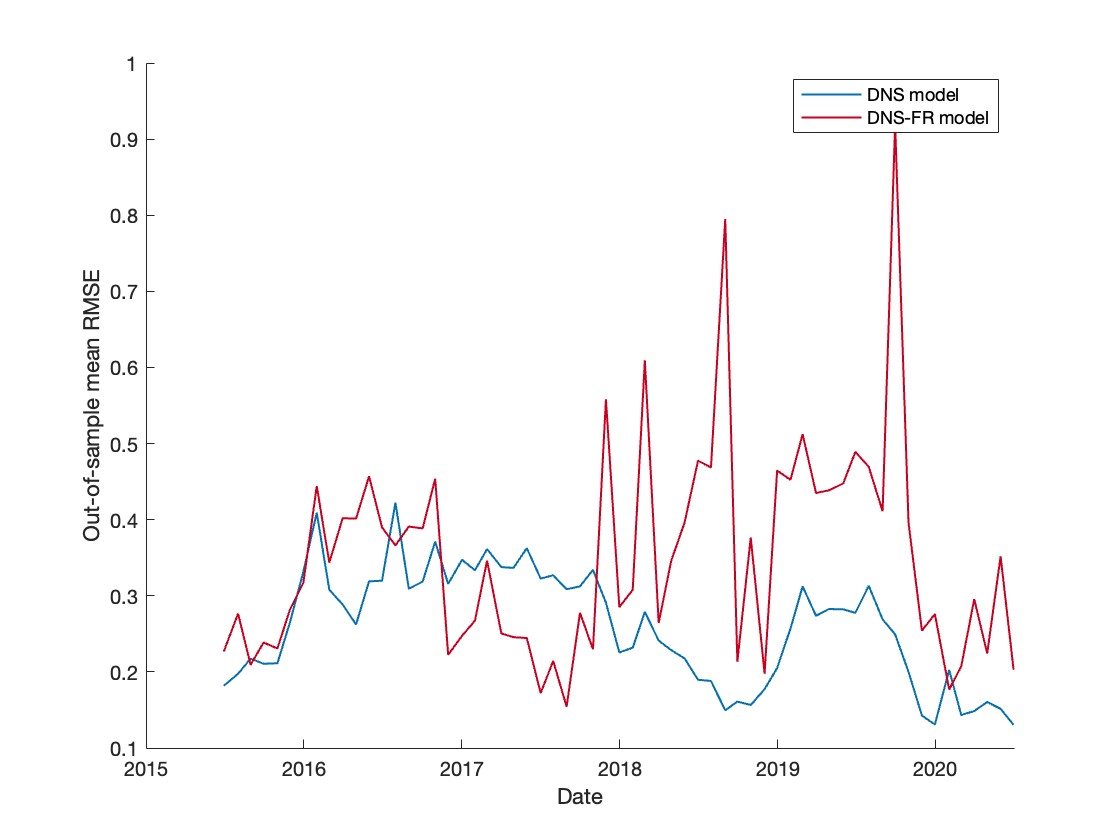}
        \caption{Out-of-sample mean RMSE.}
    \end{subfigure}
    \caption{In-sample and out-of-sample mean RMSE for EU yields using a 5-year moving window, move forward for 1 month each time.}
\end{figure} 

\FloatBarrier

\section{Time Series of US Treasury Yields for Different Stress Testing Scenarios}
\label{app:time_series_st}

\begin{figure}[h]
    \centering
    \includegraphics[width=0.6\textwidth]{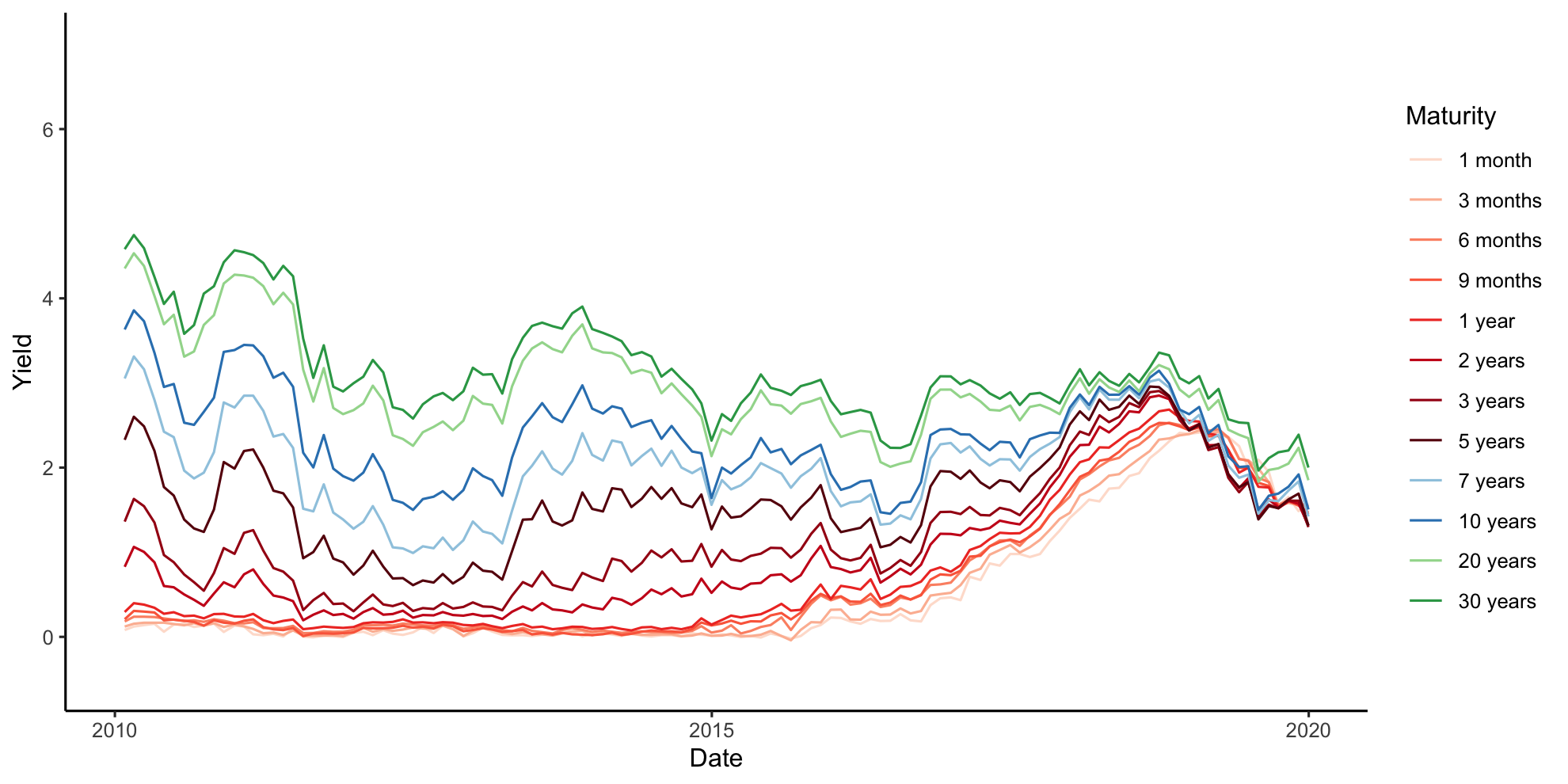}
    \caption{Time series of original US Treasury bond yields. }
\end{figure}

\begin{figure}[h]
    \centering
    \begin{subfigure}{0.45\textwidth}
        \includegraphics[width=\textwidth]{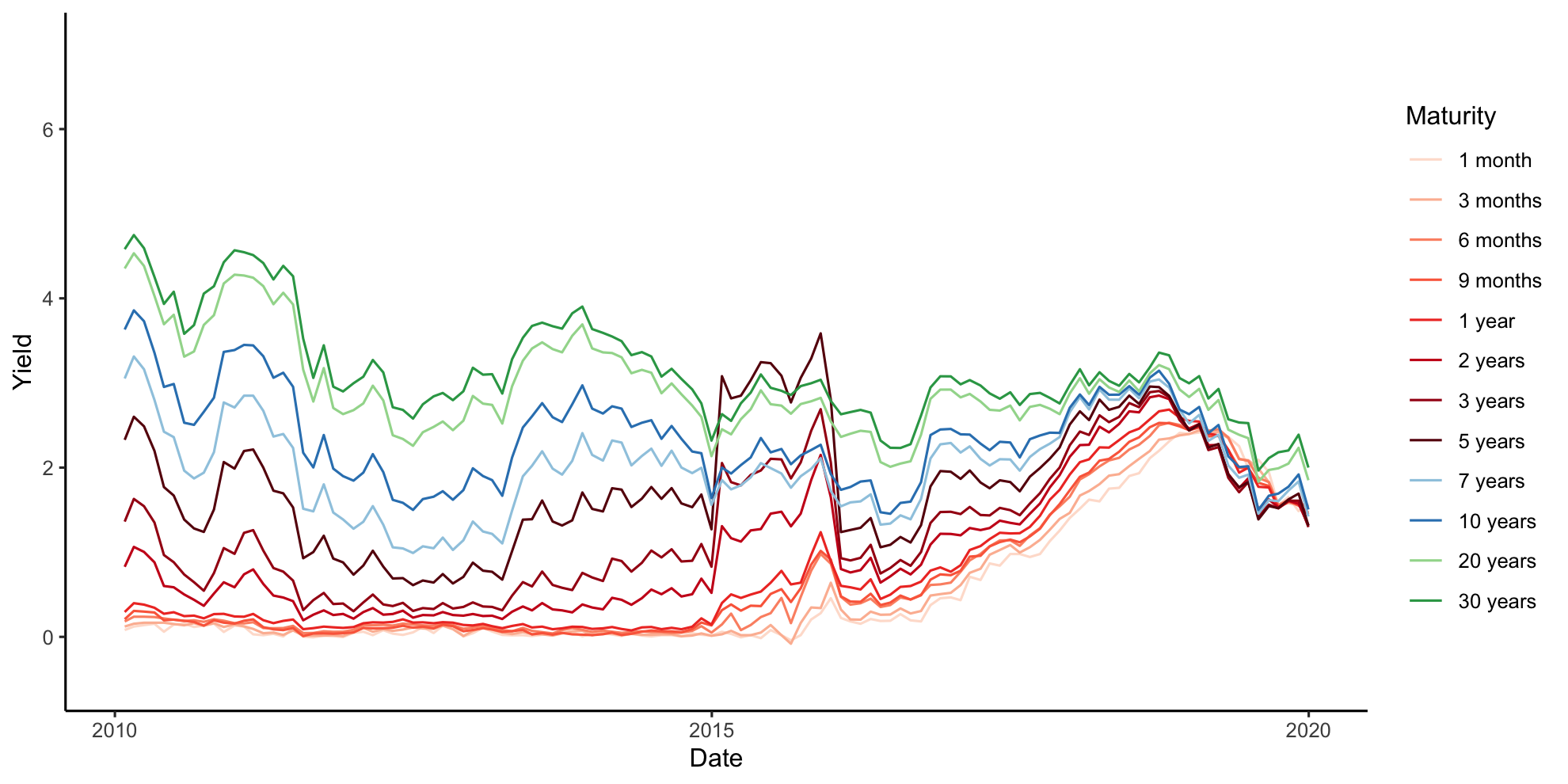}
        \caption{Case 1.1}
    \end{subfigure}
    \hfill
    \begin{subfigure}{0.45\textwidth}
        \includegraphics[width=\textwidth]{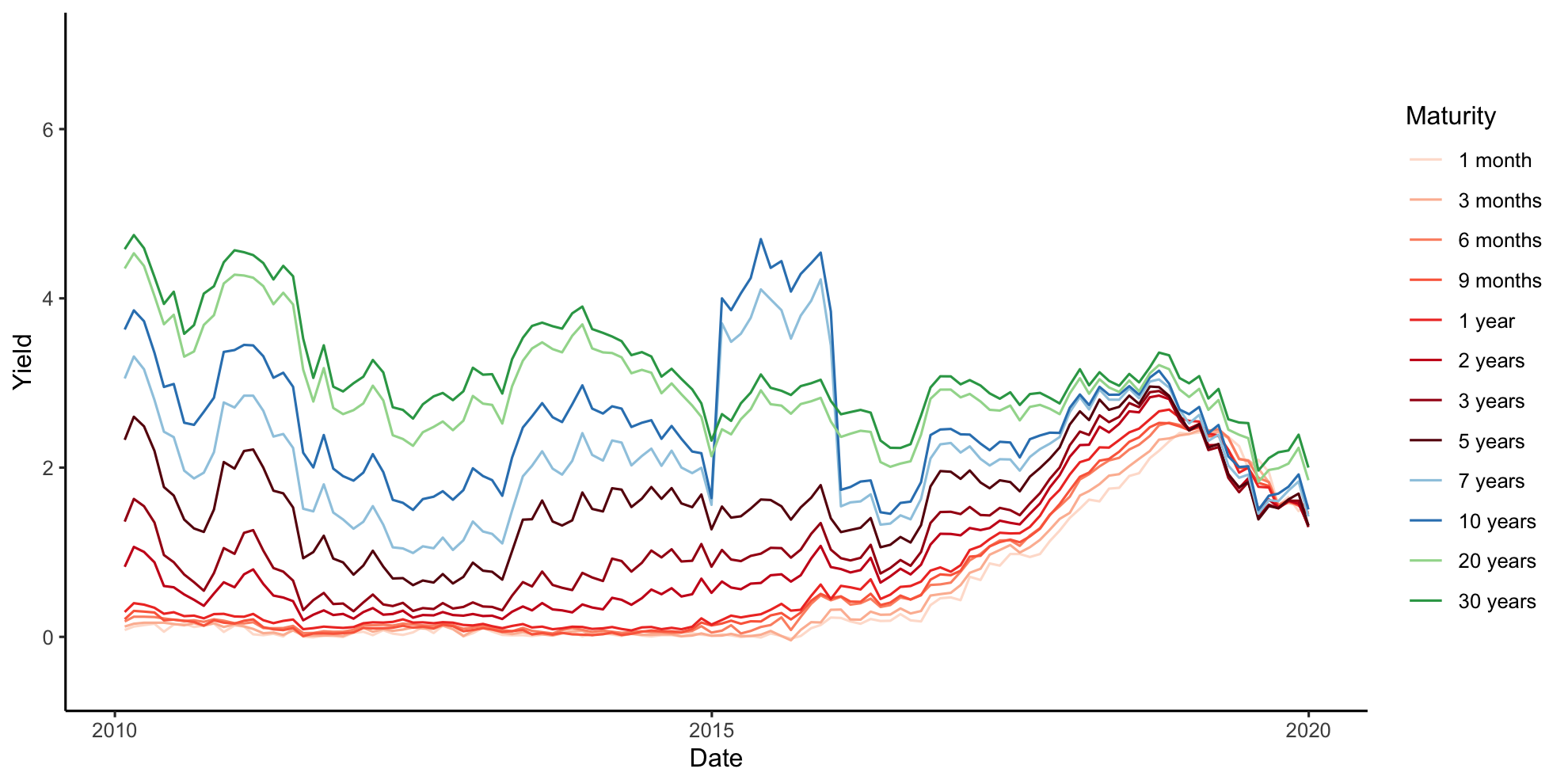}
        \caption{Case 1.2}
    \end{subfigure}
    \hfill
    \begin{subfigure}{0.45\textwidth}
        \includegraphics[width=\textwidth]{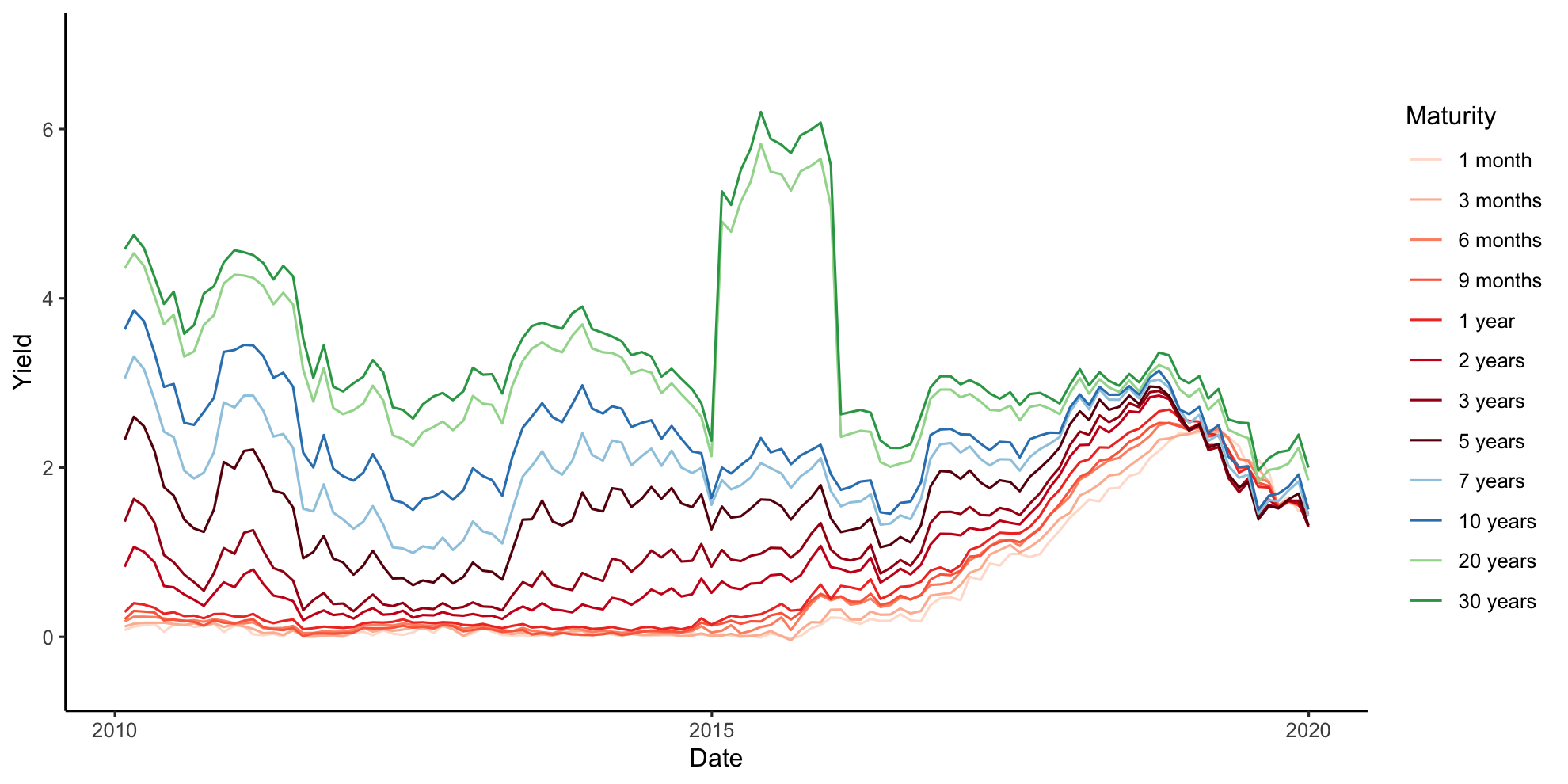}
        \caption{Case 1.3}
    \end{subfigure}
    \hfill
    \begin{subfigure}{0.45\textwidth}
        \includegraphics[width=\textwidth]{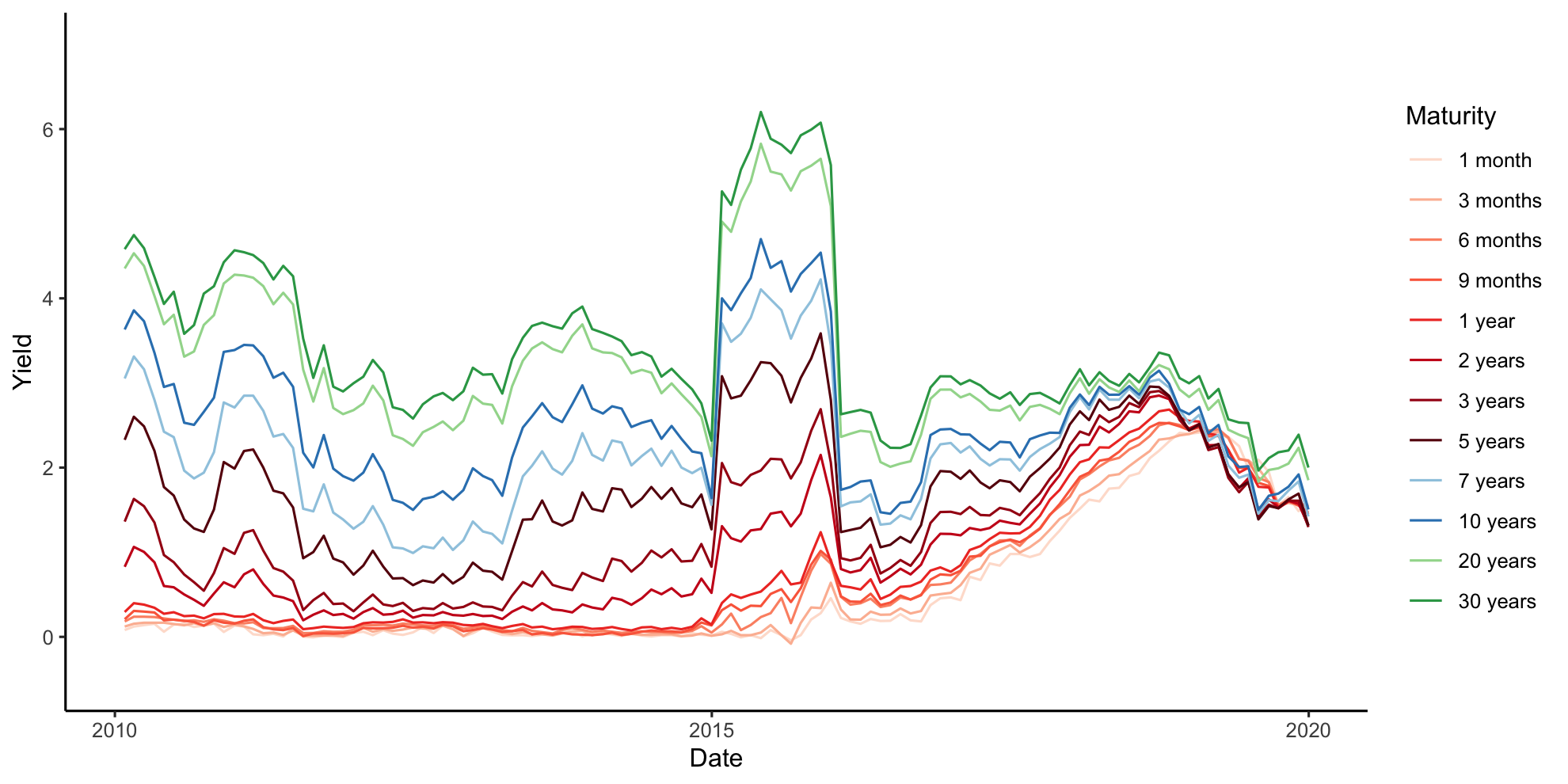}
        \caption{Case 1.4}
    \end{subfigure}      
    \caption{Time series of US Treasury bond yields for stress testing scenario 1.}
\end{figure}

\begin{figure}[h]
    \centering
    \begin{subfigure}{0.45\textwidth}
        \includegraphics[width=\textwidth]{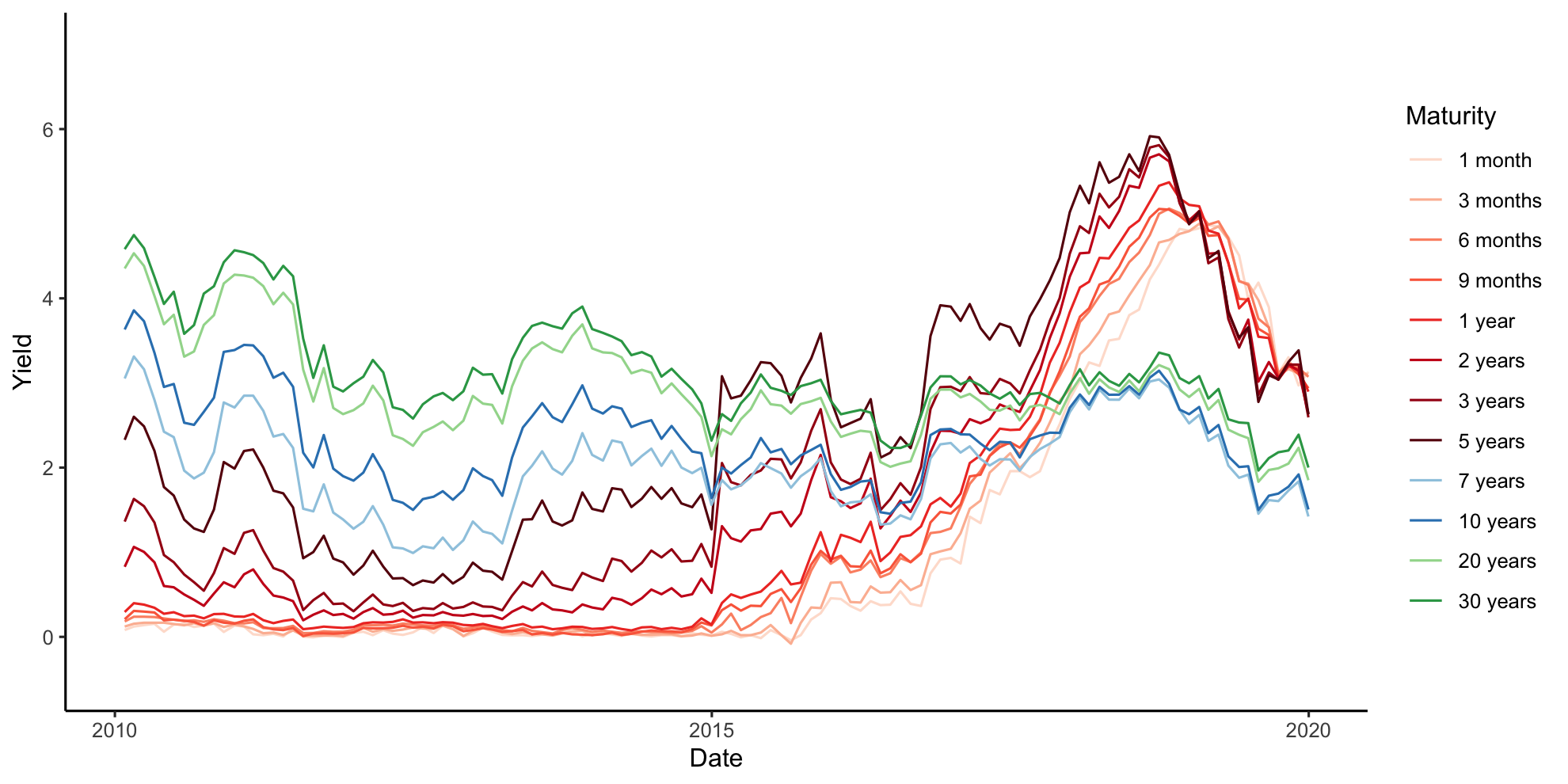}
        \caption{Case 2.1}
    \end{subfigure}
    \hfill
    \begin{subfigure}{0.45\textwidth}
        \includegraphics[width=\textwidth]{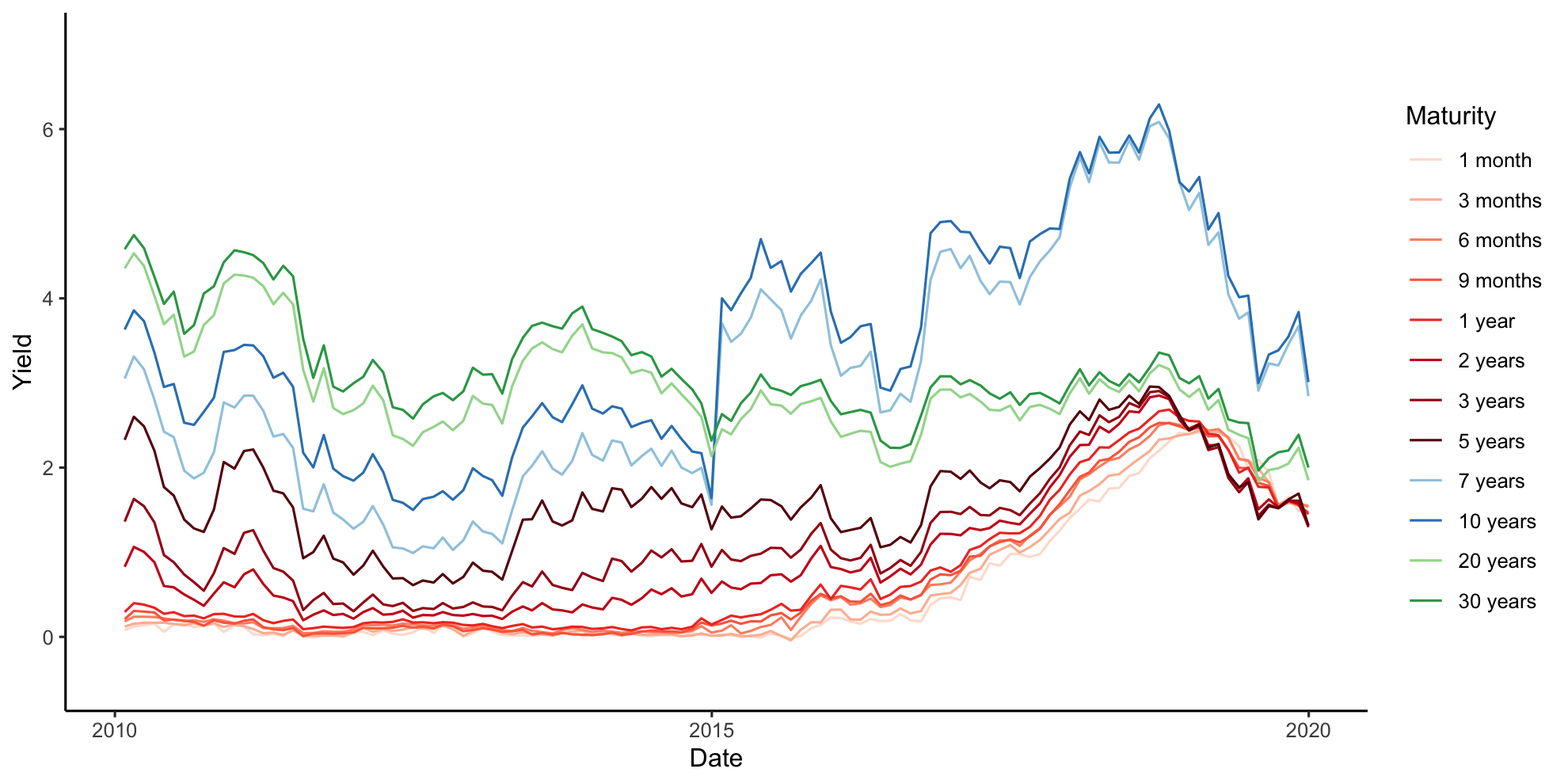}
        \caption{Case 2.2}
    \end{subfigure}
    \hfill
    \begin{subfigure}{0.45\textwidth}
        \includegraphics[width=\textwidth]{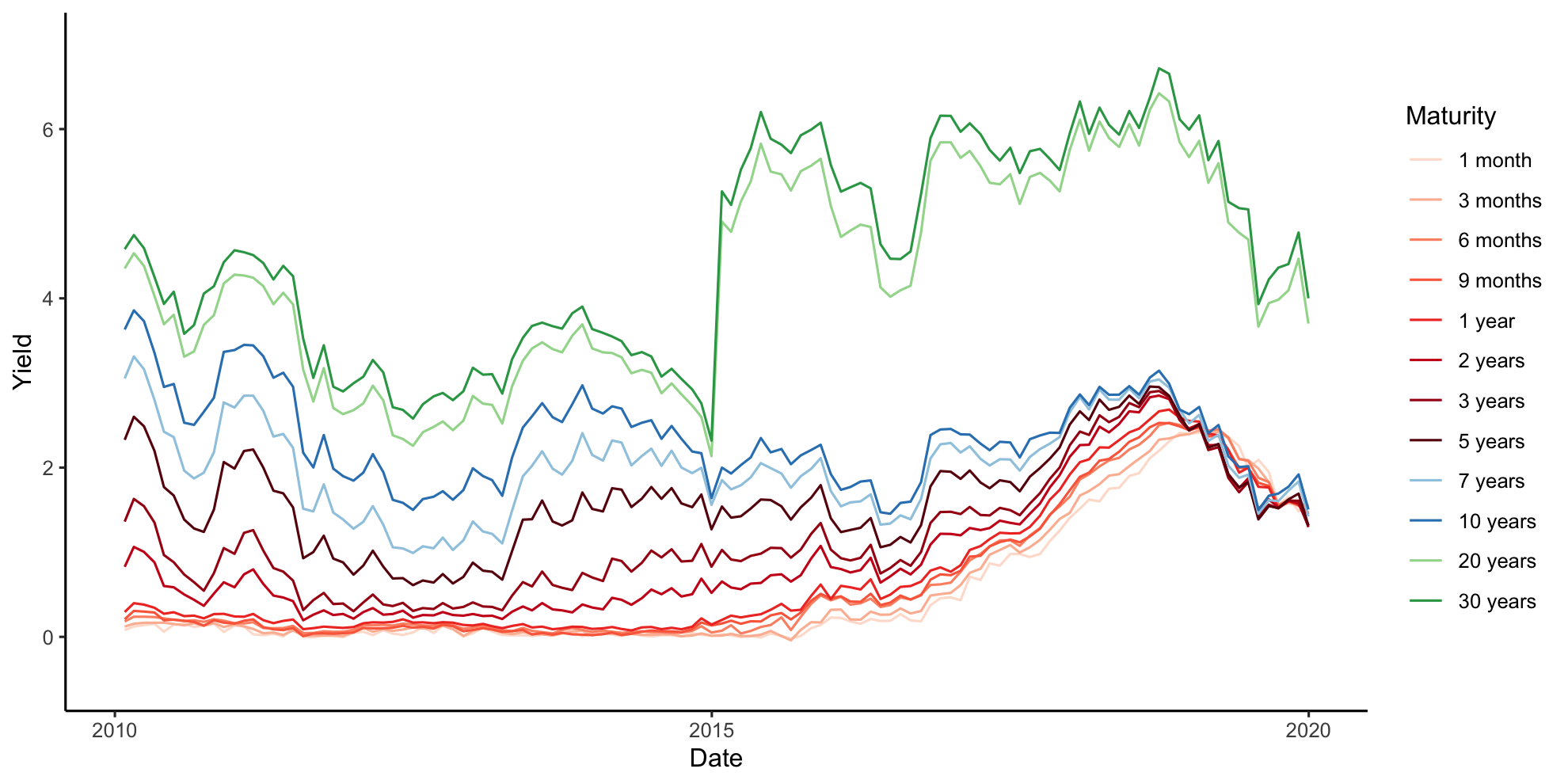}
        \caption{Case 2.3}
    \end{subfigure}
    \hfill
    \begin{subfigure}{0.45\textwidth}
        \includegraphics[width=\textwidth]{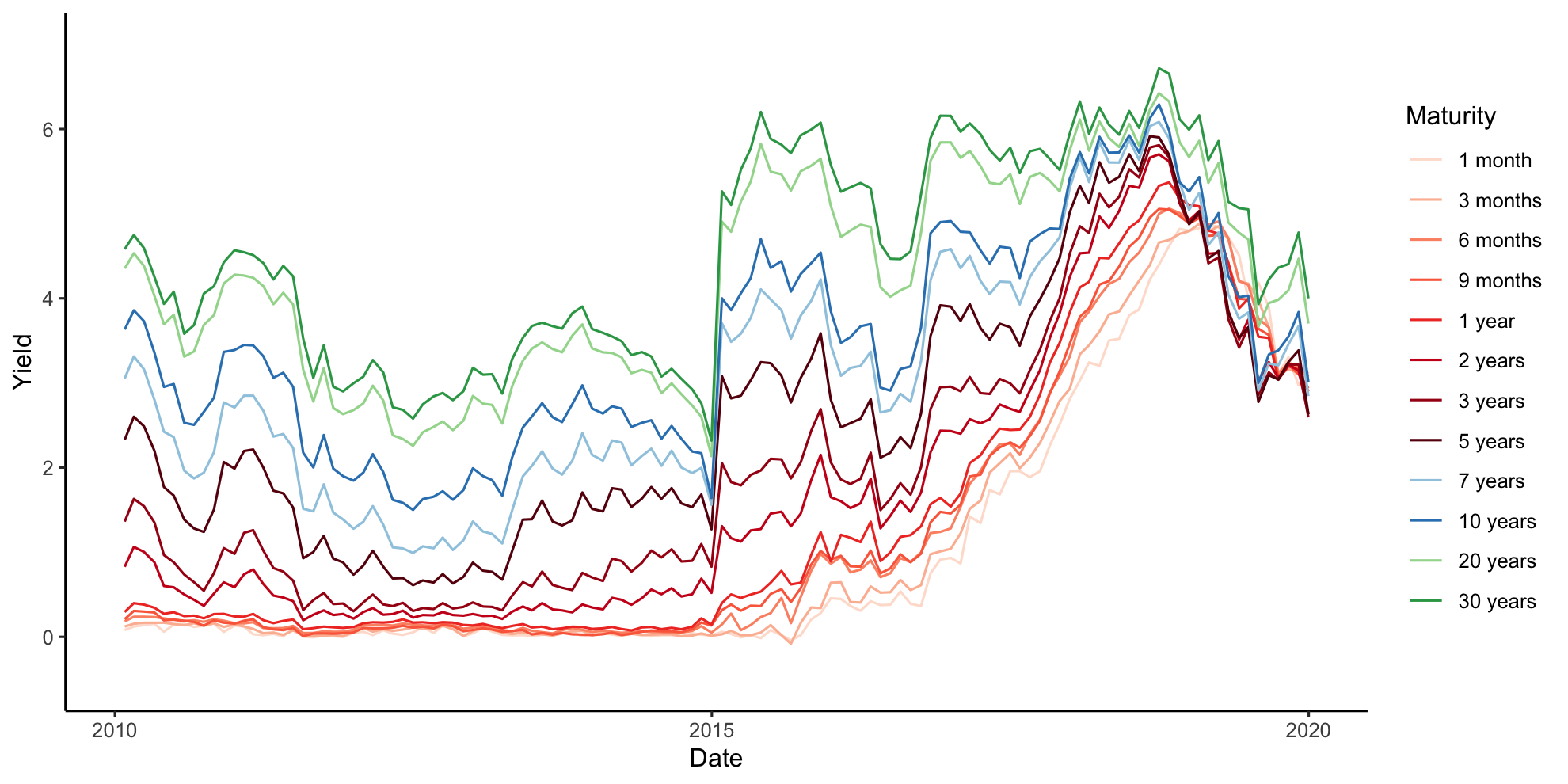}
        \caption{Case 2.4}
    \end{subfigure}      
    \caption{Time series of US Treasury bond yields for stress testing scenario 2.}
\end{figure}

\FloatBarrier

\printcredits

\section*{Acknowledgements}

This study was partially presented at the 68th Euro Working Group for Commodity and Financial Modelling. We would like to thank all the audiences and organisers for their
valuable feedback and suggestions. 

\section*{Declaration of generative AI and AI-assisted technologies in the writing process}

While preparing this work, the authors used ChatGPT to improve the manuscript's readability and language. After that, the authors thoroughly reviewed and edited the content, taking full responsibility for the manuscript's content.

\bibliographystyle{vancouver}

\bibliography{references}


\end{document}